\documentclass[11pt]{article}
\pdfoutput=1

\usepackage{jheppub}
\usepackage[T1]{fontenc}
\usepackage[normalem]{ulem}
\usepackage{jheppub}
\usepackage{url,comment}
\usepackage{times}
\usepackage{latexsym}
\usepackage{graphicx, graphics, hyperref, amsmath, amssymb, slashed, color, bbm,amsthm, array}
\usepackage[usenames,dvipsnames]{xcolor}
 \usepackage{subfigure}
\usepackage{mdwlist}
\usepackage{multirow}
\usepackage[e]{esvect}

\usepackage{cancel}
\usepackage{pstricks}
\usepackage[toc,page]{appendix}
\usepackage{enumerate}

\theoremstyle{definition}

\newcommand{\dd}{\mathrm{d}}

\subheader{\hfill \rm KEK-TH-2435}

\title{First Evaluation of Meson and $\tau$ lepton Spectra and Search for Heavy Neutral Leptons at ILC Beam Dump}
\author{Mihoko M. Nojiri$^{1,2,3}$, Yasuhito Sakaki$^{4,2}$, Kohsaku Tobioka$^{1,5}$, and Daiki Ueda$^6$}
\affiliation{\vspace{2mm} $^1$Theory Center, High Energy Accelerator Research Organization (KEK), 1-1 Oho, Tsukuba, Ibaraki
305-0801, Japan \\
$^2$The Graduate University of Advanced Studies (Sokendai), 1-1 Oho, Tsukuba, Ibaraki 305-0801, Japan \\
$^3$Kavli IPMU (WPI), University of Tokyo, 5-1-5 Kashiwanoha, Kashiwa, Chiba 277-8583, Japan \\
$^4$ Radiation Science Center, High Energy Accelerator Research Organization (KEK), Ibaraki
305-0801, Japan
$^5$Department of Physics, Florida State University, 77 Chieftan Way, Tallahassee, FL 32306, USA \\
$^6$Center for High Energy Physics, Peking University, Yiheyuan Road, Beijing 100871, China}
\emailAdd{nojiri@post.kek.jp}
\emailAdd{sakakiy@post.kek.jp}
\emailAdd{ktobioka@fsu.edu}
\emailAdd{ueda@pku.edu.cn}

\abstract{
A beam dump experiment can be seamlessly added to the {proposed} International Linear Collider (ILC)  program because the high energy electron beam should be dumped after the collision point. The ILC beam dump experiment will provide an excellent opportunity to search for new long-lived particles. Since many of them can be produced by a rare decay of standard model particles, we evaluate spectra of the mesons and $\tau$ lepton at the decay based on the PHITS and PYTHIA8 simulations. As a motivated physics case, we study the projected sensitivity of heavy neutral leptons at the ILC beam dump experiment. The heavy neutral leptons can also be produced via deep inelastic scattering and $Z$ boson decay at the ILC main detector, which we include in the projection.  With the multi-track signal, the reach would be greatly extended in mass and coupling, even compared with the other proposed searches.
}
\begin{document}

\maketitle

\section{Introduction}\label{sec:intro}

The large hadron collider (LHC) has discovered the  Higgs boson and measured its properties consistent with the precision measurements of the Standard Model (SM). The discovery of the last particle of the standard model strengthened the foundation of the SM. However, overwhelming evidence and hints require physics beyond the Standard Model (BSM). While the intensive searches for new particles at the weak scale or heavier have been performed experimentally, we have not found any convincing evidence of such new particles yet. In these circumstances, there is growing attention to feebly interacting light particles from both theoretical and experimental points of view.

New light particles with mass less than tens of GeV are compelling addition to the SM as they can directly resolve the central issues of the SM, such as the strong CP problem and the finite neutrino masses,  explain  many flavor physics anomalies, and be a portal to the dark sector accommodating the dark matter and the baryon asymmetry  of the universe. On the experimental grounds, the light particles, if they exist, need to be feebly interacting with the SM sector due to various experimental constraints. However, one should be aware that the bounds are strongly model-dependent because the signature depends on the nature of the light degrees of freedom. This situation is contrasted with heavy new particle exploration with mass beyond the reach of current collider energies, which can be probed through  Effective Field Theories~(EFT). Therefore, there is growing interest in various scenarios based on the current experimental data and the potential for future high-intensity experiments. The status of recent progress is well-summarized in the report of {\it Physics Beyond Collider}~\cite{Beacham:2019nyx}.

In this paper, we study the physics potential of the dump facilities of the International Linear Collider (ILC) to hunt new light particles. The ILC is a proposed future $e^+e^-$ linear collider with beam-energy of 125~GeV, which can be upgraded to 500~GeV. The main goal of the ILC is to study events of $e^+e^-$ collision to perform high precision measurements of the Higgs bosons and search for new particles produced by the electroweak interaction.   In the linear collider, the beam has to go into the beam dump after the collision, and it provides an excellent opportunity for a high-intensity experiment to produce feebly interacting light particles. The number of electrons on the beam dump is enormous, $N_{\rm EOT}=4\times 10^{21}/\rm$~per~year, and the electromagnetic (EM) shower also leads to a high-intensity photon carrying $\cal O$(10\%) of the beam energy. The setup is equivalent to the other fixed target experiments if a detector is placed downstream of the beam dump. Hereafter we refer to such setup as {\it the ILC beam dump experiment}. One can search for light particles produced in the dump, fly beyond the muon shield, then decay to the SM particles or scatter the detector material. The potential of the ILC beam dump project has been investigated in Refs~\cite{Izaguirre:2013uxa,Kanemura:2015cxa,Sakaki:2020mqb,Asai:2021ehn,Asai:2021xtg,Moroi:2022qwz}.

In the previous works of the electron beam dump experiments, the searches for light particles  dominantly interacting with electron or photon were mainly studied since they are produced by  the primary electron and positron or the secondary shower particles. In the ILC beam dump experiment, the initial beam energy is much higher than those of the past  electron beam dump experiments. In addition to the EM showers, heavy mesons and $\tau$ lepton can be produced by the shower photon  hitting nuclei. The decay of the produced SM particles is another promising source of light particles. This paper examines the production yield and spectrum of the mesons and $\tau$ lepton at the ILC beam dump setup for the first time.

As an application of this study, we investigate a projected sensitivity of heavy neutral leptons  (HNLs, called sterile neutrinos in some literature). The HNLs mix with the SM neutrinos with an angle of $U_\ell \ (\ell=e,\mu, \tau)$, resulting in a suppressed weak interaction to the SM. Therefore, it is very natural to consider the HNL production from the meson and $\tau$ lepton decays where  the weak interaction dominates. The current experimental constraints are mostly for mixing with $\nu_e$ and $\nu_\mu$, and the high intensity of mesons would further reach the under-explored parameter space. Also the sensitivity to the $\tau$ neutrino mixing angle $U_\tau$ can be significantly improved because $\tau$ lepton is accessible.

The phenomenology of the HNLs is well-reviewed in Refs~\cite{Alekhin:2015byh, Dasgupta:2021ies}, and see references therein. We stress that feebly interacting HNLs in a GeV mass range are a motivated and well-defined physics target. The seesaw mechanism can explain the neutrino masses observed by the  neutrino oscillations with at least two HNLs.  If the two HNLs are almost degenerate,  the sum of the mixing angles have the lower bound, approximately $U^2\equiv \sum_l |U_l|^2\gtrsim m_{\rm atm}/m_N \sim 10^{-11} (m_N/{\rm 1~GeV})^{-1}$~\cite{Shaposhnikov:2006nn,Alekhin:2015byh}. Furthermore, the degenerate HNLs in the early universe can produce the baryon asymmetry via the HNL oscillations~\cite{Akhmedov:1998qx, Asaka:2005pn}. The feeble interactions are necessary for the departure of the thermal equilibrium which is one of Sakharov's conditions for generating the baryon asymmetry. Together with the neutrino mass constraints, the interesting parameter space is in a range of $10^{-11}\lesssim U^2\lesssim10^{-6}$.~\footnote{The benchmark values depend on the number of HNLs involved in the seesaw mechanism. Three or more HNLs will allow a smaller value of $|U_l|^2$. However, it is important to examine the parameter space of the minimal scenarios.}

This paper is organized as follows. In Sec.~\ref{sec:dumpsetup}, we briefly review the ILC beam dump experiment, and in the following section, we evaluate the spectra of mesons and $\tau$ lepton. In Sec.~\ref{sec:HNL}, we study the HNLs at the ILC beam dump experiment. In addition to  the HNL production from the SM particle decays, we  consider an HNL  production via deep inelastic scattering and another production from $Z$ decays at the interaction point.  We finish with the discussion in Sec.~\ref{sec:discussion}.

\section{ILC Beam Dump Experiment}\label{sec:dumpsetup}
\begin{figure}
\centering
\includegraphics[width=0.95\textwidth]{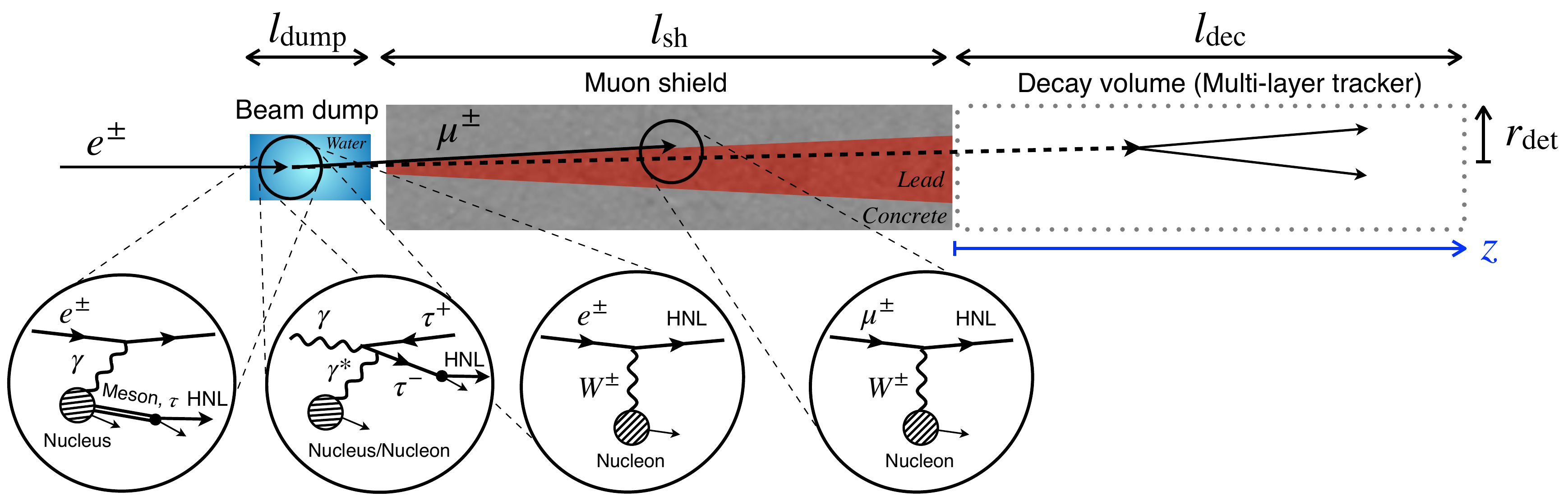}
\caption{A setup for ILC beam dump experiments. It consists of the main beam dump, a muon shield, and a decay volume. We assume a multi-layer tracker is placed in the decay volume so that the charged tracks are measured.}
\label{fig:exp}
\end{figure}

The ILC  main beam dump has to absorb 2.6~MW ($125~{\rm GeV}\times 21~{\rm \mu A}$) of the $e^\pm$ beam energy for 125~GeV in the initial stage and 13.6~MW ($500~{\rm GeV}\times 27.3~{\rm \mu A}$) for 500~GeV in an upgrade stage. Following a water dump designed with the length of $l_{\rm dump}=11~{\rm m}$ and the diameter of 1.8~m, a muon shield of length $l_{\rm sh}=70~{\rm m}$ would be placed as proposed in \cite{Sakaki:2020mqb,Asai:2021ehn}, to remove the secondary muon background. The cylindrical decay volume of length $l_{\rm dec}=50~{\rm m}$ and radius $r_{\rm det}=3$~m would lie between the muon shield and the downstream detector. The schematic view is seen in Fig.~\ref{fig:exp}, and the similar design of the setup can be found in \cite{Sakaki:2020mqb,Asai:2021ehn}. Here, we additionally assume a multi-layer tracker in the decay volume. We consider a time frame of 10-year run  for both ILC-250 and ILC-1000 where the  beam energy is 125~GeV and 500~GeV, respectively.

The EM shower (photons, electrons, and positions) starts in the beam dump. The ILC beam dump experiment is unique compared to the past electron beam dump experiments with regards to the higher beam energy and the high intensity, i.e., the number of electrons on the beam dump  $N_{\rm EOT}$ is about $4\times 10^{21}$ per year\footnote{In ILC-1000, we assume to be $N_{\rm EOT}=4\times 10^{21}$ per year in numerical calculations.}.
The energetic beam  creates photons at $\cal O$(1-10)~GeV energy scale (Fig.~9 of \cite{Asai:2021ehn}), and the secondary interaction between the photon and the nucleus can  produce light mesons ($\pi$ and $K$), heavy mesons ($D$, $B$, and even $B_c$) and $\tau$ lepton.
They loose the energy in the beam dump and finally decay, and some of them produce the HNLs.  The yield and energy spectrum of the SM particles at the decay is therefore important to study the sensitivity of the ILC beam dump experiment.
We show the evaluation in Sec.~\ref{sec:spectrum}.

The secondary muons are stopped in the muon shield at ILC-250, but their penetration behind the shield cannot be neglected for $E_{\rm beam}=500$~GeV at ILC-1000. In this case, an additional active muon shield behind the muon shield would be necessary, and we assume that the muon shield consists of the lead shield ($l^{\rm lead}_{\rm sh}=10$~m) and the active shield ($l^{\rm active}_{\rm sh}=70~{\rm m}-l^{\rm lead}_{\rm sh}$). The HNL with a dominant mixing with the muon neutrino can be produced inside the muon shield by  scattering of the shower muon, and we approximate that the HNLs produced behind the lead shield do not contribute to the signal events. In Appendix~\ref{app:muonshield}, we study how  different depth of the muon shield affects the sensitivity for the HNL at ILC-1000.

\section{Meson and $\tau$ lepton Spectra}\label{sec:spectrum}

This section presents the meson and $\tau$ lepton spectrum obtained by Monte-Carlo simulation at the ILC beam dump. We use PHITS~3.25~\cite{Sato:2018imy} for production and transport of particles other than heavy mesons. PHITS (Particle and Heavy Ion Transport code System) is a general-purpose Monte Carlo particle transport simulation code developed under collaboration between JAEA, RIST, KEK, and several other institutes. PHITS can transport most particle species for a given geometry of the materials, and it is tested thoroughly by benchmarks studies~\cite{iwamoto2017benchmark,matsuda2011benchmarking}. For heavy mesons production, we implement their differential production cross sections obtained by PYTHIA8.3~\cite{Bierlich:2022pfr} into PHITS. More details are given in the following.

\subsection{Light Mesons}

\begin{figure*}
\centering 
\includegraphics[width=0.49\textwidth]{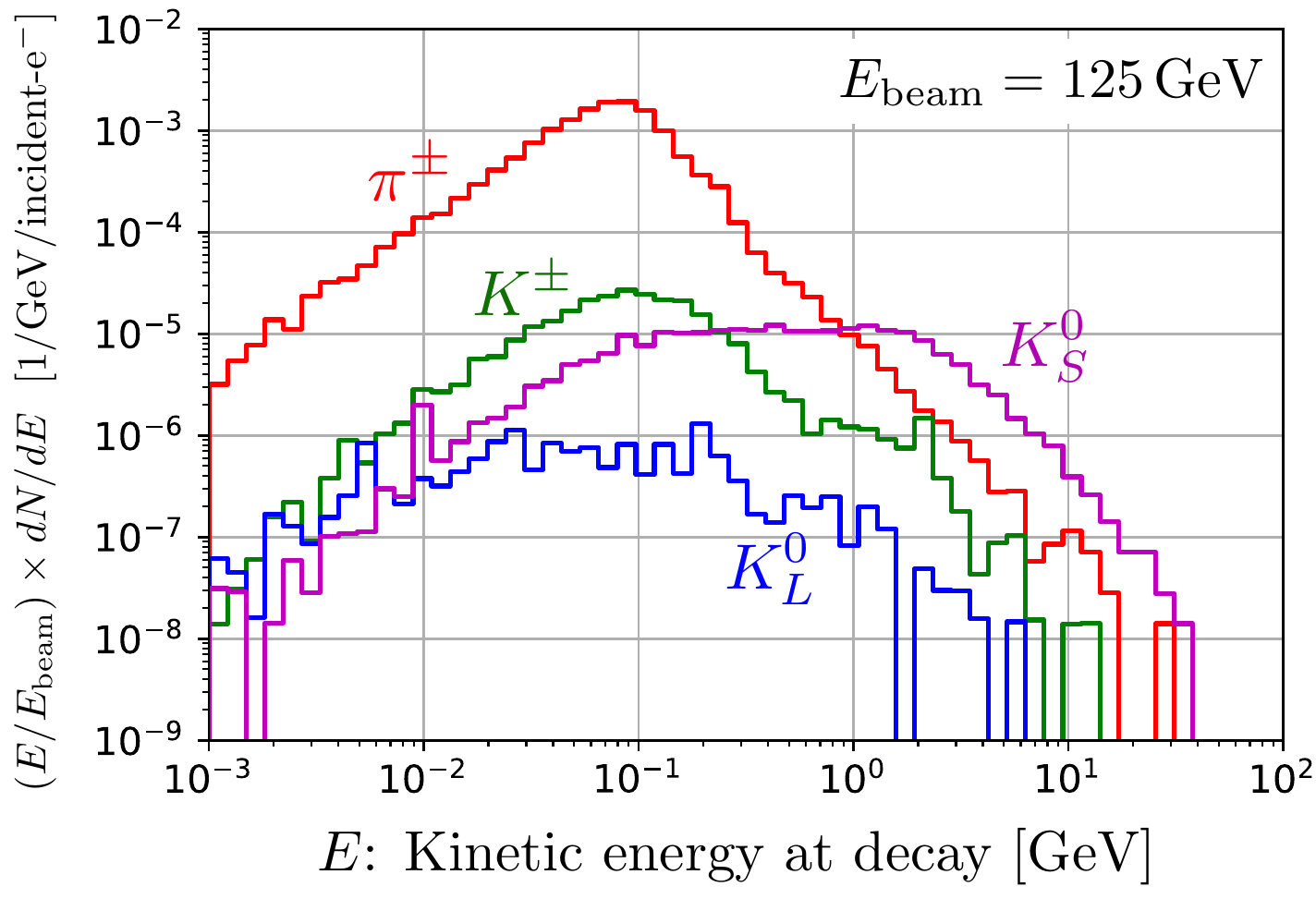}~~
\includegraphics[width=0.49\textwidth]{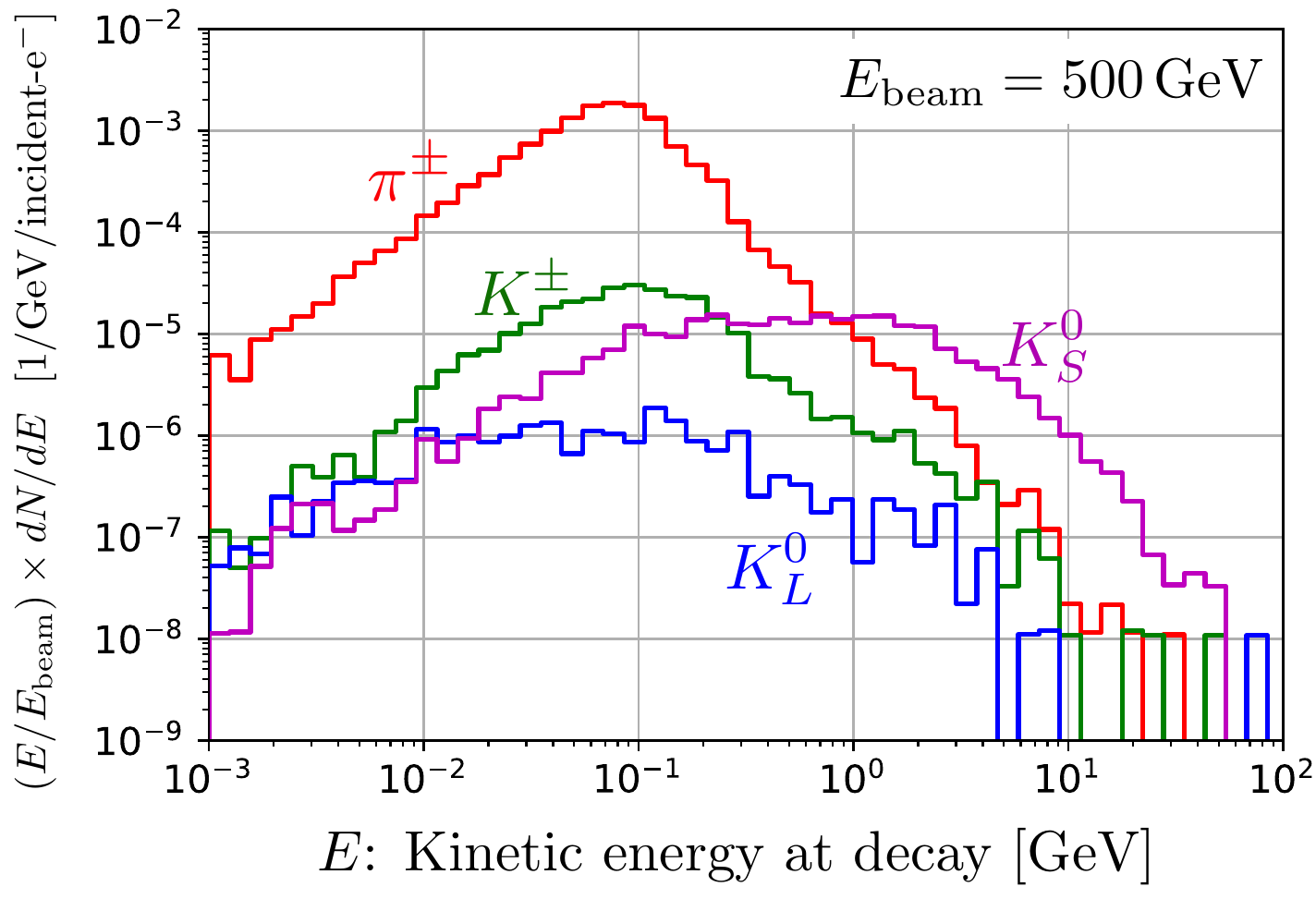}
\caption{The kinetic energy distribution of light mesons when they decay, where $\pi^\pm$ and $K^\pm$ indicates the sum of charged pions and kaons.
We consider two beam energies, $E_{\rm beam}=125,~500$~GeV at ILC-250 and ILC-1000, respectively.}
\label{fig:decay}
\end{figure*}

The light mesons are mainly produced by the interaction of real photons in the electromagnetic shower with the nucleons in the beam dump. If the decay length of the produced light mesons is in the same order of magnitude of or greater than the mean free path in the material, the particles reduce their energy or change into different flavors by  (in-)elastic scattering or  multiple scattering. We use the following codes and models which are available in PHITS to simulate the electromagnetic shower and the production and transport of the light mesons. For the electromagnetic shower, the simulation is performed by EGS5~\cite{Hirayama:2005zm}. For the light meson production and transport, the JAM~\cite{Nara:1999dz} and JQMD~\cite{PhysRevC.52.2620} models modified for photoproduction (photonuclear interaction) are used. In addition to these models, INCL4.6~\cite{PhysRevC.87.014606} is also employed to calculate the interaction of the mesons with nuclei during transport. The energy loss of the charged particles due to multiple scattering is evaluated by ATIMA~\cite{ATIMA}.

Fig.~\ref{fig:decay} shows the kinetic energy distribution of light mesons when they decay. The decay energy distribution is more important than the production energy distribution because the kinematics of new particles is determined by the parent particle distribution at the decay. For reference, the production energy distribution of light mesons is shown in Fig.~\ref{fig:production}.

\subsection{Heavy Mesons}

We use PYTHIA8 to calculate the differential cross sections of the  $\gamma p(n)\to B(D)+X$ process for the heavy mesons.\footnote{We thank the Pythia team, especially to Ilkka Helenius for helping us understand the latest photoproduction feature of PYTHIA~8.3.}
We have checked that the sum of the direct and non-diffractve cross section of the $D$ meson production agrees very well with the photoproduction data~\cite{SLACHybridFacilityPhoton:1985baw,TaggedPhotonSpectrometer:1989bpi}, see Appendix~\ref{app:Dmeson}. Therefore, we regard the sum of the two cross sections as the total cross section, 
$\sigma_{\rm total}(\gamma p(n))$
$=\sigma_\text{non-diff}(\gamma p(n))$
$+ \sigma_{\rm direct}(\gamma p(n))$. The total cross section of atomic neucleus is obtained by taking into account the shadowing effect \cite{Caldwell:1978ik,Kopeliovich:2012kw}, 
\begin{equation}
\sigma_{\rm total}(\gamma A)=  A^{l} \left(\frac{Z}{A}\sigma_{\rm total}(\gamma p)+\frac{A-Z}{A}\sigma_{\rm total}(\gamma n)\right) ,
\label{eq:atomic}
\end{equation}
where $l=0.92$~\cite{Kopeliovich:2012kw}.

\begin{figure*}
\centering
\includegraphics[width=0.45\textwidth]{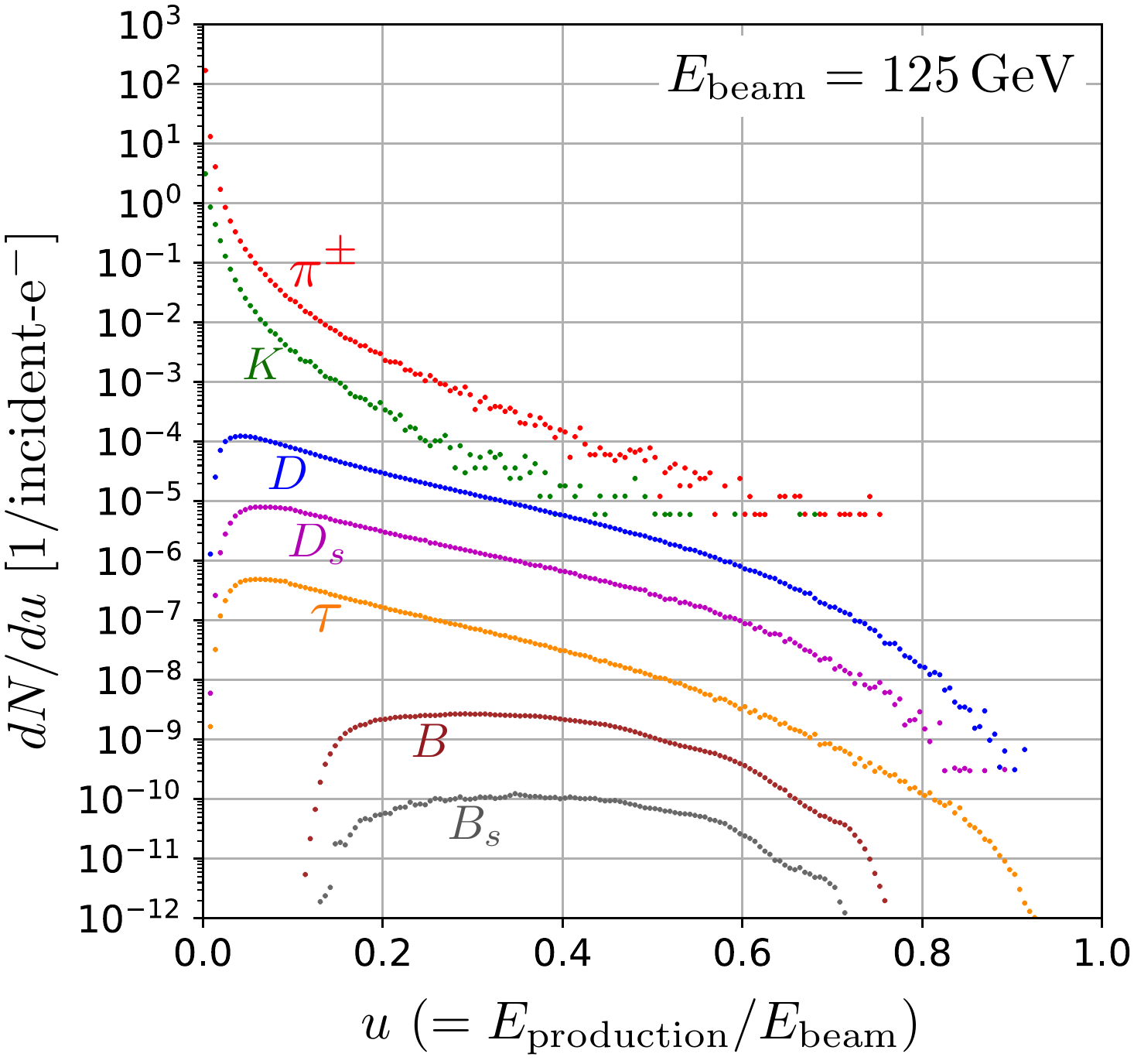}
\includegraphics[width=0.45\textwidth]{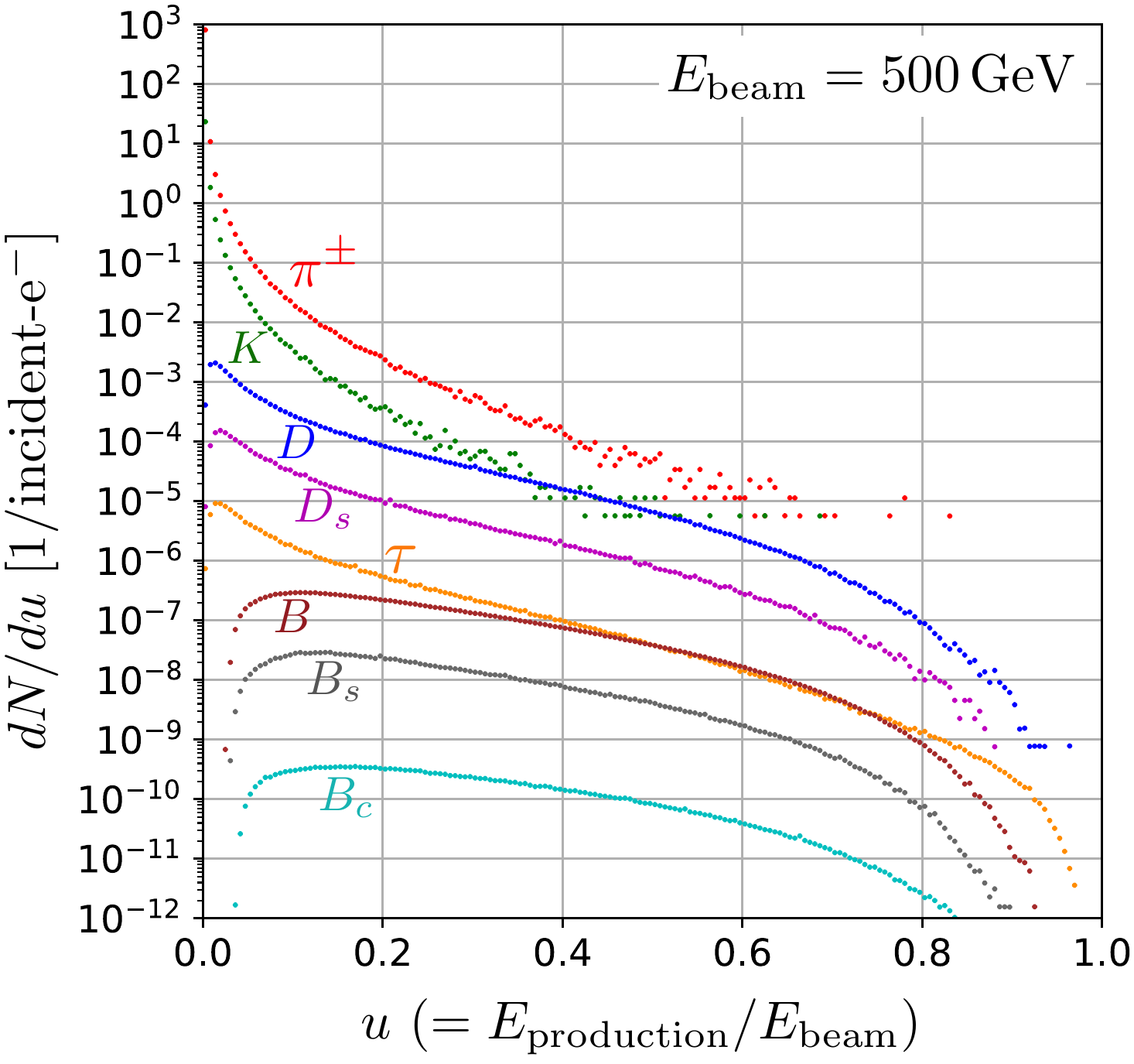}
\caption{The production rate per electron injection in the beam dump for mesons and $\tau$ lepton with respect to the kinetic energy at production, where the energy is normalized by the beam energy.
The results for $\pi^{\pm}(\pi^+, \pi^-)$, $K(K^+, K^-, K_S^0, K_L^0)$, $D(D^+, D^-, D^0, \overline{D}^0)$, $D_s(D_s^+, D_s^-)$, $B(B^+, B^-, B^0, \overline{B}^0)$, $B_s(B_s^0, \overline{B}_s^0)$, $B_c(B_c^+, B_c^-)$, and $\tau(\tau^+, \tau^-)$ produced in the beam dump are shown, which represent the sum of the particles in the parenthesis.
We consider two beam energies, $E_{\rm beam}=125,~500$~GeV at ILC-250 and ILC-1000, respectively.}
\label{fig:production}
\end{figure*}

The differential production cross sections for heavy meson photoproduction are obtained by PYTHIA8~\cite{Bierlich:2022pfr} and implemented in PHITS. Since the decay lengths of the heavy mesons are much shorter than the mean free path in the material, the spectra at their production and decay are similar. In Fig.~\ref{fig:production}, we show the production rate per electron injection in the beam dump for mesons and $\tau$ lepton with respect to the kinetic energy at production, where the energy is normalized by the beam energy. The results for $\pi^{\pm}(\pi^+, \pi^-)$, $K(K^+, K^-, K_S^0, K_L^0)$, $D(D^+, D^-, D^0, \overline{D}^0)$, $D_s(D_s^+, D_s^-)$, $B(B^+, B^-, B^0, \overline{B}^0)$, $B_s(B_s^0, \overline{B}_s^0)$, $B_c(B_c^+, B_c^-)$, and $\tau(\tau^+, \tau^-)$ produced in the beam dump are shown, which represent the sum of the particles in the parenthesis. We consider two beam energies, $E_{\rm beam}=125,~500$~GeV at ILC-250 and ILC-1000, respectively. The overall yield of the heavy meson production increases as the beam energy gets higher, and $B_c$ becomes accessible at ILC-1000. For the sake of comparison, we also include $\pi$, $K$ distributions at production.

\subsection{$\tau$ lepton}
As discussed in the previous literature~\cite{2018}, the primary source of $\tau$ lepton is the $D_s$ decay with approximately 5\% branching ratio. PHITS simulates $D_s$ meson propagation and decay, which accounts for the $\tau$ lepton production. The sub-dominant source of $\tau$ lepton is the $\tau$ lepton pair production, $\gamma + {\rm nucleus/nucleon} \to \tau^+ + \tau^- + X$. We implement a complete differential cross section calculated with the Born approximation in QED in the PHITS code and generate events for the process~\cite{Tsai:1973py,Sakaki:2020cux}. The form factors for coherent (nucleus elastic), quasi-elastic and inelastic interactions are included. The spectrum of $\tau$ lepton is shown in Fig.~\ref{fig:production}.

We find that the number of $\tau$ leptons from the pair production is about 20 times smaller than those from the decay of $D_s$. However, the pair production process becomes dominant in the high-energy region where the kinetic energy of $\tau$ lepton is above 65\% of the beam energy. So the pair production process will be necessary when considering the physics of high-energy $\tau$ leptons or $\tau$ neutrinos at the ILC beam dump.

\section{Heavy Neutral Leptons}\label{sec:HNL}

If gauge singlet fermions $N$ exist in the BSM sector, a renormalizable interaction with the SM sector is possible
\begin{align}
{\cal L} = -\lambda_{\ell I} (\bar L_\ell \tilde{H}) N_I -\frac{1}{2}M_{I} {\bar{N}_I^c} N_I +\rm h.c. \label{eq:HNL}, 
\end{align}
where $L_\ell$ is the SM lepton doublet of flavor $\ell=e,\mu,\tau$, and $I$ is the index of $N$. 
If $M_I\gg \lambda_{iI} v$, the standard seesaw mechanism   make the SM neutrino mass light, $m_{\nu, \ell\ell'}\sim  \sum_I (\lambda_{\ell I}\lambda_{\ell'I}) v^2/M_I$. For $\mathcal{O}(1)$  Yukawa coupling, the singlet fermion $N$ has to be extremely heavy to satisfy the bounds on the neutrino mass, $m_{\nu} \gtrsim 0.05~{\rm eV}$~\cite{ParticleDataGroup:2020ssz}. On the other hand,  if the size of Yukawa coupling $\lambda$ is small, $M$ can be in MeV-GeV mass scale to satisfy the same condition. Such particles can be searched directly in laboratories, and they are often called  heavy neutral leptons (HNLs) or sterile neutrinos. The active neutrinos $\nu$ from $L$ doublet and the HNLs $N$ are almost in the mass eigenstates up to a small admixture between $\nu_i$ and $N_I$ characterized by the mixing angle
\begin{align}
U_{\ell I} = \frac{v \lambda_{\ell I}}{M_I}.  \label{eq:mixing}
\end{align}
Since the mixing and mass determine the HNL interactions with the SM particles, we use the mixing parameter $U_{\ell I}$ instead of the Yukawa couplings to discuss the HNL phenomenology.

\begin{figure*}[tt]
\centering
\includegraphics[width=0.6\textwidth]{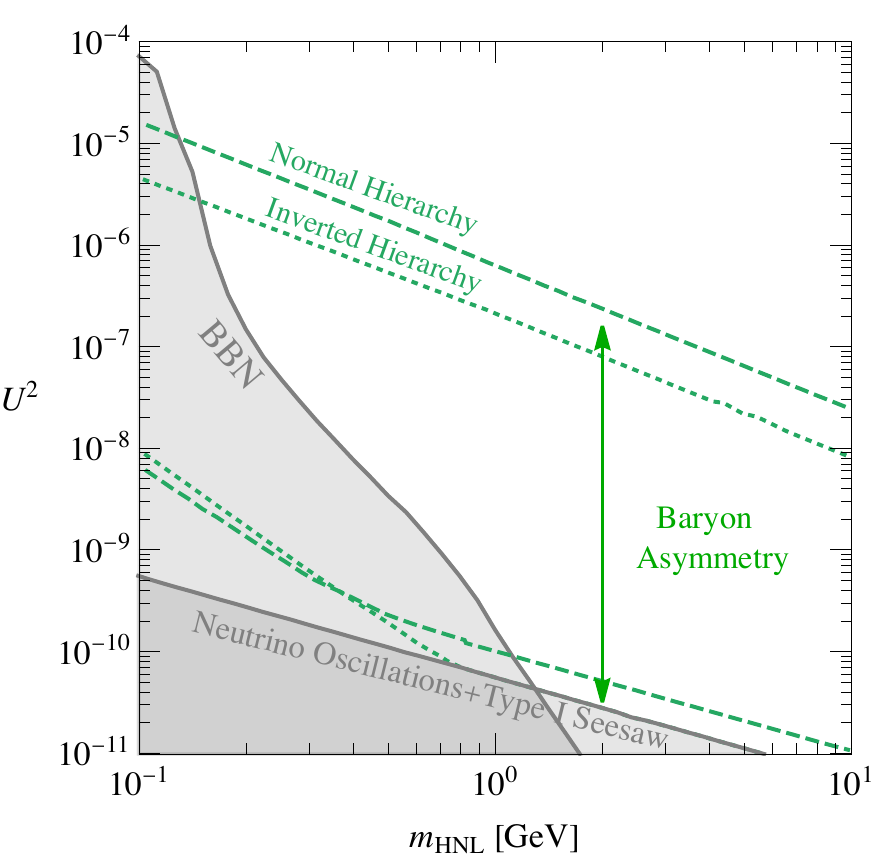}
\caption{
The target parameter space in a scenario of two degenerate HNLs with respect to the HNL mass and the sum of the mixing squared defined in Eq.\eqref{eq:mixingsum}. The region between dashed (dotted) green lines is favored as the HNL can generate the baryon asymmetry of the universe~\cite{Akhmedov:1998qx, Asaka:2005pn}, and the lines are adopted from Fig.~4.17 of \cite{Alekhin:2015byh}. 
The bottom shaded region cannot explain the neutrino oscillation data in Type-I seesaw mechanism with the two degenerate HNLs. Another shaded region with the BBN label is excluded by that the long-lived HNLs affect the successful big bang nucleosynthesis. The bound is obtained with respect to $U_e$, $U_\mu$, or $U_\tau$ in \cite{Boyarsky:2020dzc}, and we take $U^2<\min_{\ell=e,\mu,\tau}[|U_{\ell}^{(\rm BBN)}|^{2}]$ for this plot. 
}
\label{fig:bench}
\end{figure*}

The HNLs in the GeV mass range have another exciting aspect other than explaining the mass of active neutrinos and testability. They can be responsible for the baryon asymmetry by leptogenesis via HNL oscillation~\cite{Akhmedov:1998qx, Asaka:2005pn}. The effective leptogenesis occurs for fast $N_I$ oscillation, and therefore, two degenerate HNLs are an excellent benchmark model to investigate.

In the minimalistic scenario with two quasi-degenerate HNLs, the target parameter space is well-defined by the baryon asymmetry and the neutrino mass. We schematically show it in Fig.~\ref{fig:bench}. Note that the vertical axis is the sum of the mixing angles  over both the active neutrinos and the HNLs, 
\begin{align}
U^2\equiv \sum_{i=e,\mu,\tau \ I=1,2} |U_{\ell I}|^2 
\label{eq:mixingsum}
\end{align}
We also include a bound on  $U^2$ from Big Bang Nucleosynthesis (BBN). The small mixing angle is disfavored as the HNLs are sufficiently long-lived to decay during or slightly before BBN and  affect the ratio of the neutron and proton number densities~\cite{Boyarsky:2020dzc}.

It is essential to probe all flavor mixings to cover the target parameter space for baryon asymmetry characterized by $U^2$. The sensitivity to $U_{e I}$ and $U_{\mu I}$ of the current and proposed experiments is high, but the $\tau$ neutrino mixing is poorly constrained. In the ILC beam dump experiment, the relevant region of $U_{\tau I}$ can be probed because  $\tau$ leptons can be copiously produced from $D_s$ meson decay thanks to the higher beam energy.

In many phenomenological studies of HNLs, one assumes a single HNL (say $N_1$) in the low energy  because having two HNLs will have little impact on  the search sensitivity as long as the two HNLs are degenerate. In the following, we deal with one HNL for simplicity, and thus the index $I$ is omitted. Furthermore we turn on only one $U_\ell$, a mixing with one of the active neutrinos in the flavor eigenstate ($\nu_e, \nu_\mu, \nu_\tau$), at a time, which helps us to understand which underlying process matters. Under these assumptions, the phenomenology is well-described by the HNL mass and the single mixing in each benchmark model. Also, these benchmarks are commonly used in the literature, which allows us to compare our results with the previous works.

\subsection{Sensitivities at ILC beam dump\label{sec:sensitivity}}

In this subsection, we evaluate the sensitivity of the ILC beam dump experiment to the HNLs. We consider the following two production mechanisms of HNL:
\begin{enumerate}[(i)]
 \item Productions from meson and $\tau$ lepton decays; 
 \item Direct productions from electrons and muons in EM shower interacting with nucleons.
\end{enumerate}
In both cases, we consider the HNLs decaying 
inside the decay volume as the signal. We adopt the decay widths of HNL to the SM particles based on Ref.~\cite{2018}, and most decay patterns of the HNL would leave multiple tracks, which is distinguishable from the background. 
We then require that the HNL decays inside the decay volume with two or more charged tracks, i.e., the decay modes of $\rm HNL\to \nu\nu\bar\nu, $ and $\pi^0\nu$ are excluded.

We estimate that the background events are limited. One of the possible backgrounds is hardons produced by high-energy neutrinos, such as $K_{S,L}$ decaying to the charged pions. This type of background was estimated with GIANT~4 in the SHiP setup~ \cite{Bonivento:2013jag,SHiP:2015vad}. They found that the relevant hadrons are always produced at the edge of the muon shield and can be significantly reduced by requiring the charged tracks consistent with the HNL kinematics. Comparing the number of produced neutrinos between SHiP and the ILC beam dump experiment, we can infer from the number of expected background events. The SHiP experiment expects $7\times 10^{17}$ neutrinos per $2\times 10^{20}$ protons on target, which leads to $\sim 10^7$ neutrino interactions and $\sim 10^4$ events with two tracks of opposite charge. 99.4\% of the two-track events can be discarded by the simple topology cuts, which leads to $50$ remaining events. Furthermore, the veto system of SHiP can reduce the background events by $10^{-4}$ to the level of 0.07 events. At the ILC 250 (1000), we estimate $8\times 10^{16}$ ($3\times 10^{17}$) neutrinos for the 10-year run, and, assuming the similar reduction factors as in SHiP, we expect about 5 (20) background events after the two-track requirement and the simple topology cuts. 
Although the veto system could further reduce the background, the reduction factor strongly depends on the detector setup. Therefore, 
we conservatively take a pessimistic scenario such that 5 and 20 background events are expected at ILC-250 and 1000 for 10-year statistics. The corresponding upper bounds on the signal events are respectively 5.5 and 9.1.

As another source of background, cosmic muons may produce beam-unrelated events. However, the deep underground location of the detector (100 m from the ground surface), direction of tracks, and coincidence time window based on the pulsed electron beam significantly reduce the backgrounds. The detector setup and event selections to suppress the background further are beyond the scope of this paper.

In the HNL production process (i), evaluating the position and 4-momentum of the HNL is not straightforward because it involves various intermediate particles with possible higher multiplicity. Therefore, it is suitable to  estimate the sensitivity using Monte-Carlo simulation. We include the SM particle decays to an HNL  which are calculated in Refs.~\cite{Shrock:1980vy,Shrock:1980ct,Shrock:1981wq,Johnson:1997cj,Gorbunov:2007ak,2018} and summarized in Appendix~\ref{sec:Br} and Ref.~\cite{2018}.

For the other production (ii), associated physical processes are  tractable   without simulation. Therefore, we evaluate its sensitivity by the numerical integration and provide relevant formulae for it. To help the quantitative understanding, we also provide an approximate formula of the signal rate of the process (i) that can be integrated numerically. The signal rate of the numerical integration will be compared with the one obtained by the Monte-Carlo simulation.

In the following, we describe the detail of each method.

\subsubsection{Monte-Carlo method} 
We simulate particle production and transport by PHITS with the help of PYTHIA8 for electron injection as described in Sec.~\ref{sec:spectrum}. We also modify the decay tables of mesons and $\tau$ lepton of PHITS  to include decay modes to an HNL as discussed in Appendix~\ref{sec:Br}. It is essential to perform the Monte-Carlo simulation to track how the system evolves from the incident electron since the intermediate steps are very involved in the production (i). The result of the Monte-Carlo simulation also becomes the guiding post for the coarse-grained integration method which is described in Sec.~\ref{sec:nummethod}.

However, a naive Monte-Carlo simulation of the long-lived particle would suffer from technical challenges in obtaining sufficient statistics, in particular in the following cases:
\begin{enumerate}[(a)]
 \item Small cross section of new physics process or photoproduction of mesons.
 \item Small decay branching ratio from a SM particle to the new particle.
 \item The decay length of the new particle much shorter than the shield length.
 \item The decay length of the new particle much longer than the length of the experimental setup.
\end{enumerate}

In cases (a) and (b), the processes resulting in the signal process are so rare that it is difficult to obtain a sufficient number of events. The issue can be solved in PHITS-based simulation using the {\it biasing technique}. In this technique, PHITS provides a biasing parameter for photoproduction. In this technique, the biased process occurs more frequently according to the biasing parameter, and an appropriate weight of the produced particle lower than that of the incident particle is assigned. We obtain the correct expected value of physical quantities by adding up the weights rather than adding up the number of particles produced.

In case (c), the new particles generated by the simulation predominantly decay inside the shield, so  sufficient signal statistics are not easily obtained. This issue can be avoided by using the {\it importance sampling technique} which PHITS supports. This technique allows us to assign an {\it importance} to different regions of the simulation geometry. When a particle passes through the boundary between different regions with increasing importance, several copies of the particle are created in an event, and their weights are reduced depending on the importance ratio between the regions. We divide the region of the shield in the direction of the beam axis and increase their importance value exponentially so that a short lifetime particle passes through the shield more efficiently without exponential loss.

The importance technique is also useful in the opposite case (d) as we can set a large importance value in the decay volume to increase the event sampling in the relevant region. In addition, {\it forced-decay technique} is more useful when the decay length of the long-lived particle $X$ is extremely longer than the length of the shield and decay volume. In this technique, we introduce a {\it maximum decay length}, $l_{\rm max}$. When the decay length of $X$, $l_X$, is sufficiently longer than the typical length of experiment, $l_{\rm exp} (\sim l_{\rm dump}+l_{\rm sh}+l_{\rm dec})$, the differential decay rate of $X$ with respect to the flight distance $z$ is
\begin{align}
\left. \frac{dP}{dz}\right\vert_{l_X}
= \frac{1}{l_X} {\rm e}^{-\frac{z}{l_X}}
\simeq
\frac{1}{l_X},~~(z<l_{\rm exp}).
\end{align}
This is a very small value, so it's hard to get enough Monte-Carlo statistics for the decay of $X$ in the decay volume.  To deal with this problem, we multiply the decay probability by $b = l_X / l_{\rm max} (\,>1)$ and the weight of $X$ by $1/b$. The rescaled probability is
\begin{align}
\left. b \frac{dP}{dz}\right\vert_{l_X}
\simeq
\frac{b}{l_X}
\simeq
\left. \frac{dP}{dz}\right\vert_{l_{\rm max}}.
\end{align}
This corresponds to the decay rate when the lifetime of $X$ is multiplied by $1/b$. In summary, when $l_{\rm max}<l_X$, multiplying the lifetime and weight of $X$ by $1/b$ forces a higher probability of decay within the decay volume and reduces the computational cost by a factor $b$. This approach is an approximation, and the higher order correction to weight is $\mathcal{O}(l_X^2/l_{\rm max}^2)$.  We typically use $l_{\rm max}\sim 10^{3}$~m, so that we can safely ignore the error. We count the number of the HNL decays to the visible SM particles in the decay region taking into account all the weight factors appropriately.

\subsubsection{Coarse-grained integration method}
\label{sec:nummethod}
In this subsection, we describe the approximate estimation for the HNL productions (i) and (ii). Here, we apply suitable approximations to obtain simplified expressions involving multiple integrals for the signal yield, and then perform the integrals numerically. We refer this calculation scheme to as the coarse-grained integration (CGI) method. The same scheme was used in Ref.~\cite{Sakaki:2020mqb,Asai:2021ehn} in which the ALPs and dark photon productions by the EM shower were studied.

For the production mechanism (i), we consider processes where a photon from the EM shower interacts with a nucleus in the beam dump and produces the SM particle $k$. 
The number of signal events is schematically given by
\begin{align}
N_{\rm signal}^{\rm (i)}&=N_{\rm EOT}\times\sum_{k=\pi, K, D, B,\tau} \left( l_{\gamma}\times n_{\rm nucleus}\times  \sigma_{\gamma~{\rm nucleus}\to k X}\right)
\notag
\\ 
& \times{\rm Br}(k \to {\rm HNL}~X) \times {\rm Br}_{\rm vis}\times {\rm Acc}^{\rm (i)}({\rm HNL}), \label{eq:sig}
\end{align}
where  $l_{\gamma}$ is  the photon track length in the beam dump, $n_{\rm nucleus}$ is the number density of beam dump nucleus and the production cross section of particle $k$ is denoted by $\sigma_{\gamma~{\rm nucleus}\to k X}$. The branching ratio ${\rm Br}(k \to {\rm HNL}~X)$ and ${\rm Br}_{\rm vis}={\rm Br}({\rm HNL} \to {\rm visible~SM})$, where the visible SM denotes the decay modes except for ${\rm HNL}\to \nu\nu\bar{\nu}$, and $\pi^0\nu$, are summarized in Ref.~\cite{2018} and also Appendix~\ref{sec:Br}. ${\rm Acc}^{\rm (i,ii)}({\rm HNL})$ denotes the detector acceptance depending on the production process. The distribution of photon track length  $l_{\gamma}$ in the beam dump can be  obtained by the Monte-Carlo simulation, and the result is given in Ref.~\cite{Asai:2021ehn} and also Appendix~\ref{app:track}.

In the production mechanism (ii), an incoming lepton $\ell=e,\mu$ from the EM shower interacts with a nucleon and would produce an HNL. The number of signal events is schematically expressed as
\begin{align}
N_{\rm signal}^{\rm (ii)}&=N_{\rm EOT}\times l_{{\ell}^{\pm}}\times n_{N}^{\rm dump/shield}\times \sigma_{{\ell}^{\pm}~{\rm nucleon}\to {\rm HNL}~ X}\times{\rm Br}_{\rm vis}\times {\rm Acc}^{\rm (ii)}({\rm HNL}),\label{eq:sig2}
\end{align}
where the track length $l_{\ell^{\pm}}$ of the lepton in the beam dump is provided in Ref.~\cite{Asai:2021ehn} and Appendix~\ref{app:track}, and $n_{N}^{\rm dump/shield}$ is the number density of nucleon in the beam dump or muon shield. For the incoming muon, we ignore the HNL production in the beam dump. The production cross section of  HNL is denoted by $\sigma_{{\ell}^{\pm}~{\rm nucleon}\to {\rm HNL}~X}$, which is summarized in Appendix~\ref{app:DIS}.

We elaborate the schematic formulae Eqs.~\eqref{eq:sig} and \eqref{eq:sig2} in the following. Given that the differential distribution of  mesons and $\tau$ lepton were obtained by PHITS and PYTHIA8, as the results are shown in Sec.~\ref{sec:spectrum}, we fit the differential distribution of the number of SM particle $N_k$ as the function of $E_k$. This quantity is expressed by the more fundamental quantity as follows,
\begin{align}
\frac{d N_k}{dE_k}=\int dE_{\gamma}\frac{d l_{\gamma}}{dE_{\gamma}}\times n_{\rm nucleus}\times \int d \Pi _X  \frac{d\sigma_{\gamma~{\rm nucleus}\to k X}}
{dE_k d \Pi_X}.
\end{align}
This corresponds to the schematical terms in the parentheses of Eq.~\eqref{eq:sig}. Note that ${d l_{\gamma}}/{dE_{\gamma}}$ and $\sigma_{\gamma~{\rm nucleus}\to k X}$ is absorbed in ${d N_k}/{dE_k}$.

By using $dN_k/dE_k$ the production rate approximately expressed as follows, 
\begin{align}
    N^{\rm (i)}_{\rm signal}&\sim N_{\rm EOT}\sum_{k=\pi, K, D, B,\tau}\int d E_k \frac{d N_k}{dE_k}\cdot {\rm Br}(k \to {\rm HNL}~X)\cdot {\rm Br}_{\rm vis}\cdot{\rm Acc}_{\rm dump}^{\rm (i)}({\rm HNL}), \label{eq:N1approx}
\end{align}
where we assume momentum of the SM particle $k$ is aligned to the electron beam direction. The HNL production via the SM particle decay is the leading production channel at the ILC beam dump experiment for the most accessible parameter space of mass and mixing angle. As we will see later, this production mode even contributes to the sensitivity of $U_\tau$.

Additionally, the direct production initiated by the charged leptons of the EM shower takes an essential role to  explore the higher mass region of the HNL in the presence of $U_e$ or $U_\mu$. The direct production from the electron and positron occurs inside the beam dump, while the one from the muon occurs dominantly in the muon shield. Several direct production processes contribute to the HNL productions~\cite{Formaggio:2012cpf}, such as quasi-elastic scattering, resonance production, and deep inelastic scattering~(DIS). Among those, DIS would be a significant contribution to the HNL productions around $E_{e^{\pm},\mu^{\pm}}\geq 10~{\rm GeV}$ which is favored by the acceptance of the HNL production angle. Therefore, we focus on the DIS process, and the formulae of the number of signal events are expressed as follows, 
\begin{align}
    N^{{\rm (ii)},e^{\pm}}_{\rm signal}&=N_{\rm EOT}\sum_{i=e^-,e^+}\int d E_{i}\frac{1}{2}\frac{dl_{i}}{d E_{i}}
      \sum_{N=n,p}n_{N}^{\rm dump}
\int dx \int dy \frac{d^2\sigma (i N\to {\rm HNL}~X)}{dx dy}
\nonumber\\ &\times
{\rm Br}_{\rm vis}\cdot {\rm Acc}_{\rm dump}^{{\rm (ii)},e^{\pm}}({\rm HNL}),\label{eq:DISel}
    \\
    N^{{\rm (ii)},\mu^{\pm}}_{\rm signal}&=N_{\rm EOT}\sum_{i=\mu^-,\mu^+}\int d E_{i 0}\frac{1}{2}\frac{d Y_{i 0}}{dE_{i 0}}\int dE_{i}\frac{1}{2}\frac{dl_{i}}{d E_{i}} \int d\theta_{\rm MCS} \frac{d P_{\rm MCS}}{d\theta_{\rm MCS}} \notag
    \\
&\times \sum_{N=n,p} n_{N}^{\rm shield}\int dx \int dy \frac{d^2\sigma (i N\to {\rm HNL}~X)}{dx dy}\cdot {\rm Br}_{\rm vis}\cdot{\rm Acc}_{\rm shield}^{{\rm (ii)},\mu^{\pm}}({\rm HNL}).\label{eq:DISmu}
\end{align}
The energy spectrum of the muon yield per incident electron behind the beam dump $dY_{\mu 0}/d E_{\mu 0}$ where $E_{\mu^{\pm}0}\geq E_{\mu^{\pm}}$ is evaluated in Ref.~\cite{Sakaki:2020cux}, {also see Eq.(A.3) of Ref.~\cite{Sakaki:2020mqb}} and Appendix~\ref{app:track}, and the track length of muons $l_{\mu^{\pm}}$ in the muon shield is provided in Ref.~\cite{Sakaki:2020mqb} and Appendix~\ref{app:track}. The angular distribution stemming from the multiple Coulomb scattering (MCS)  of the muon $dP_{\rm MCS}/d\theta_{\rm MCS}$ is given in Ref.~\cite{Lynch:1990sq,ParticleDataGroup:2020ssz}. The integral variable $x\equiv$ $Q^2/[2M_N(E_{i}-E_{\rm HNL})]$ denotes the Bjorken scaling, where $E_{i}$ is the incoming lepton energy at the HNL production, and $E_{\rm HNL}$ is the HNL energy in the nucleon rest frame. The other integral variable $y\equiv(E_{i}-E_{\rm HNL})/E_{i}$ is the fraction of the lepton energy loss in the rest frame.

Finally, we give the formula we used to estimate the acceptance ${\rm Acc}_{\rm dump,shield}^{(\rm i,ii)}({\rm HNL})$ as follows,
\begin{align}
{\rm Acc}_{\rm dump}^{{\rm (i)}}({\rm HNL}) &=\int_0^{l_{\rm dec}}dz \frac{dP_{\rm dec}^{\rm dump}}{dz}\cdot \int_0^{r_{\rm det}/(l_{\rm dump}+l_{\rm sh})}d\theta_{\rm HNL}^{\rm dec} \frac{dP_{\rm ang}}{d\theta_{\rm HNL}^{\rm dec}}\cdot \Theta \left(r_{\rm det}-r_{\perp,{\rm dump}}^{{\rm (i)}}\right),\label{eq:acci}
\\
{\rm Acc}_{\rm shield}^{{\rm (ii)},e^{\pm}}({\rm HNL}) &=\int_0^{l_{\rm dec}}dz \frac{dP_{\rm dec}^{\rm shield}}{dz}\cdot \Theta \left(r_{\rm det}-r_{\perp,{\rm shield}}^{{\rm (ii)},e^{\pm}}\right),\label{eq:acciie}
\\
{\rm Acc}_{\rm shield}^{{\rm (ii)},\mu^{\pm}}({\rm HNL}) &=\int_0^{l_{\rm dec}}dz \frac{dP_{\rm dec}^{\rm shield}}{dz}\cdot \Theta \left(r_{\rm det}-r_{\perp,{\rm shield}}^{{\rm (ii)},\mu^{\pm}}\right)\times
\begin{cases}
\Theta \left(l_{\rm sh}-\delta_{\mu}\right)~~~{\rm for}~{\rm ILC\text{-}250}
\\
\Theta \left(10~{\rm m}-\delta_{\mu}\right)~~~{\rm for}~{\rm ILC\text{-}1000}
\end{cases},\label{eq:accmuon}
\end{align}
where, as in the Monte-Carlo simulation, all the HNL decays inside the decay volume are assumed to be the visible signal except for the decay modes of ${\rm HNL}\to \nu\nu\bar{\nu}$, and $\pi^0\nu$.
Here $z$ is the decay position of HNL in the decay volume, and $P_{\rm dec}$ denotes the decay probability of HNL given by
\begin{align}
\frac{d P_{\rm dec}^{\rm dump}}{dz}&=\frac{1}{l^{({\rm lab})}_{\rm HNL}}{\rm exp}\left(-\frac{l_{\rm dump}+l_{\rm sh}+z}{l^{({\rm lab})}_{\rm HNL}}\right),\label{eq:decP}
\\
\frac{d P_{\rm dec}^{\rm shield}}{dz}&=\frac{1}{l^{({\rm lab})}_{\rm HNL}}{\rm exp}\left(-\frac{l_{\rm sh}-\delta_{\mu}+z}{l^{({\rm lab})}_{\rm HNL}}\right), 
\end{align} 
where $l^{({\rm lab})}_{\rm HNL}$ denotes the lab-frame decay length of HNL given by
\begin{align}
l^{(\rm lab)}_{\rm HNL}=\frac{p_{\rm HNL}}{m_{\rm HNL}}\frac{1}{\Gamma_{\rm HNL}}
\end{align}
with $p_{\rm HNL}$ being the momentum, $m_{\rm HNL}$ the mass, and $\Gamma_{\rm HNL}$ the total decay width of HNL evaluated in \cite{2018}.
The angular distribution of the HNLs is denoted as $dP_{\rm ang}/d \theta^{\rm dec}_{\rm HNL}$, where $\theta^{\rm dec}_{\rm HNL}$ is the decay angle of HNL from the particle $k$ with respect to the direction of $k$, and we use an approximated formula as follows,
\begin{align}
    \frac{dP_{\rm ang}}{d\theta_{\rm HNL}^{\rm dec}}=\sin \theta_{\rm HNL}^{\rm dec} \frac{1}{2}\left(\frac{m_k}{(E_k+m_k)-p_k\cos\theta_{\rm HNL}^{\rm dec}}\right)^2,
\end{align}
where $m_k$ is the mass of the SM particle $k$, and $p_k$ denotes the momentum of $k$ in the lab-frame.
Then, the energy of the HNL is approximately given by
\begin{align}
    E_{\rm HNL}=\frac{1}{2}\frac{m_k^2}{(E_k+m_k)-p_k\cos\theta_{\rm HNL}^{\rm dec}}.
\end{align}
The distance that a muon passes through the muon shield before emitting the HNL is the function of $E_{\mu0}$ and $E_{\mu}$~\cite{Sakaki:2020mqb}, 
\begin{align}
    \delta_{\mu}=\frac{E_{\mu 0}-E_{\mu}}{{\langle dE/dx\rangle}_{\rm Lead}},
\end{align}
with the energy independent stopping power: ${\langle dE/dx\rangle}_{\rm Lead}=0.02~{\rm GeV}/{\rm cm}$. 
For $E_{\rm beam}=500$~GeV, the additional active shield behind the muon shield may be necessary due to their strong penetrating power, and the muon's trajectory becomes non-trivial after the active shield. Then we account for only the HNLs produced before the active shield, which leads to  the Heaviside step function in Eq.~\eqref{eq:accmuon}. Regarding the impact of varying the active shield length, we study how  different depth of the muon shield affects the sensitivity for the HNL at ILC-1000 in Appendix~\ref{app:muonshield}. The angular acceptance is expressed by the Heaviside step function associated with the transverse radius $r_\perp (z)$ in Eqs.~\eqref{eq:acci}, \eqref{eq:acciie}, and \eqref{eq:accmuon}. The typical deviation of the visible SM particles emitted from HNL from the beam axis is estimated as
\begin{align}
r_{\perp,{\rm dump}}^{\rm (i)}&=|\theta_{\rm HNL}^{\rm dec}|\cdot (l_{\rm dump}+l_{\rm sh}+z),\label{eq:ri}
\\
r_{\perp,{\rm dump}}^{{\rm (ii)},e^{\pm}}&=\sqrt{(\theta_{e}^{\rm shower})^2\cdot (l_{\rm dump}+l_{\rm sh}+z)^2+(\theta_{\rm HNL}^{\rm prod})^2\cdot (l_{\rm dump}+l_{\rm sh}+z)^2},
\\
r_{\perp,{\rm dump}}^{{\rm (ii)},\mu^{\pm}}&=\sqrt{(\theta_{\mu}^{\rm shower})^2\cdot (l_{\rm sh}-\delta_{\mu}+z)^2+(\theta_{\rm HNL}^{\rm prod})^2\cdot (l_{\rm sh}-\delta_{\mu}+z)^2},
\end{align}
where $\theta_e^{\rm shower}$ ($\theta_{\mu}^{\rm shower}$) is the angle of shower electrons and positrons (muons) with respect to the beam axis, and $\theta_{\rm HNL}^{\rm prod}$ is the production angle of HNL.
For the DIS process, the production angle is expressed as
\begin{align}
    \cos\theta_{\rm HNL}^{\rm prod}=\frac{1}{|p_{e,\mu}|\cdot|p_{\rm HNL}|}\left(E_{e,\mu}E_{\rm HNL}-\frac{1}{2}\left(Q^2+m_{e,\mu}^2+m_{\rm HNL}^2\right) \right),
\end{align}
where $p_{e,\mu}$ is the momentum of incoming electron and muon, and $Q^2=2 M_N E_{e,\mu} xy$ with the mass of nucleus. 
As provided in Refs.~\cite{Sakaki:2020mqb} and \cite{Asai:2021ehn}, $\theta_{e}^{\rm shower}$ is estimated by the Monte-Carlo simulation, and we use the mean value $\theta_{e}^{\rm shower}=16~{\rm mrad}\cdot {\rm GeV}/E_{e^{\pm}}$.
For the incoming muons, we use $\theta_{\mu}^{\rm shower}=\theta_{\rm MCS}$. Although for the production mechanism (i), the angle of shower photon and the production angle of the particle $k$ are precisely calculated in the Monte-Carlo simulation, we omit them in the coarse-grained integration method. This is because the heavy mesons are produced by high-energy photons and the mean value of the angle of shower photons $\theta^{\rm shower}_{\gamma}=8~{\rm mrad}\cdot {\rm GeV}/E_{\gamma}$~\cite{Sakaki:2020mqb,Asai:2021ehn} is suppressed.

\begin{figure*}[t!]
\centering
\includegraphics[width=0.6\textwidth]{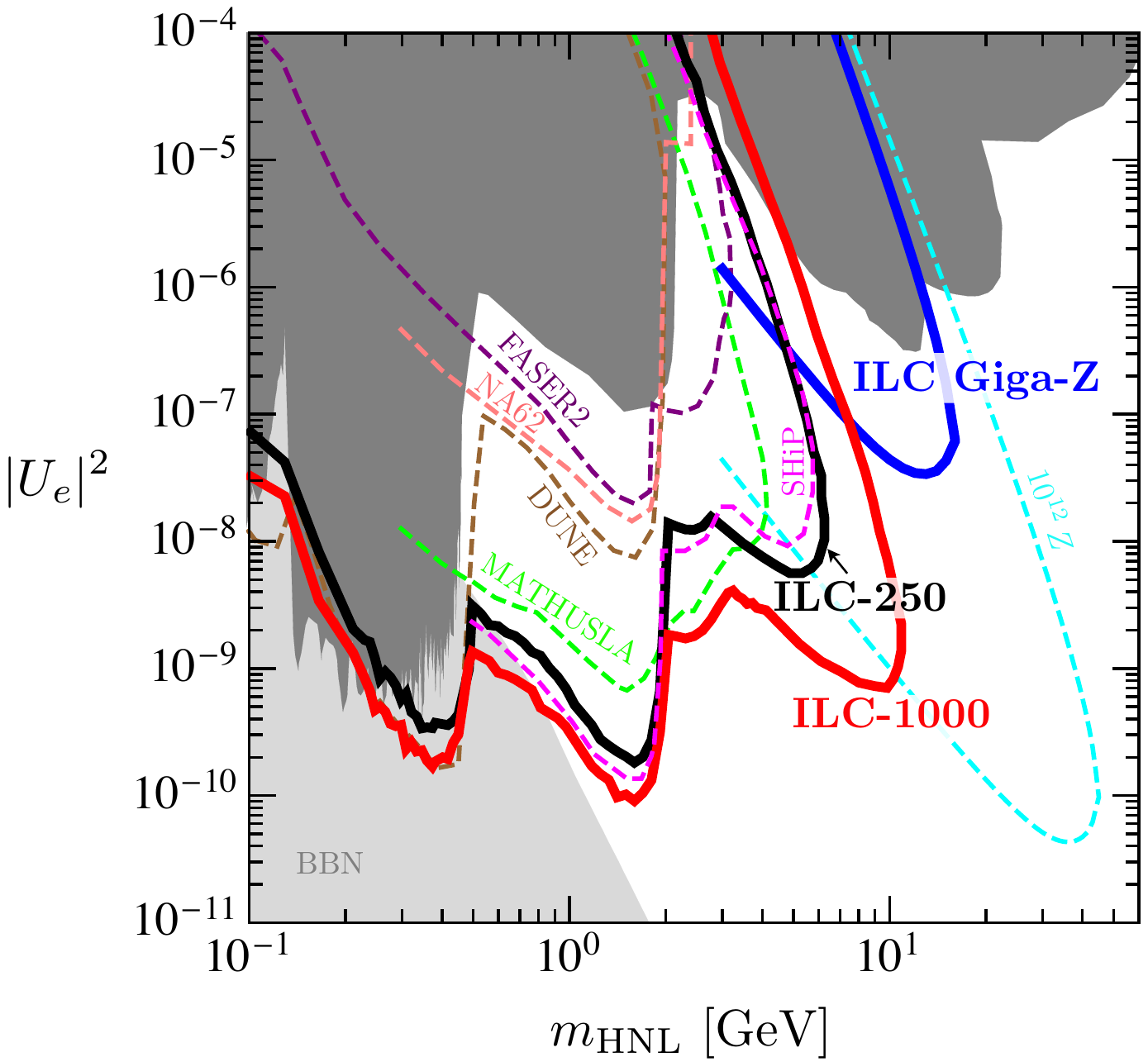}
\caption{
Sensitivity reach of ILC beam dump experiment to HNLs mixing with the electron neutrino in the mass and mixing plane, assuming 10 year run at ILC-250 (black solid) and ILC-1000 (red solid).  
The number signal events more than 5.5 (9.1) is required at ILC-250(1000), which corresponds to the $95\%$ C.L. sensitivity. The discussion about the background is in Sec.~\ref{sec:sensitivity}. 
The current exclusion bounds are shown in the gray region, see Sec.~\ref{sec:bounds}. The darker grey region is from the laboratory bounds, and the lighter gray region is the BBN bound for the HNL, which is roughly $\tau_{\rm HNL}>0.02$~s~\cite{Boyarsky:2020dzc}. Sensitivity reach through $10^ 9$ $Z$-decays  at ILC is shown as a blue solid line. See Sec.~\ref{sec:ZatILC}. For a comparison, dashed lines show a sensitivity reach of the DUNE experiment~\cite{Coloma:2020lgy} (brown), the FASER2 experiment~\cite{Kling:2018wct} (purple), the NA62 experiment~\cite{Beacham:2019nyx} (orange), and the SHiP experiment~\cite{Alekhin:2015byh} (magenta), the  MATHUSLA experiment\cite{Curtin:2018mvb} (green), and $10^{12}$ $Z$-decays that could be realized at the FCC-ee experiment~(cyan). \label{fig:sensitivityUe}
}
\end{figure*}

\subsubsection{Projected sensitivities}
We give the prospects of the ILC beam dump experiment in ILC-1000 and ILC-250 for 10 year run. We assume the parameter region with more than  5.5 (9.1) events can be probed at ILC-250 (1000), corresponding to the expected $95\%$ C.L. exclusion sensitivity. After independently evaluating the signal events by the productions (i) and (ii), we combine the plots of the sensitivity of ILC.

\subsection*{$U_e$ dominance}

\begin{figure*}[t!]
\centering
\includegraphics[width=0.6\textwidth]{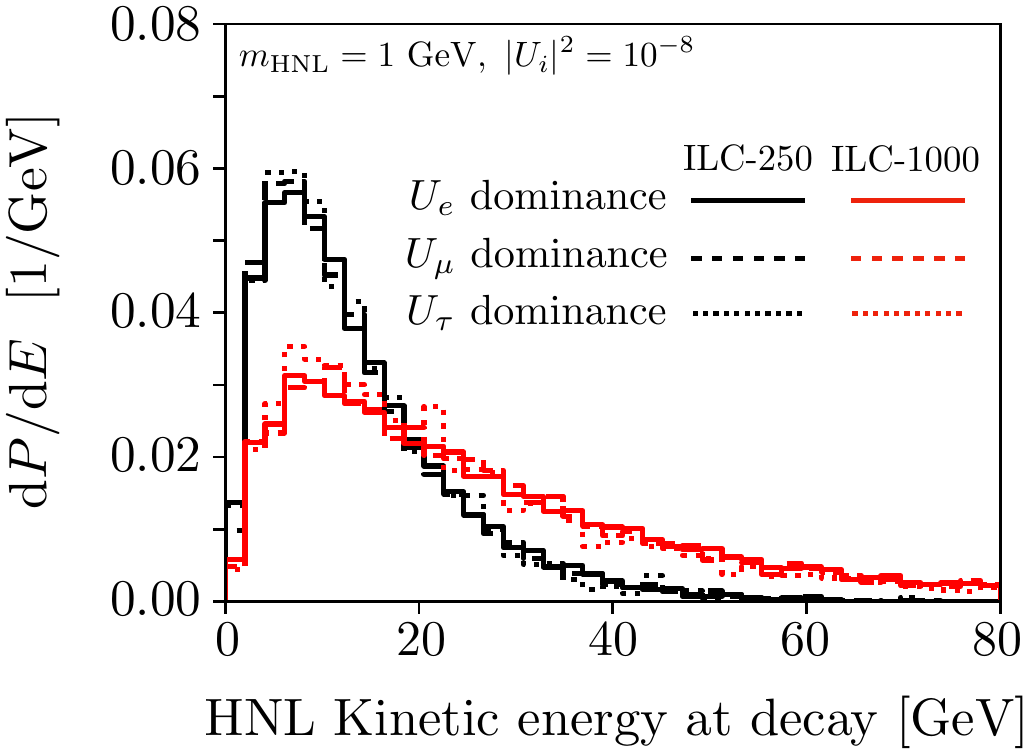}
\caption{
The normalized kinetic energy distribution of HNLs when they decay inside the decay volume. The mixing angle squared is $10^{-8}$, and the mass is 1~GeV. The red (black) lines are for ILC-1000(250). 
\label{fig:kinetic}
}
\end{figure*}

Fig.~\ref{fig:sensitivityUe} shows the sensitivity of ILC for the HNL whose mixing to active neutrinos is dominated by the $U_e$ mixing. The region above the red (black) curves is the region {expected $95\%$ C.L. exclusion sensitivity} that can be searched by ILC-1000 (ILC-250) with 10-year statistics. The gray-shaded regions are constrained from past experiments which is discussed in Sec.~\ref{sec:bounds}. 
Also, for a single benchmark point, we show the kinetic energy distribution of the decayed HNLs inside the decay volume in Fig.~\ref{fig:kinetic}.

Let us explore the results in Fig.~\ref{fig:sensitivityUe}, highlighting each of the HNL production processes.
\begin{enumerate}[(a)]
\item Meson and $\tau$ lepton decays

In the small $|U_e|^2$ regions where the lifetime of HNL becomes longer, the decay probability in Eq.~\eqref{eq:decP} is approximately $d P_{\rm dec}^{\rm dump}/dz \simeq 1/l^{\rm (lab)}_{X}$. Depending on the mass of HNL, the HNLs are produced by the decay of mesons and $\tau$ lepton, see Figs.~\ref{fig:brUeli}, \ref{fig:brUeD}, \ref{fig:brUeB}, and \ref{fig:brUetau} in Appendix~\ref{sec:Br}.
Below the kaon mass $m_{K}\simeq 0.5~{\rm GeV}$, the HNLs are mainly produced by the kaon decay. For $m_{\rm HNL}\leq m_{D,\tau}-m_e\sim 2~{\rm GeV}$, the $D(D_s)$ meson decay dominates the HNL productions. In the region of $1~{\rm GeV}\lesssim m_{\rm HNL}\lesssim 2~{\rm GeV}$ there are many thresholds of production mode, such as $D\to K+e+{\rm HNL}$, $D_s\to \eta+e+{\rm HNL}$, $D\to \pi+e+{\rm HNL}$ and $\tau\to \nu_{\tau}+e+{\rm HNL}$, see Fig.~\ref{fig:brUeD}.

The limit obtained by the Monte-Carlo simulation can be checked using the approximate formula Eq.~\eqref{eq:N1approx}. As explained in Appendix \ref{app:Com}, the approximate formula agrees well with the results of the Monte-Carlo simulation. For the electron beam energy $E_{\rm beam}=500~{\rm GeV}$, the number of events by the leptonic $D$ meson decays $D^\pm\to e^\pm +{\rm HNL}$ is given by 
\begin{align}
N_{\rm signal}^{\rm (i)}&\sim \left(\frac{N_{\rm EOT}}{4\times 10^{22}}\right)  \left(\frac{l_{\rm dec}}{50~{\rm m}}\right)\left(\frac{r_{\rm det}}{3~{\rm m}}\right)^2\left(\frac{81~{\rm m}}{l_{\rm dump}+l_{\rm sh}}\right)^2\left(\frac{|U_e|^2}{ 10^{-10}}\right)^2  \left(\frac{m_{\rm HNL}}{1~{\rm GeV}}\right)^4,
\end{align}
where we used follwing approximations:
\begin{align}
 {\rm Br}(D^{\pm}\to e^{\pm}+{\rm HNL})\propto |U_e|^2 , \quad   
 {\rm Br}_{\rm vis}\simeq 1, \quad 
 (l^{\rm lab}_X)^{-1}\propto |U_e|^2  m_{\rm HNL}^4/E^{\rm lab}_{\rm HNL} \ . 
\end{align}
Near the kinematic threshold, the number of events is rapidly decreased by a suppression of the branching ratio ${\rm Br}(D^{\pm}\to e^{\pm}+{\rm HNL})$. For $m_D-m_e \lesssim m_{\rm HNL} \lesssim m_B-m_e$, the $D$ meson decay channels are closed, and the $B$ meson decay channels become dominant in the HNL productions up to around $m_{\rm HNL}\sim 3$ GeV.

\item Direct production

Above $m_{\rm HNL}\sim 3$ GeV, the direct production channel with the incoming electrons and positrons become significant. 
In particular, the production with the high-energy primary electron is essential to extend the mass reach of the $U_e$ dominant scenario.
In the small $|U_e|^2$ regions, the decay probability in Eq.~\eqref{eq:decP} is approximately $d P_{\rm dec}^{\rm dump}/dz \simeq 1/l^{\rm (lab)}_{X}$, and the energy of the produced HNL is approximately $E^{\rm lab}_{\rm HNL}\simeq E_{e^{\pm}}$. Then, the number of events for $E_{\rm beam}=500~{\rm GeV}$ is given by
\begin{align}
N_{\rm signal}^{\rm (ii),e^{\pm}}\sim \left(\frac{N_{\rm EOT}}{4\times 10^{22}}\right)\left(\frac{l_{\rm dec}}{50~{\rm m}}\right)\left(\frac{r_{\rm det}}{3~{\rm m}}\right)^2\left(\frac{81~{\rm m}}{l_{\rm dump}+l_{\rm sh}}\right)^2\left(\frac{|U_e|^2}{10^{-10}}\right)^2 \left(\frac{m_{\rm HNL}}{10~{\rm GeV}}\right)^6,
\end{align}
where we use, 
\begin{align}
    &dl_{e^{\pm}}/d E_{e^{\pm}}\simeq \left(dl_{e^{\pm}}/d E_{e^{\pm}}\right)_{\rm primary}\propto 1/E_{e^{\pm}},\quad
    d^2\sigma(e^{\pm} N\to {\rm HNL})/dx dy \propto |U_e|^2 E_{e^{\pm}},
    \nonumber\\
    &(l^{(\rm lab)}_X)^{-1}\propto |U_e|^2 m_{\rm HNL}^6/E^{\rm lab}_{\rm HNL},\quad
    {\rm Br}_{\rm vis}\simeq 1,\quad
    y\lesssim x^{-1}E_{e^{\pm}} r^2_{\rm det}(l_{\rm dump}+l_{\rm sh})^{-2}.
\end{align}

In the larger $|U_e|^2$ regions, where $l^{\rm (lab)}_{\rm HNL}\ll l_{\rm dump}+l_{\rm sh}$, the shape of the upper contour lines in Fig.~\ref{fig:sensitivityUe} is determined by the probability to decay inside the decay volume given in Eq.~\eqref{eq:decP}. 
The contour is characterized by the exponent of the decay probability
\begin{align}
\frac{m_{\rm HNL}\Gamma_{\rm HNL}}{E^{\rm lab}_{\rm HNL}}(l_{\rm dump}+l_{\rm sh})\sim {\rm const}.\label{eq:diupp}
\end{align} 
Combining $E^{\rm lab}_{\rm HNL}\simeq E_{\rm beam}$ and $\Gamma_{\rm HNL}\propto |U_e|^2\cdot m_{\rm HNL}^5$, Eq.~\eqref{eq:diupp} becomes $|U_e|^2\propto m_{\rm HNL}^{-6}$ , which is consistent with the parameter dependence of Fig.~\ref{fig:sensitivityUe}. 

\end{enumerate}

\begin{figure*}[t!]
\centering
\includegraphics[width=0.6\textwidth]{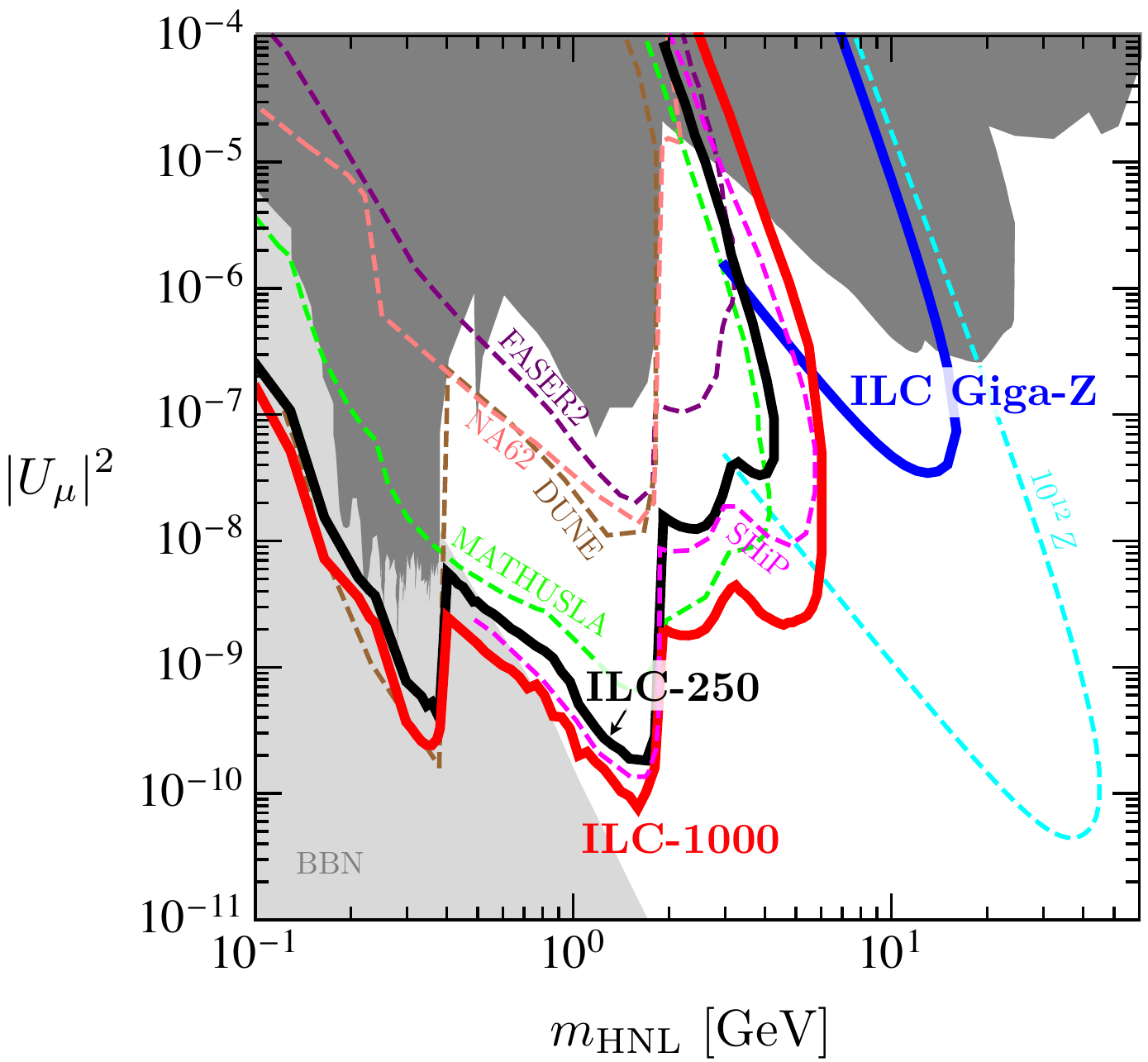}
\caption{
Sensitivity reach of ILC beam dump experiment to HNLs mixing with the mu-neutrino in the mass and mixing
plane. For the description of this plot, see the caption of Fig.~\ref{fig:sensitivityUe}.  \label{fig:sensitivityUmu}
}
\vspace{20pt}
\includegraphics[width=0.6\textwidth]{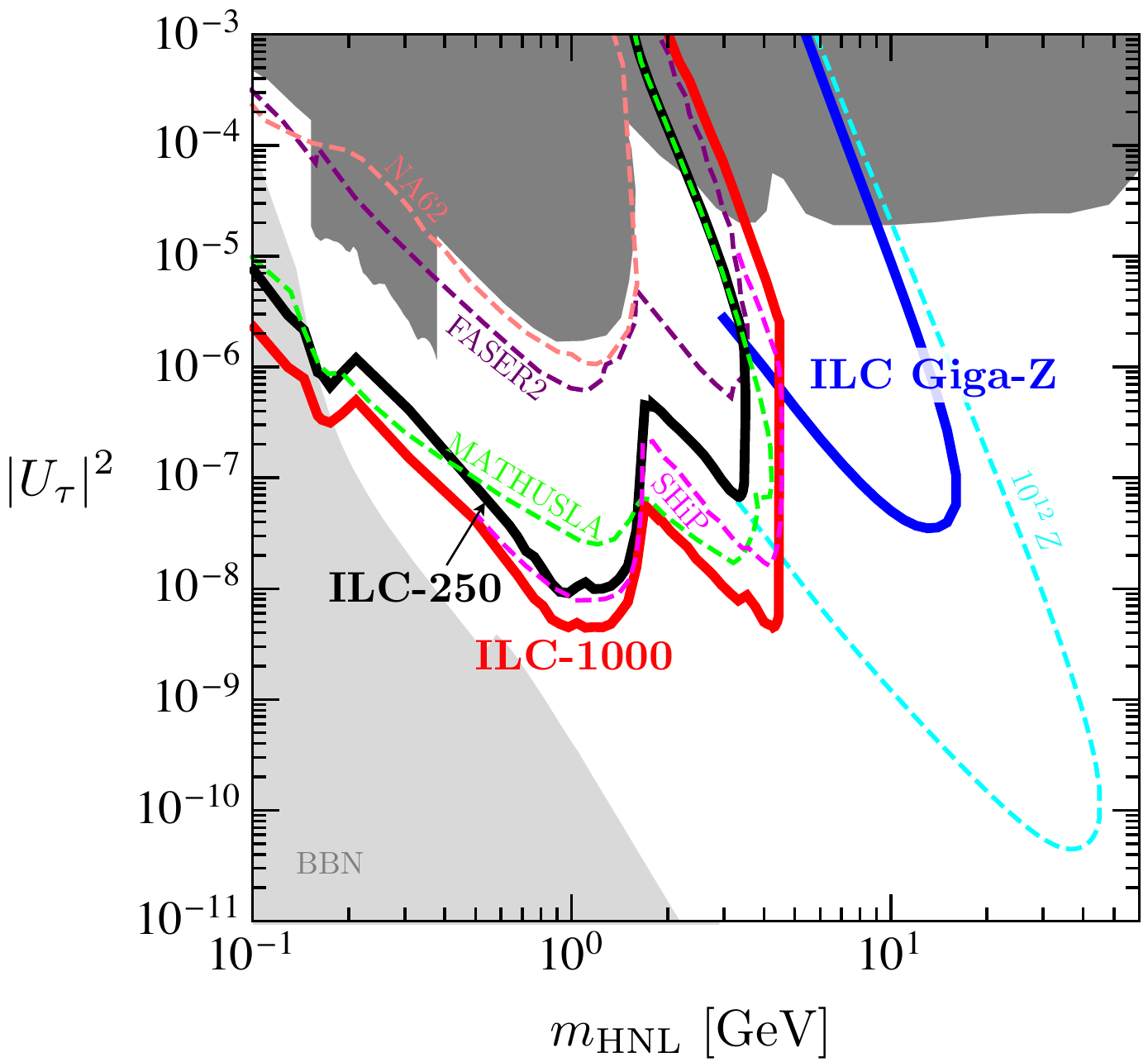}
\caption{
Sensitivity reach of ILC beam dump experiment to HNLs mixing with the tau-neutrino in the mass and mixing
plane. For the description of this plot, see the caption of Fig.~\ref{fig:sensitivityUe}. 
\label{fig:sensitivityUtau}
}
\end{figure*}

\subsection*{$U_\mu$ dominance}
Fig.~\ref{fig:sensitivityUmu} summarizes the prospects of the ILC beam dump experiment in ILC-1000 and ILC-250 for the HNL mixing dominantly with $\nu_{\mu}$.
The red (black) curves show the expected sensitivity for $95\%$ C.L. exclusion of ILC-1000 (ILC-250) with 10-year statistics given by the meson and $\tau$ lepton decays, and the DIS process for incoming muons from the EM shower. 
We provide approximated formulae of the number of signal events for each HNL production mode for ease of understanding.

\begin{enumerate}[(a)]
\item Meson and $\tau$ lepton decays

The production process from meson and $\tau$ lepton decays of the HNL that mix dominantly with $\nu_{\mu}$ can be calculated in parallel with $U_e$ dominant case except for minor threshold difference.
The mass dependence of the dominant HNL production process is the same as that of $U_e$ dominant case with $e$ replaced with $\mu$.

As shown in Figs.~\ref{fig:brUmuD} and \ref{fig:brUmuB}, for $m_D-m_{\mu} \lesssim m_{\rm HNL} \lesssim m_B-m_{\mu}$, 
the $B$ meson decay gives a significant contribution to the HNL production.
The zigzag curves near $m_{\rm HNL}\sim 3~{\rm GeV}$ correspond to the threshold of $B\to D+{\mu}+{\rm HNL}$.
Above $m_{\rm HNL}\simeq 3~{\rm GeV}$, the leptonic $B$ meson decay such as $B^{\pm}\to \mu^{\pm} +{\rm HNL}$ dominates the HNL productions.

\item Direct production

As shown in Fig.~\ref{fig:exp}, incident real photons produce muon pairs in the beam dump through the electromagnetic interaction with the nucleus or nucleon~\cite{Sakaki:2020cux}, and the HNL can be generated by the DIS process in the muon shield.
The muon pair production cross section is about $(m_{\mu}/m_e)^2\simeq 10^{5}$ times smaller than that of the electron pair production.

In the small $|U_{\mu}|^2$ regions, the projected sensitivity scales with the ratio of numbers of muon and electron in the 
shower, and the projected sensitivity via the DIS process is less significant than the 
$U_e$ dominant case.
On the other hand in the large $|U_{\mu}|^2$ regions, the number of signal events  is mostly determined by the muon shield length and  less sensitive to the number of muon pairs because the lifetime is shorter.
The shape of the upper side of contours in Fig.~\ref{fig:sensitivityUmu} is determined by the exponential factor in Eq.~\eqref{eq:decP}, which is characterized by 
\begin{align}
    \frac{m_{\rm HNL} \Gamma_{\rm HNL}}{E^{\rm lab}_{\rm HNL}}(l_{\rm sh}-\delta_{\mu})\sim {\rm const}.\label{eq:larUmu}
\end{align}
Combining $E^{\rm lab}_{\rm HNL}\simeq E_{\rm beam}$ and $\Gamma_{\rm HNL}\propto |U_{\mu}|^2\cdot m^5_{\rm HNL}$, Eq.~\eqref{eq:larUmu} becomes $|U_{\mu}|^2\propto m^{-6}_{\rm HNL}$. 
Since the DIS process with a muon happens in the muon shield,  the shorter distance from the HNL production point to the decay volume enhances the acceptance compared the DIS process in the beam dump.
  
\end{enumerate}

\subsection*{$U_\tau$ dominance}

The projected sensitivities in the $U_\tau$ dominant scenario are shown in Fig.~\ref{fig:sensitivityUtau} with a similar notation as in the $U_{e,\mu}$ dominance.
The red (black) curves show the expected $95\%$ C.L. exclusion sensitivity with 10-year statistics in ILC-1000 (ILC-250) by $B$ and $D_s$ mesons and $\tau$ lepton decay production. The direct (DIS) production is absent in this case due to very limited flux of $\tau$ leptons.

As shown in Figs.~\ref{fig:production}, \ref{fig:brUtauD}, \ref{fig:brUtauB}, and \ref{fig:brUtautau}, below the threshold of $m_{D_s}-m_{\tau}\simeq 0.2~{\rm GeV}$,
the main HNL production channel is the $D_s\to \tau+{\rm HNL}$ decay.
For $0.2~{\rm GeV}\lesssim m_{\rm HNL}\leq m_{\tau}-m_e\simeq 1.8~{\rm GeV}$, the production of HNL is dominated by $\tau$ lepton decay.
Most of the $\tau$ leptons are produced by $D_s$ meson decays, and the $\tau$ pair production in the electromagnetic showers are subdominant, which is therefore dropped in the HNL study. 
The branching ration of $D_s\to \tau +\nu_{\tau}$ is $5.48\%$, and an energy dependence of $\tau$ lepton spectra is determined by that of $D_s$ meson spectra in Fig.~\ref{fig:production}.
The main decay channels of $\tau$ are $\tau\to \rho +{\rm HNL}$, $\tau\to \mu+\nu+{\rm HNL}$, $\tau\to e+\nu+{\rm HNL}$, $\tau\to \pi+{\rm HNL}$, $\tau\to K^{\ast}+{\rm HNL}$, and $\tau\to K+{\rm HNL}$, see Fig~\ref{fig:brUtautau}.

In the small $|U_{\tau}|$ regions with the longer lifetime of HNL, the decay probability in Eq.~\eqref{eq:decP} is approximated by $d P_{\rm dec}^{\rm dump} /dz\simeq 1/l^{\rm (lab)}_X$.
For the electron beam energy $E_{\rm beam}=500~{\rm GeV}$, the number of events by the $\tau^{\pm}\to \rho^{\pm} +{\rm HNL}$ is approximately given by
\begin{align}
    N_{\rm signal}^{\rm (i)}&\sim \left(\frac{N_{\rm EOT}}{4\times 10^{22}}\right)\left(\frac{l_{\rm dec}}{50~{\rm }}\right)\left(\frac{r_{\rm det}}{3~{\rm m}}\right)^2 \left(\frac{81~{\rm m}}{l_{\rm dump}+l_{\rm sh}}\right)^2 \left(\frac{|U_{\tau}|^2}{5\times 10^{-8}}\right)^2 \left(\frac{m_{\rm HNL}}{0.5~{\rm GeV}}\right)^4,\label{eq:nsigapp} 
\end{align}
where we use
\begin{align}
    {\rm Br}(\tau^{\pm}\to \rho^{\pm}+{\rm HNL})\propto |U_{\tau}|^2,\quad
    {\rm Br}_{\rm vis}\simeq 0.01,\quad
    (l^{\rm lab}_X)^{-1}\propto |U_{\tau}|^2 \cdot m^4_{\rm HNL} /E^{\rm lab}_{\rm HNL}.
\end{align}
For $1.8~{\rm GeV}\lesssim m_{\rm HNL}$, the $B$ meson decays dominate the HNL productions, see Figs.~\ref{fig:brUtauD}, \ref{fig:brUtauB}, and \ref{fig:brUtautau}.

\subsection{Sensitivities from $Z$~decays at ILC and FCC-ee}\label{sec:ZatILC}

The future $e^+ e^-$ colliders can be seen as a $Z$ boson factory, such as Giga-$Z$ program of the ILC and Tera-$Z$ program of the CERN Future $e^+ e^-$ Circular Collider, dubbed FCC-ee. The HNLs can leave a clear signal with displaced tracks at $e^+ e^-$ colliders once  they are produced via $Z\to N \bar\nu, \bar N \nu$. This type of signal was examined at DELPHI detector of the Large Electron-Positron collider (LEP) \cite{DELPHI:1996qcc}. We briefly study the future sensitivities of the HNL search using the displaced tracks from the $Z$ decay, which is complementary to the sensitivities of the ILC beam dump experiment.

The proposed ILC detector \cite{Behnke:2013lya} has three layers of vertex detector (VTX) starting from 1.5~cm to 19.5~cm surrounding the beam,  Time Projection Chambers (TPC) as the main tracking detector spreading from 33~cm to 170~cm, and  two layers of silicon strip detectors (SIT) arranged between TPC and VTX. We assume the ILC detector is sensitive to the HNL signal with negligible background if the HNL decays between 3~cm and 170~cm from the collision point. We adopt  the radius-dependent track-detection efficiency $\epsilon_{\rm trk}$ linearly decreasing from $r = 3$~cm ($\epsilon_{\rm trk}=100\%$) to $r = 170$~cm ($\epsilon_{\rm trk}=0\%$)~\cite{Bertholet:2021hjl}. Then, we assume $10^9$ $Z$ decays and  zero background after requiring the displaced tracks and consider all the HNL decays except for the fully invisible mode, $N\to \nu\nu\bar\nu$. The $95\%$~C.L. future sensitivity corresponds to three HNL decays with the multi-displaced tracks in the fiducial volume, and the result is included in Figs.~\ref{fig:sensitivityUe},~\ref{fig:sensitivityUmu},~and~\ref{fig:sensitivityUtau}.

Similarly, we study the future projection at the FCC-ee. For the simplicity, we take the same detector setup of the ILC and rescale the statistics assuming $10^{12}$ $Z$ decays. This is also shown in Figs.~\ref{fig:sensitivityUe},~\ref{fig:sensitivityUmu},~and~\ref{fig:sensitivityUtau}. Ref.~\cite{Blondel:2014bra} studied also the projected sensitivities at the FCC-ee, but our estimate is different with respect to the detector coverage and the signal efficiency. 


\subsection{Existing constraints on HNL and projected sensitivities at other experiments}\label{sec:bounds}

The presence of HNL can have drastic consequences on early-universe observables and has been explored extensively in the literature. The HNL is thermally produced in the early universe, and its late decay can disrupt the standard  Big Bang Nucleosynthesis. The typical constraint is for $\tau_{\rm HNL}\lesssim 1\rm s$, see \cite{Abazajian:2012ys} and references therein. In addition, Ref ~\cite{Boyarsky:2020dzc}  studies that the pions from the HNL hadronic decay alter the decoupling of neutrons from the prediction of the standard BBN, which affects the ratio of neutrons to protons. The upper limit on the HNL lifetime is obtained from the $^4$He abundance as $\tau_{\rm HNL}\lesssim 0.02$~s (see also other recent studies \cite{Alonso-Alvarez:2022uxp, Bondarenko:2021cpc}).

The long-lived HNLs are also extensively searched in the accelerator-based experiments. The HNLs may be produced directly in the beam-target collisions, or in the decays of secondary particles such as $B$, $D$ mesons and $\tau$ leptons. The searches are carried out either by detecting the displaced decay of the HLN or the significant missing momentum of the events. In the following, we briefly list the existing constraints included in Fig.~\ref{fig:sensitivityUe}, \ref{fig:sensitivityUmu}, and \ref{fig:sensitivityUtau}.

\begin{itemize}
\item CHARM  proton beam dump experiment --- 
Searches for HNLs produced at a proton beam dump operating at the 400 GeV CERN SPS are  sensitive to all the scenarios we consider~\cite{CHARM:1983ayi, CHARM:1985nku}. Recently, the result of the $U_\tau$ dominant scenario was reanalyzed, and the bound was improved for  $ 0.3~{\rm GeV}\lesssim m_{\rm HNL}\lesssim 1.5~{\rm GeV}$~\cite{Boiarska:2021yho}.

\item Other beam damp/neutrino beam experiments --- 
In addition to the CHARM experiment, the HNL searches were conducted at a variety of fixed target and neutrino experiments, such as NuTeV~\cite{NuTeV:1999kej}, BEBC~\cite{WA66:1985mfx}, and PS191~\cite{Bernardi:1987ek}. They are mostly sensitive to the $U_\mu$ mixing, while PS191 placed a strong bound on the $U_e$ dominant scenario as well.

\item Long-baseline neutrino experiment --- 
At the T2K experiment, the near detector ND280 was utilized to search for the HNL~\cite{T2K:2019jwa}. The detection principle is similar to the one at  PS191. 
We include the bounds of ``single-channel'' analysis where only one of  $U_\ell$ is non-zero at a time. In the future, this type of search can be performed at the near detector of the DUNE experiment~\cite{Coloma:2020lgy}.

\item Super-Kamiokande --- 
The HNLs can be copiously produced from kaon and pion decays in atmospheric showers and decay in the Super-Kamiokande detector. In~\cite{Coloma:2019htx}, the authors analyzed the Super-Kamiokande data~\cite{Super-Kamiokande:2017yvm}, and obtained bounds on the HNL in all the scenarios. The bound on $U_{\tau}$ is as strong as the T2K one.

\item Pion decay --- 
Precision measurements of charge pions can test $\pi^+ \to \ell^+ +\rm HNL$  with no subsequent HNL decay which is sensitive to  the $U_e$ and $U_{\mu}$ mixings \cite{PIENU:2017wbj, PIENU:2019usb}. It gives a unique bound in the low mass region $m_{\rm HNL}\sim 0.1~{\rm GeV}$ of the $U_{e}$ dominant scenario.

\item  Kaon decay ---  
Similar to the search with the charged pion, the HNL searches have been conducted by $K^+\to \ell^+ +\rm HNL(invisible)$ at NA62~\cite{NA62:2020mcv, NA62:2021bji}, E949~\cite{E949:2014gsn}, and KEK\cite{Yamazaki:1984sj}.  Typically they give the best limit near the threshold of $m_{\rm HNL}\lesssim m_{K^+}-m_{\ell^+}$.

\item $B$ meson and $\tau$ lepton decay --- 
The heavier HNL can be looked for from the $B$ meson decays. The HNL production by $B\to X \ell +{\rm HNL}\ (\ell=e,\mu)$  with a subsequent displaced decay of ${\rm HNL}\to \ell \pi$ was examined using the Belle data in~\cite{Belle:2013ytx}.  The $B$-factories also produce $\tau$ leptons  copiously, and  the search of long-lived HNL from $\tau$ lepton decays at Belle and Babar placed a bound on $U_{\tau}.$ \cite{Dib:2019tuj}

\item $Z$ boson decay --- 
The search of $Z$ boson decaying into HNL using the DELPHI data at the LEP collider constraints all the scenarios. It sets strongest limit  in mass range around $m_{\rm HNL}\sim {\cal O}(10)~\rm GeV$ \cite{DELPHI:1996qcc}. See Sec.~\ref{sec:ZatILC} for more detail and  the  projection at the future $e^+e^-$ colliders.

\item Large Hadron Collider --- 
At the ATLAS and CMS detectors of LHC, the $W$ boson mediated process efficiently produces the HNL. Depending on the final states and decay length, different analyses were performed at ATLAS~\cite{ATLAS:2019kpx, ATLAS:2022atq} and CMS~\cite{CMS:2018jxx, CMS:2018iaf, CMS:2022fut}. In particular, the search for the displaced decays of HNL~\cite{CMS:2022fut, ATLAS:2022atq} excludes the large parameter space.

\end{itemize}

Among studies of the future HNL searches, we include projected sensitivities at FASER2~\cite{FASER:2018eoc}, NA62~\cite{Beacham:2019nyx}, DUNE~\cite{Coloma:2020lgy}, SHiP~\cite{Alekhin:2015byh}, and MATHUSLA~\cite{Curtin:2018mvb}  in Figs.~\ref{fig:sensitivityUe}, \ref{fig:sensitivityUmu}, and \ref{fig:sensitivityUtau} to compare  to the sensitivities at the ILC beam dump experiment. Other experiments including LHCb~\cite{Antusch:2017hhu, Cvetic:2019shl}, Codex-b~\cite{Aielli:2019ivi}, Belle~II~\cite{Dib:2019tuj}, Dark-Quest~\cite{Batell:2020vqn}, IceCube~\cite{Coloma:2017ppo},  ATLAS and CMS  at HL-LHC ~\cite{ATLAS:2022atq, CMS:2022fut, Cheung:2020buy}, and FCC-hh~\cite{Boyarsky:2022epg}  can also search for the HNLs beyond the current experimental constrains. 

In Figs.~\ref{fig:sensitivityUe}, \ref{fig:sensitivityUmu}, and \ref{fig:sensitivityUtau}, the estimated sensitivity contours of ILC is better than the other proposed searches. 
The ILC sensitivity is very close to the one of SHiP in the low mass region ($m_{\rm HNL}<2$~GeV). Above 2~GeV, the sensitivity of the ILC beam dump experiment is better than SHiP because the initial electron energy is high enough to produce $B$ mesons at a higher rate. Thanks to the higher HNL mass, the HNL decay products are more clearly separated from the background so that the neutrino background would be less critical in the region.


\section{Discussions}\label{sec:discussion}

The ILC beam dump experiment is a seamless extension of the ILC program, which provides a unique opportunity to test feebly interacting light particles. The electron beam energy of ILC is much higher than the ones at the current and past electron beam dump experiments, and therefore, not only the light mesons but also the heavier SM particles, such as heavy flavor mesons and $\tau$ lepton, can be produced in the beam dump. Moreover, a large number of electrons on target are expected thanks to the high-intensity beam. In many BSM scenarios, new particles can be efficiently produced by decays of the SM particles. Therefore, it is essential to estimate the production rate of the SM particles at the ILC beam dump experiment. In this paper, we evaluate the light and heavy mesons and $\tau$ lepton spectra at the decay for the first time using  PHITS and PYTHIA8. PHITS is responsible for producing and transporting light SM particles, and we incorporate the heavy meson productions calculated by PYTHIA8 into PHITS. The main results are in Fig.~\ref{fig:decay} for the light mesons and in Fig.~\ref{fig:production} for the heavy mesons and $\tau$ lepton.

The spectra of SM particles can be used to estimate the  yield of BSM particles from SM particle decay. As a demonstration, we studied the projected sensitivity of the heavy neutral leptons at the ILC beam dump experiment. Figs.~\ref{fig:sensitivityUe}, \ref{fig:sensitivityUmu}, and \ref{fig:sensitivityUtau} show that  ILC would explore the HNLs with the heavier mass and small mixing and cover the large parameter space motivated by the baryon asymmetry of the Universe. For the reader's convenience, we show the sensitivities after one year of operation in Appendix~\ref{app:one-year}.
We use the Monte-Carlo simulation to evaluate the HNL signal from the SM particle decay, and the results are well reproduced by the coarse-grained integration method convolving the SM particle spectra given in Figs.~\ref{fig:decay} and \ref{fig:production}.

Apart from the SM particle decays, the HNLs can be produced through various processes at ILC. The EM shower can directly create an HNL via DIS, and we account for this production by the coarse-grained integration method. Moreover, copious $Z$ decays expected in the primary detector of ILC would realize  a different search method for the mildly long-lived HNLs.

Other than the HNLs, many motivated long-lived particles can be predominantly produced by the SM particle decay. For example, dark scalars such as the QCD axion and the Higgs portal scalar can be efficiently produced via flavor-changing decays of meson (especially $B\to K X$ and $K\to \pi X$ where $X$ denotes a long-lived particle).  The heavy mesons produced from the high energy electron beam allow us to probe long-lived particles heavier than several GeV. Another advantage of the abundant heavy mesons is that the experiment is sensitive to a new particle that preferably couples to the third generation fermions ($b$ quark or $\tau$ lepton). One can estimate the sensitivity of these particles at the ILC beam dump experiment based on our results in Figs.~\ref{fig:decay}, \ref{fig:production} and the approximate formula like Eqs.~\eqref{eq:sig}.

\section*{Note added}\label{sec:note}
While completing this work, we became aware of \cite{Giffin:2022rei}, which considers related topics. A main difference is that Ref.~\cite{Giffin:2022rei} does not include meson production through the EM shower photons, which is the dominant channel of meson production in our work. 
Also, our setup of the experiment is slightly different from theirs. We added projections for one year operation as assumed in Ref.~\cite{Giffin:2022rei}.

\section*{Acknowledgement}\label{sec:ackn}
MN is supported by 	 Grant-in-Aid for Scientific Research on Innovative Areas(16H06492) JSPS KAKENHI 22K03629, YS is supported by JSPS KAKENHI JP21H05466.  KT is supported by in part the US Department of Energy grant DE-SC0010102 and JSPS KAKENHI 21H01086. 

\appendix
\section{Track lengths of electromagnetic showers and muon spectra behind beam dump}
\label{app:track}
For convenience, we summarize the track lengths of electromagnetic showers and the muon spectra behind the beam dump we used to estimate the number of signal events.
The details of the calculations are explained in Ref.~\cite{Sakaki:2020mqb,Sakaki:2020cux,Asai:2021ehn}.
The track lengths of electromagnetic showers in a thick beam dump $(\gtrsim 10 X_0)$ filled with water are expressed as follows~\cite{Asai:2021ehn},
\begin{alignat}{2}
u\frac{\dd \hat l_{\gamma}}{\dd u}&=
   \frac{0.572}{u}+0.067\ln(1-\sqrt{u})
&&\text{(photon from $e^\pm$ beam)},\\
u\frac{\dd \hat l_{e}}{\dd u}&=
\biggl(u\frac{\dd \hat l_{e}}{\dd u}\biggr)_{\rm primary}
+
\biggl(u\frac{\dd \hat l_{e}}{\dd u}\biggr)_{\rm shower}
\quad&&\text{($e^\pm$ from $e^\pm$ beam)},
\\
u\frac{\dd \hat l_{e}}{\dd u}&=
\biggl(u\frac{\dd \hat l_{e}}{\dd u}\biggr)_{\rm shower}
&&\text{($e^\mp$ from $e^\pm$ beam)},
\end{alignat}
with
\begin{align}
\biggl(u\frac{\dd \hat l_{e}}{\dd u}\biggr)_{\rm primary}
 &= 0.581 + 0.131 \left( \frac{u}{1-u} \right)^{0.7},\\
\biggl(u\frac{\dd \hat l_{e}}{\dd u}\biggr)_{\rm shower}&=
  \frac{1-u}{u}(0.199-0.155u^2),
\end{align}
where $u=E_i/E_{\rm beam}$ with $E_{\rm beam}$ being the energy of the injected beam.
Here, $\hat{l}_{i}$ denotes a dimensionless normalized track lengths of a particle $i$ and is defined by
\begin{align}
    \hat{l}_i =\frac{\rho}{X_0}l_i,
\end{align}
where $X_0=36.08~\mathrm{g/cm^2}$ and $\rho=1.00~\mathrm{g/cm^3}$ are the radiation length of water and the density of water, respectively. 

The energy distribution of the muon yield per incident electron behind the beam dump is given by~\cite{Sakaki:2020mqb,Sakaki:2020cux}
\begin{align}
    \frac{dY_{\mu 0}}{dE_{\mu 0}}=\frac{0.572 E_{\rm beam}}{\ln (183 Z^{-1/3})}\left(\frac{m_e}{m_{\mu}}\right)^2\left(\frac{1}{E^2_{\mu 0}}-\frac{1}{E^2_{\rm beam}}\right),
\end{align}
with the atomic number of water $Z=7.5$. 
The track length of muons in lead is given by
\begin{align}
    \frac{dl_{\mu}}{dE_{\mu}}=\frac{1}{{\langle dE/dx\rangle}_{\rm lead}},
\end{align}
with the stopping power for lead ${{\langle dE/dx\rangle}_{\rm lead}}=0.02~{\rm GeV/cm}$.

\section{Evaluation of $D$~meson production}\label{app:Dmeson}

\begin{figure}[t]
    \centering
    \includegraphics[width=0.7\textwidth]{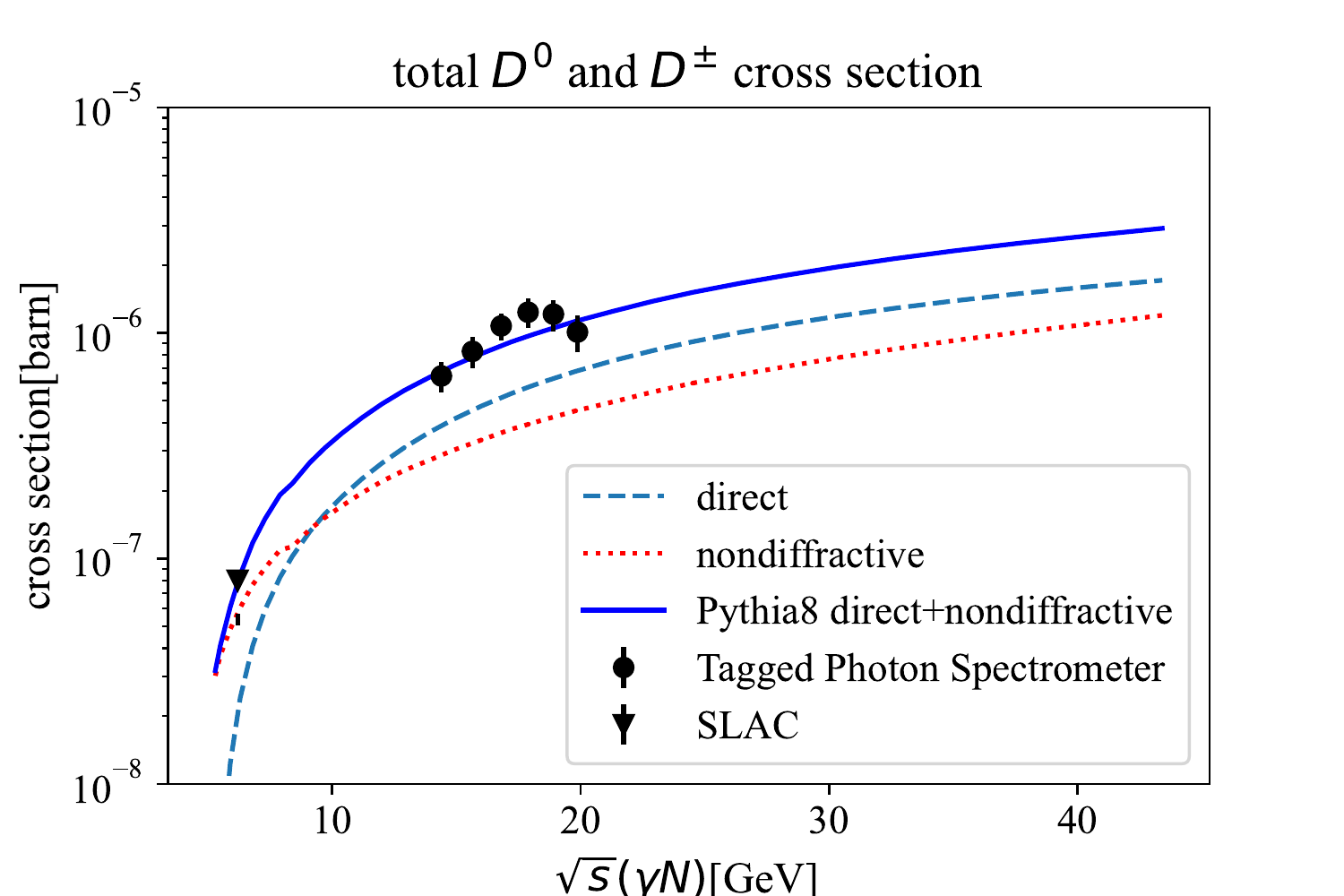}
    \caption{The photoproduction cross section of $D$ mesons ($D^{\pm}$, $D^0$ and $\overline{D}^0$), and the PYTHIA8 prediction.  The dotted and long dashed lines correspond to non-diffractive and direct production cross section of charm quark. The solid line is the total charm meson production cross section. They are consistent with the measured cross section of Tagged Photon Spectrometer (circle) \cite{TaggedPhotonSpectrometer:1989bpi} and SLAC (triangle) \cite{SLACHybridFacilityPhoton:1985baw}.
    }
    \label{fig:my_label}
\end{figure}

PHITS ignores charm and bottom quark production because the production cross section is minor relative to the one of the light meson. However, they are important sources of the heavy neutral leptons. Therefore, we estimate the probability of the $D$ and $B$ meson productions from $\gamma p$ and $\gamma n$ interactions by PYTHIA8. We ignore energy loss of the heavy mesons in the material because they decay immediately after the production. 

The photoproduction cross section used in this paper consists of the non-diffractive  and direct production cross sections of PYTHIA8.\footnote{The diffractive cross section is small and therefore ignored.} The non-diffractive cross section is empirically estimated by the MPI  (multi-parton interaction) model of PYTHIA8. 
We found that the photons of energy at $E_{\gamma}\sim 30~\rm GeV$ contribute to the production of HNL through the $D$ meson decay.  For $E_{\gamma}\gg 30$~GeV, the $D$ meson flux is suppressed mildly because of the reduction of average track length  $dl_\ell/dE_i$ of photons at the high energy. See  Fig.  9 of \cite{Asai:2021ehn}.
The photon differential track length for a thick beam dump ($>X_0$) simulated by  PHITS is summarized in Appendix~\ref{app:track}.
At low $E_{\gamma}$, the $D$ meson flux is suppressed kinetically, while the $\hat{l}_{\gamma}$ increases in proportion to $1/u^2$. 
The charmed meson photoproduction cross section is measured by the Tagged Photon Spectrometer Collaboration~\cite{TaggedPhotonSpectrometer:1989bpi} and the SLAC Hybrid Facility Photon Collaboration~\cite{SLACHybridFacilityPhoton:1985baw}.

In Fig \ref{fig:my_label}, we show the measured cross section(the dots and the triangle ) and the non-diffractive  and direct production cross sections of PYTHIA8 as a function of $\sqrt{s}$. The sum of the two cross sections (the solid line) is consistent with the measured cross sections.\footnote{The prediction of the non-diffractive cross section shows a minor discontinuity around 7~GeV, but the effect in our numerical calculation is negligible.}

\section{Production cross sections of the HNLs by deep inelastic scattering}
\label{app:DIS}
We consider the production of the HNLs through deep inelastic scatterings (DIS) in addition to the SM particle decays.
For convenience, we summarize the production cross sections used in our numerical evaluations; using Eqs.~\eqref{eq:DISel} and \eqref{eq:DISmu}, one can estimate the number of signal events.
 
The cross sections for the DIS on the unpolarized nucleon are given by five structure functions~ \cite{Albright:1974ts,Kamal:1979ev,Anselmino:1993tc,Formaggio:2012cpf,Grover:2018ggi,ParticleDataGroup:2020ssz,Kretzer:2002fr}
\begin{align}
    &\frac{d^2 \sigma(\ell^{-}  p\to {\rm HNL}~X)}{dx dy}
    \notag\\
    &= |U_{\ell}|^2\frac{G_F^2 M_N E_{\ell^-}}{\pi \left(1+Q^2/M_W^2\right)^2}\Bigg[&
    \left(xy^2 +\frac{y \left( m_{\ell}^2 +m^2_{\rm HNL}\right)}{4 M_N E_{\ell^-}}-\frac{(m_{\ell}^2-m_{\rm HNL}^2)^2}{8 x M_N^2 E_{\ell^-}^2}\right) F_{1,p}^{W^-}(x,Q^2)\notag
\\
&&+\bigg(1-y-\frac{M_N x y}{2 E_{\ell^-}}+\frac{4y (m^2_{\ell}-m^2_{\rm HNL})-y^2 (3 m^2_{\ell}-m^2_{\rm HNL})}{8 xy M_N E_{\ell^-}}\notag
\\
&&+\frac{(m_{\ell}^2-m_{\rm HNL}^2)^2-4 M_N^2 x^2 (m^2_{\ell}+m^2_{\rm HNL})}{16 x^2 M^2_N E^2_{\ell^-}}\bigg)F_{2,p}^{W^-}(x,Q^2)\notag
\\
&&+\left(-\frac{y(y-2)}{2}+\frac{y (m^2_{\ell}-m^2_{\rm HNL})}{4x M_N E_{\ell^-}}\right)xF_{3,p}^{W^-} (x,Q^2)\notag
\\
&&+\frac{(m^2_{\ell}-m^2_{\rm HNL})^2+Q^2 (m^2_{\ell}+m^2_{\rm HNL})}{4 M^2_N E^2_{\ell^-}}F_{4,p}^{W^-} (x,Q^2)\notag
\\
&&+\frac{m^2_{\ell}(1-y)-m^2_{\rm HNL}}{2 M_N E_{\ell^-}}F_{5,p}^{W^-} (x,Q^2)
\Bigg],
\end{align}
\begin{align}
    &\frac{d^2 \sigma(\ell^{+} p\to {\rm HNL}~X)}{dx dy}
    \notag\\
    &= |U_{\ell}|^2\frac{G_F^2 M_N E_{\ell^+}}{\pi \left(1+Q^2/M_W^2\right)^2}\Bigg[&\left(xy^2 +\frac{y \left( m_{\ell}^2 +m^2_{\rm HNL}\right)}{4 M_N E_{\ell^+}}-\frac{(m_{\ell}^2-m_{\rm HNL}^2)^2}{8 x M_N^2 E_{\ell^+}^2}\right) F_{1,p}^{W^+}(x,Q^2)\notag
\\
&&+\bigg(1-y-\frac{M_N x y}{2 E_{\ell^+}}+\frac{4y (m^2_{\ell}-m^2_{\rm HNL})-y^2 (3 m^2_{\ell}-m^2_{\rm HNL})}{8 xy M_N E_{\ell^+}}\notag
\\
&&+\frac{(m_{\ell}^2-m_{\rm HNL}^2)^2-4 M_N^2 x^2 (m^2_{\ell}+m^2_{\rm HNL})}{16 x^2 M^2_N E^2_{\ell^+}}\bigg)F_{2,p}^{W^+}(x,Q^2)\notag
\\
&&-\left(-\frac{y(y-2)}{2}+\frac{y (m^2_{\ell}-m^2_{\rm HNL})}{4x M_N E_{\ell^+}}\right)xF_{3,p}^{W^+} (x,Q^2)\notag
\\
&&+\frac{(m^2_{\ell}-m^2_{\rm HNL})^2+Q^2 (m^2_{\ell}+m^2_{\rm HNL})}{4 M^2_N E^2_{\ell^+}}F_{4,p}^{W^+} (x,Q^2)\notag
\\
&&+\frac{m^2_{\ell}(1-y)-m^2_{\rm HNL}}{2 M_N E_{\ell^+}}F_{5,p}^{W^+} (x,Q^2)
\Bigg],
\end{align}
\begin{align}
&\frac{d^2 \sigma(\ell^{-} n\to {\rm HNL}~X)}{dx dy}\notag
\\
&= |U_{\ell}|^2\frac{G_F^2 M_N E_{\ell^-}}{\pi \left(1+Q^2/M_W^2\right)^2}\Bigg[&\left(xy^2 +\frac{y \left( m_{\ell}^2 +m^2_{\rm HNL}\right)}{4 M_N E_{\ell^-}}-\frac{(m_{\ell}^2-m_{\rm HNL}^2)^2}{8 x M_N^2 E_{\ell^-}^2}\right) F_{1,n}^{W^-}(x,Q^2)\notag
\\
&&+\bigg(1-y-\frac{M_N x y}{2 E_{\ell^-}}+\frac{4y (m^2_{\ell}-m^2_{\rm HNL})-y^2 (3 m^2_{\ell}-m^2_{\rm HNL})}{8 xy M_N E_{\ell^-}}\notag
\\
&&+\frac{(m_{\ell}^2-m_{\rm HNL}^2)^2-4 M_N^2 x^2 (m^2_{\ell}+m^2_{\rm HNL})}{16 x^2 M^2_N E^2_{\ell^-}}\bigg)F_{2,n}^{W^-}(x,Q^2)\notag
\\
&&+\left(-\frac{y(y-2)}{2}+\frac{y (m^2_{\ell}-m^2_{\rm HNL})}{4x M_N E_{\ell^-}}\right)xF_{3,n}^{W^-} (x,Q^2)\notag
\\
&&+\frac{(m^2_{\ell}-m^2_{\rm HNL})^2+Q^2 (m^2_{\ell}+m^2_{\rm HNL})}{4 M^2_N E^2_{\ell^-}}F_{4,n}^{W^-} (x,Q^2)\notag
\\
&&+\frac{m^2_{\ell}(1-y)-m^2_{\rm HNL}}{2 M_N E_{\ell^-}}F_{5,n}^{W^-} (x,Q^2)\Bigg],
\end{align}
\begin{align}
&\frac{d^2 \sigma(\ell^{+}  n\to {\rm HNL}~X)}{dx dy}\notag
\\
&= |U_{\ell}|^2\frac{G_F^2 M_N E_{\ell^+}}{\pi \left(1+Q^2/M_W^2\right)^2}\Bigg[&\left(xy^2 +\frac{y \left( m_{\ell}^2 +m^2_{\rm HNL}\right)}{4 M_N E_{\ell^+}}-\frac{(m_{\ell}^2-m_{\rm HNL}^2)^2}{8 x M_N^2 E_{\ell^+}^2}\right) F_{1,n}^{W^+}(x,Q^2)\notag
\\
&&+\bigg(1-y-\frac{M_N x y}{2 E_{\ell^+}}+\frac{4y (m^2_{\ell}-m^2_{\rm HNL})-y^2 (3 m^2_{\ell}-m^2_{\rm HNL})}{8 xy M_N E_{\ell^+}}\notag
\\
&&+\frac{(m_{\ell}^2-m_{\rm HNL}^2)^2-4 M_N^2 x^2 (m^2_{\ell}+m^2_{\rm HNL})}{16 x^2 M^2_N E^2_{\ell^+}}\bigg)F_{2,n}^{W^+}(x,Q^2)\notag
\\
&&-\left(-\frac{y(y-2)}{2}+\frac{y (m^2_{\ell}-m^2_{\rm HNL})}{4x M_N E_{\ell^+}}\right)xF_{3,n}^{W^+} (x,Q^2)\notag
\\
&&+\frac{(m^2_{\ell}-m^2_{\rm HNL})^2+Q^2 (m^2_{\ell}+m^2_{\rm HNL})}{4 M^2_N E^2_{\ell^+}}F_{4,n}^{W^+} (x,Q^2)\notag
\\
&&+\frac{m^2_{\ell}(1-y)-m^2_{\rm HNL}}{2 M_N E_{\ell^+}}F_{5,n}^{W^+} (x,Q^2)\Bigg],
\end{align}
where $|U_{\ell}|$ is the mixing angle, $E_{\ell^{\mp}}$ is the initial lepton energy in the nucleon rest frame, $M_N$ is the mass of nucleus, $M_W$ is the $W$ mass, $G_F$ is the Fermi coupling constant, and $m_{\ell}$ is the incoming charged lepton mass. 
The squared four-momentum transfer is given by $Q^2=-(k-k_{\rm HNL})^2$ where $k$ and $k_{\rm HNL}$ are the four-momenta of the incoming and outgoing leptons, respectively.
In this work, for simplicity, we impose the Callan-Gross relations~\cite{Callan:1969uq}, and the Albright-Jarlskog relations~\cite{Albright:1974ts} as follows:
\begin{align}
F_{2,p}^{W^-}(x,Q^2)&= 2 x F_{1,p}^{W^-}(x,Q^2),
\\
F_{4,p}^{W^-}(x,Q^2)&= 0,
\\
x F_{5,p}^{W^-}(x,Q^2)&=  F_{2,p}^{W^-}(x,Q^2),
\\
F_{2,p}^{W^+}(x,Q^2)&= 2 x F_{1,p}^{W^+}(x,Q^2),
\\
F_{4,p}^{W^+}(x,Q^2)&= 0,
\\
x F_{5,p}^{W^+}(x,Q^2)&=  F_{2,p}^{W^+}(x,Q^2),
\end{align}
\begin{align}
    F_{2,n}^{W^-}(x,Q^2)&= 2 x F_{1,n}^{W^-}(x,Q^2),
\\
F_{4,n}^{W^-}(x,Q^2)&= 0,
\\
x F_{5,n}^{W^-}(x,Q^2)&=  F_{2,n}^{W^-}(x,Q^2),
\\
F_{2,n}^{W^+}(x,Q^2)&= 2 x F_{1,n}^{W^+}(x,Q^2),
\\
F_{4,n}^{W^+}(x,Q^2)&= 0,
\\
x F_{5,n}^{W^+}(x,Q^2)&=  F_{2,n}^{W^+}(x,Q^2),
\end{align}
Here, the nucleon structure functions in the quark-parton model are given by \cite{Anselmino:1993tc,ParticleDataGroup:2020ssz}
\begin{align}
F_{2,p}^{W^-}(x,Q^2)&=2\left[\sum_{q=u,c,t} x q(x,Q^2)+ \sum_{q=d,s,b} x \bar{q}(x,Q^2)\right],
\\
xF_{3,p}^{W^-}(x,Q^2)&=2\left[\sum_{q=u,c,t} x q(x,Q^2)- \sum_{q=d,s,b} x \bar{q}(x,Q^2)\right],
\\
F_{2,p}^{W^+}(x,Q^2)&=2\left[\sum_{q=u,c,t} x \bar{q}(x,Q^2)+ \sum_{q=d,s,b} x q(x,Q^2)\right],
\\
xF_{3,p}^{W^+}(x,Q^2)&=2\left[-\sum_{q=u,c,t} x \bar{q}(x,Q^2)+ \sum_{q=d,s,b} x q(x,Q^2)\right],
\end{align}
\begin{align}
F_{2,n}^{W^-}(x,Q^2)&=2\left[\sum_{q=d,c,t} x q(x,Q^2)+ \sum_{q=u,s,b} x \bar{q}(x,Q^2)\right],
\\
xF_{3,n}^{W^-}(x,Q^2)&=2\left[\sum_{q=d,c,t} x q(x,Q^2)- \sum_{q=u,s,b} x \bar{q}(x,Q^2)\right],
\\
F_{2,n}^{W^+}(x,Q^2)&=2\left[\sum_{q=d,c,t} x \bar{q}(x,Q^2)+ \sum_{q=u,s,b} x q(x,Q^2)\right],
\\
xF_{3,n}^{W^+}(x,Q^2)&=2\left[-\sum_{q=d,c,t} x \bar{q}(x,Q^2)+ \sum_{q=u,s,b} x q(x,Q^2)\right],
\end{align}
where $x$ denotes the Bjorken scaling defined as $x=Q^2/2M_N(E_{\ell^{\mp}}-E_{\rm HNL})$ with the lepton energy loss in the rest frame, and $y=(E_{\ell^{\mp}}-E_{\rm HNL})/E_{\ell^{\mp}}$ is the fraction of the lepton energy loss in the rest frame. 
The squared four-momentum transfer can be expressed as $Q^2=2M_N E_{\ell^{\mp}}xy$.
Combining $0\leq x\leq 1$, $0\leq y\leq 1$, and 
\begin{align}
    \cos^2\theta_{\rm HNL}^{\rm prod} =\frac{\left(m_{\ell}^2+m_{\rm HNL}^2+2 E^2_{\ell} M_N x y\right)^2}{4 (E_{\ell}^2-m_{\ell}^2)\left(E^2_{\ell} (y-1)^2-m^2_{\rm HNL} \right)}\leq 1,
\end{align}
the $x$ and $y$ kinematic limits are obtained.
In the numerical analysis, we adopt the MSTW parton distribution functions~\cite{Martin:2009iq}.

\begin{figure*}[t]
\centering
\includegraphics[width=0.45\textwidth]{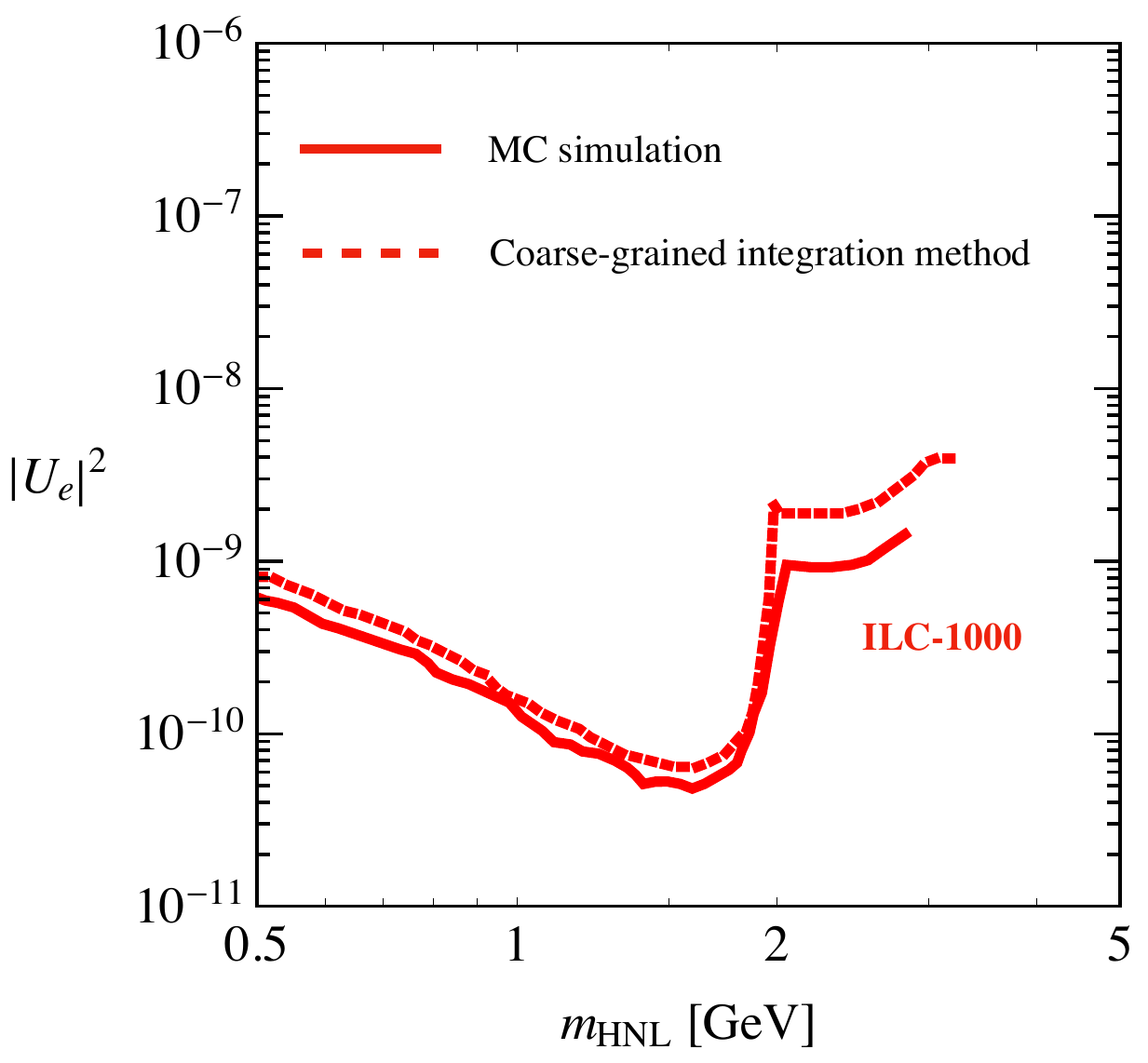}
\includegraphics[width=0.45\textwidth]{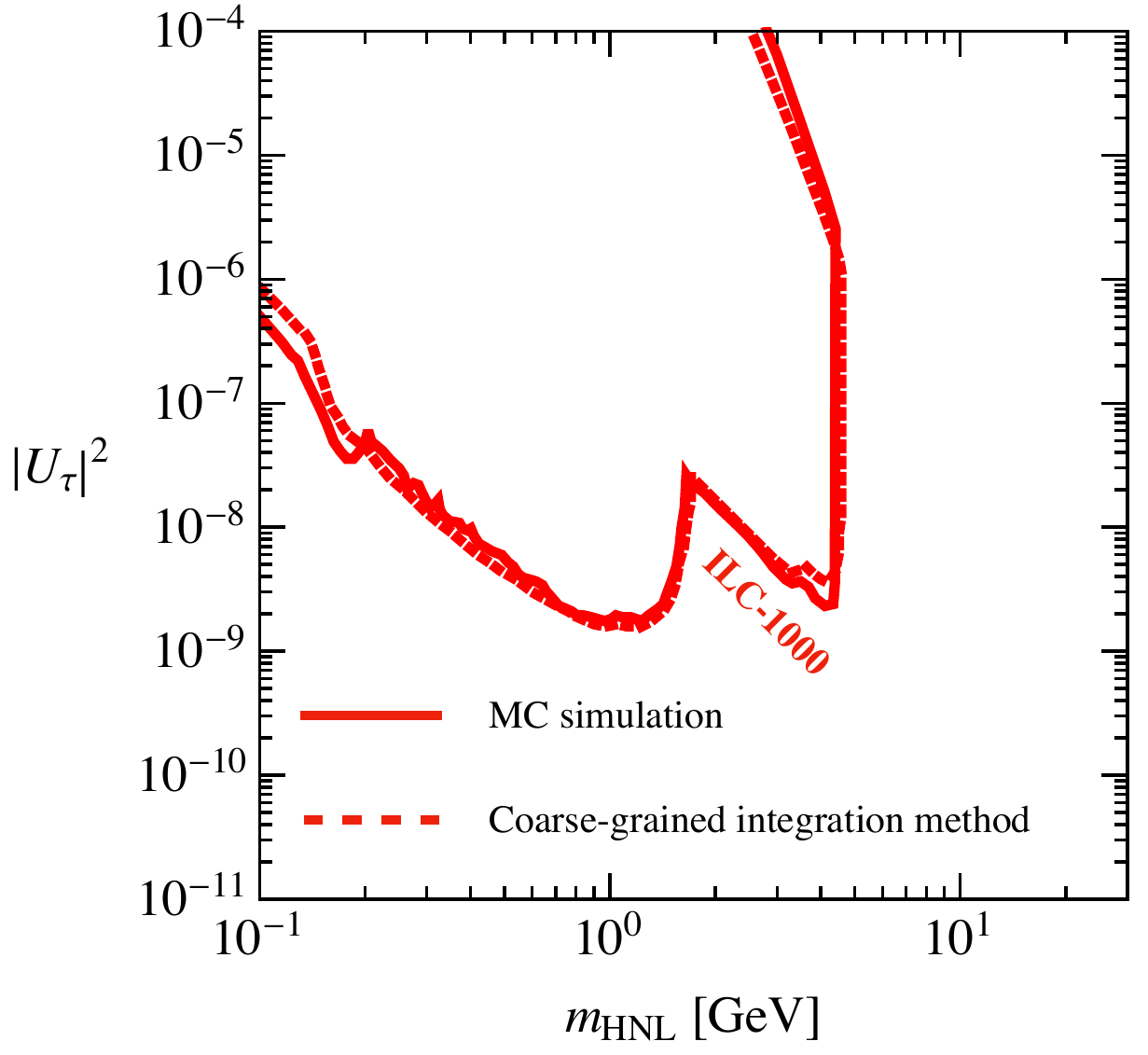}
\caption{The left (right) panel shows the curves corresponding to three signal events with $10$-year statistics for the $U_e$ ($U_{\tau}$) dominant scenario at ILC-1000.
For simplicity, we assume the signal events are the HNL decays inside the decay volume except for $\rm HNL\to \nu\nu\bar{\nu}$.
The solid (dashed) lines correspond to the limit based on the Monte-Carlo simulation (the coarse-grained integration method).
Note that the curves of the coarse-grained integration method almost reproduce that of the Monte-Carlo simulation.
}
\label{fig:compar}
\end{figure*}

\section{Comparison between Monte-Carlo method and coarse-grained integration method}\label{app:Com}
In our analyses in Sec.~\ref{sec:HNL}, we evaluated the number of signal events by the Monte-Carlo simulation based on the PHITS framework.
To explain how to use the meson and $\tau$ lepton spectra, and perform easy consistency checks, we also evaluate the number of signal events by the coarse-grained integration  method using the meson spectra in Sec.~\ref{sec:spectrum}.
The formulae of coarse-grained integration for the number of signal events are summarized in Sec.~\ref{sec:dumpsetup}.
In this section, we assume zero background and require three signal events 
that the HNL decays inside the decay volume as signal except for the invisible decay mode, $\rm HNL\to \nu\nu\bar\nu$.

In Fig.~\ref{fig:compar}, we show the sensitivity of ILC-1000 with 10-year statistics for the $U_e$ and $U_{\tau}$ dominant scenarios. 
The red solid (dotted) curves show the sensitivities based on the Monte-Carlo simulation (coarse-grained integration method).
For the $U_e$ dominance, the $D$ and $B$ meson decays dominate the HNL productions in $0.5~{\rm GeV}\lesssim m_{\rm HNL}\lesssim 2~{\rm GeV}$ and $2~{\rm GeV}\lesssim m_{\rm HNL}\lesssim 3~{\rm GeV}$, respectively. 
In $3~{\rm GeV}\lesssim m_{\rm HNL}$, the DIS process becomes significant.
For the $U_{\tau}$ dominance, $D$ and $B$ mesons dominate the HNL productions in $m_{\rm HNL}\lesssim 0.2~{\rm GeV}$ and $0.2~{\rm GeV}\lesssim m_{\rm HNL}$, respectively. 
In a wide range of the parameter space, the coarse-grained integration method almost reproduces the results of the Monte-Carlo simulation.
Note here that the coarse-grained integration method does not overestimate the number of signal events.

\begin{figure}
\centering 
\includegraphics[width=0.95\textwidth]{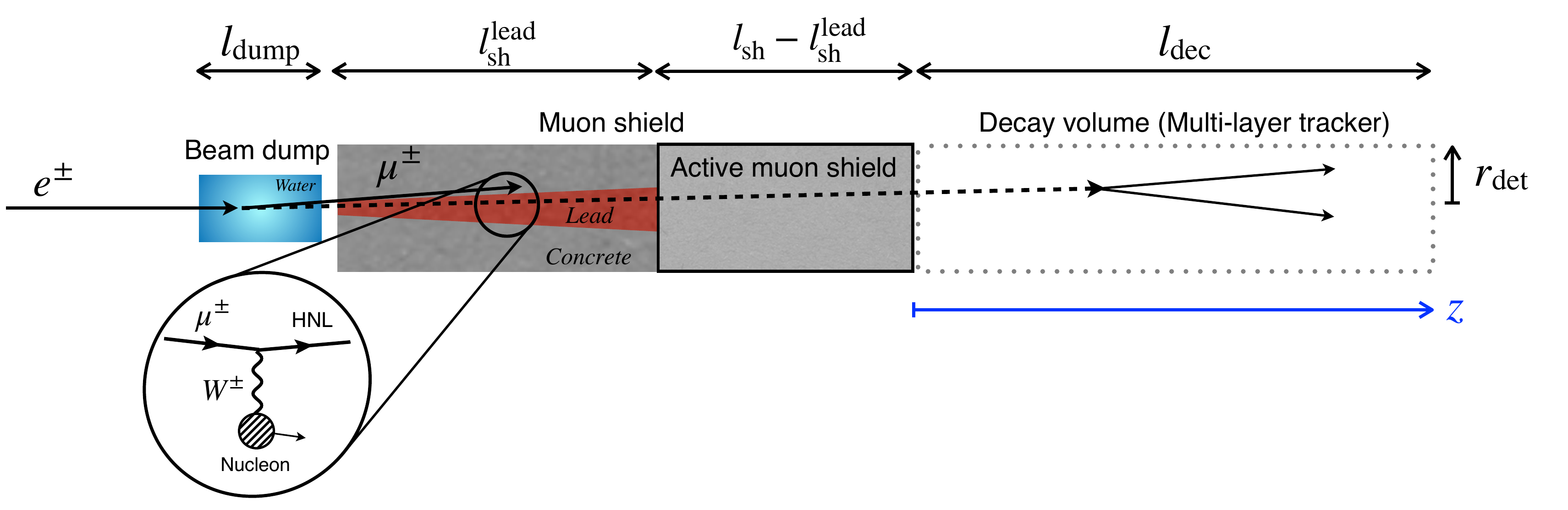}
\caption{A setup for the ILC-1000 beam dump experiment. It consists of the main beam dump, a muon lead shield, an active muon shield, a decay volume, and a detector.
The HNLs can be produced by the DIS process for incoming muons in the muon lead shield with the length of $l_{\rm sh}^{\rm lead}$.
}
\label{fig:exp_ILC1000}
\end{figure}

\begin{figure*}
\centering 
\includegraphics[width=0.6\textwidth]{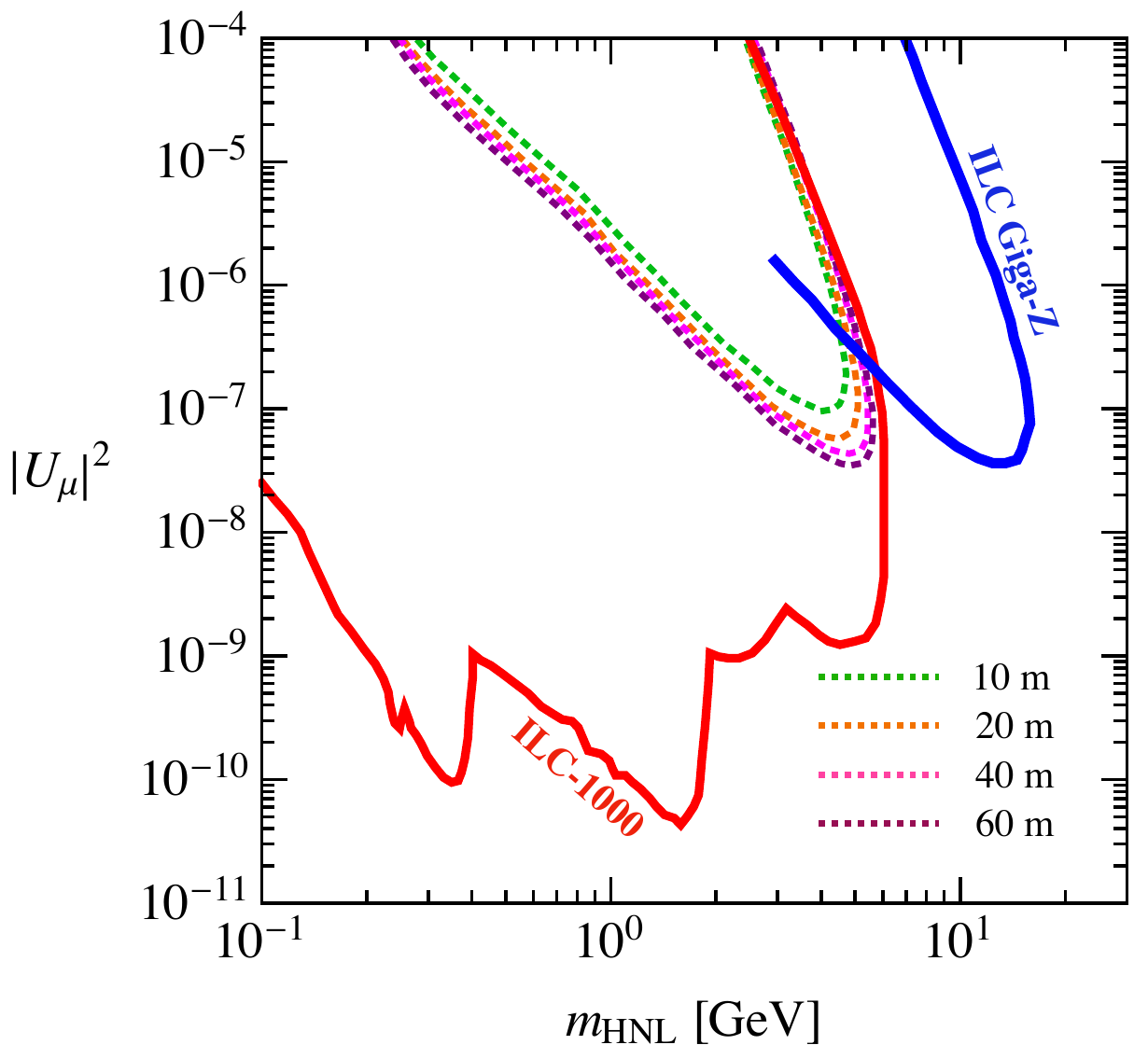}
\caption{The colored dotted curves correspond
to three signal events with 10-year statistics for the DIS process. We assume the signal events are the HNL decays inside the decay volume except for $\rm HNL\to \nu\nu\bar{\nu}$ 
We assume the signal events are the HNL decays inside the decay volume except for $\rm HNL\to \nu\nu\bar{\nu}$.
The results for the different lengths of the muon lead shield, $10, 20, 40, 60$ m are compared. 
The red curve shows the expected $95\%$ C.L. exclusion sensitivity with $10$-year statistics for the HNL productions from the SM particle decays.
The blue curve corresponds to the sensitivity of the Giga-Z program at ILC.
}
\label{fig:DISmuon}
\end{figure*}

\section{DIS process and muon shield configuration at ILC-1000}\label{app:muonshield}
For ILC-1000, high energy muons are produced in the beam dump and penetrate the muon lead shield.
One can reduce background events by placing an active muon shield behind the muon lead shield as shown in Fig.~\ref{fig:exp_ILC1000}.
In our analyses in Sec.~\ref{sec:HNL}, we assumed that the HNLs produced in the muon lead shield with the length of $l^{\rm lead}_{\rm sh}=10$~m contribute to the signal events.
Since, however, the length of the muon lead shield highly depend on the details of the detector concepts, we study the sensitivity of ILC-1000 depending on $l_{\rm sh}^{\rm lead}$. 
In this Appendix, we assume zero background and account for all the HNL decays inside the decay volume except for $\rm HNL\to \nu\nu\bar\nu$ as signals.

In Fig.~\ref{fig:DISmuon}, the sensitivities of the DIS process for incoming muons are shown for $l_{\rm sh}^{\rm lead}=10, 20, 40, 60$ m.
The constraint to the large $|U_{\mu}|^2$ region enlarges by shortening active muon shield, because more HNLs decay in the decay volume.

\section{Branching ratios of meson and $\tau$ lepton decays to an HNL}\label{sec:Br}
\begin{figure*}
\centering
\includegraphics[width=0.7\textwidth]{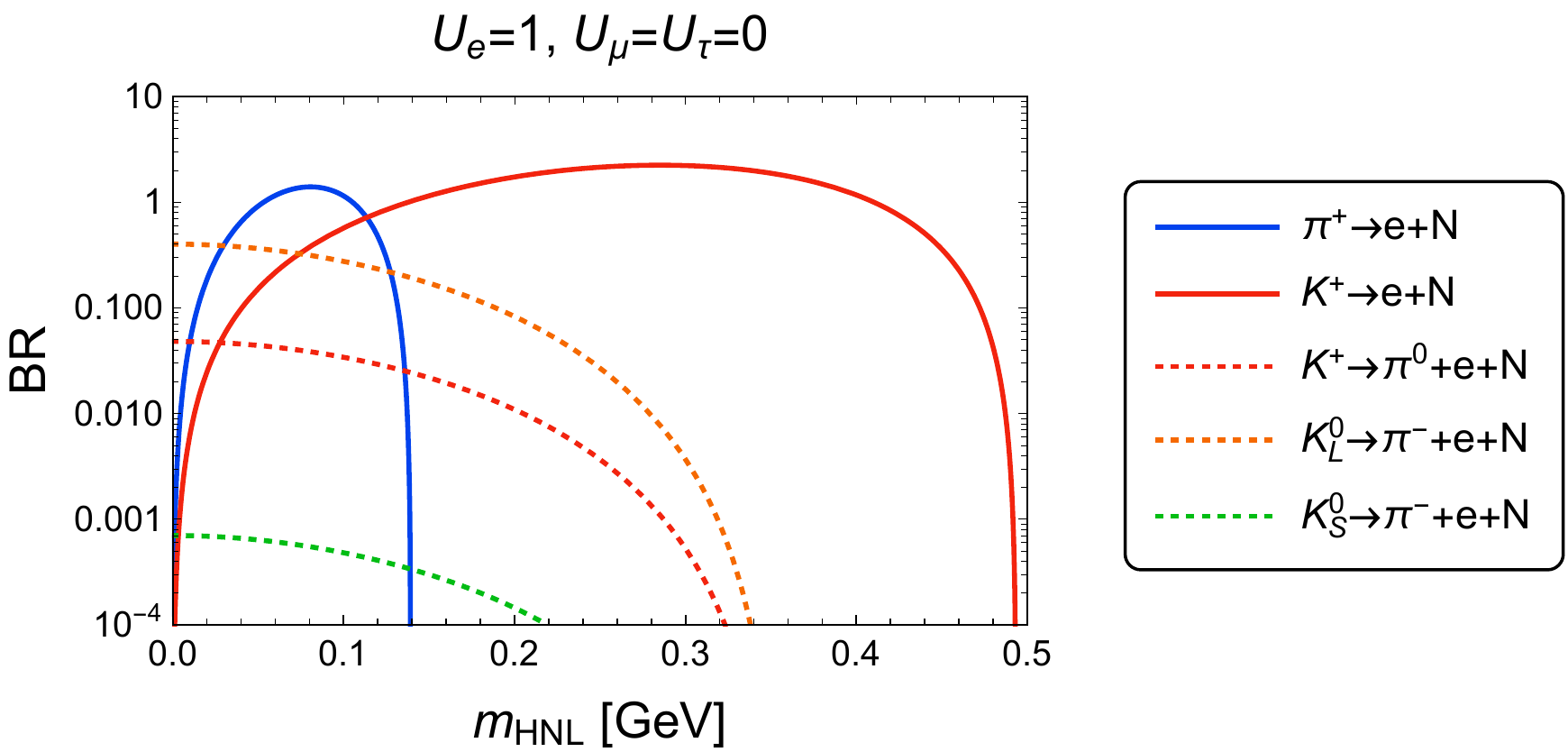}
\caption{
Dominant branching ratios of HNL production from light mesons.
In this Figure, we take $U_e=1$, $U_{\mu}=U_{\tau}=0$.
The decay width to HNLs are normalized by experimental values of the total decay width of mesons.
Below the kaon mass, the HNL productions are dominated by the leptonic decays of pions and kaons.
}
\label{fig:brUeli}
\end{figure*}

\begin{figure*}
\centering
\includegraphics[width=0.7\textwidth]{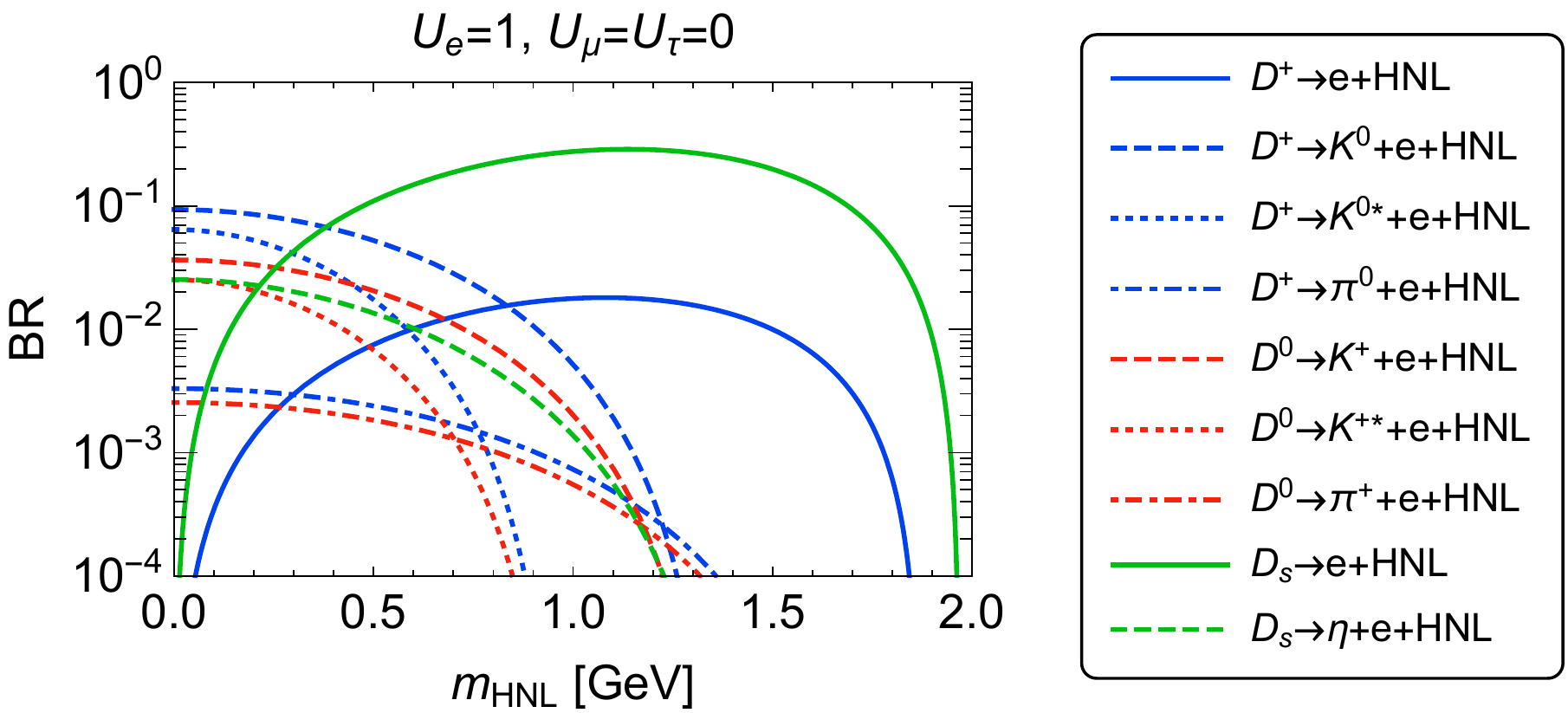}
\caption{
The same plot as Fig.~\ref{fig:brUeli} but for charmed mesons.
}
\label{fig:brUeD}
\end{figure*}

\begin{figure*}
\centering
\includegraphics[width=0.9\textwidth]{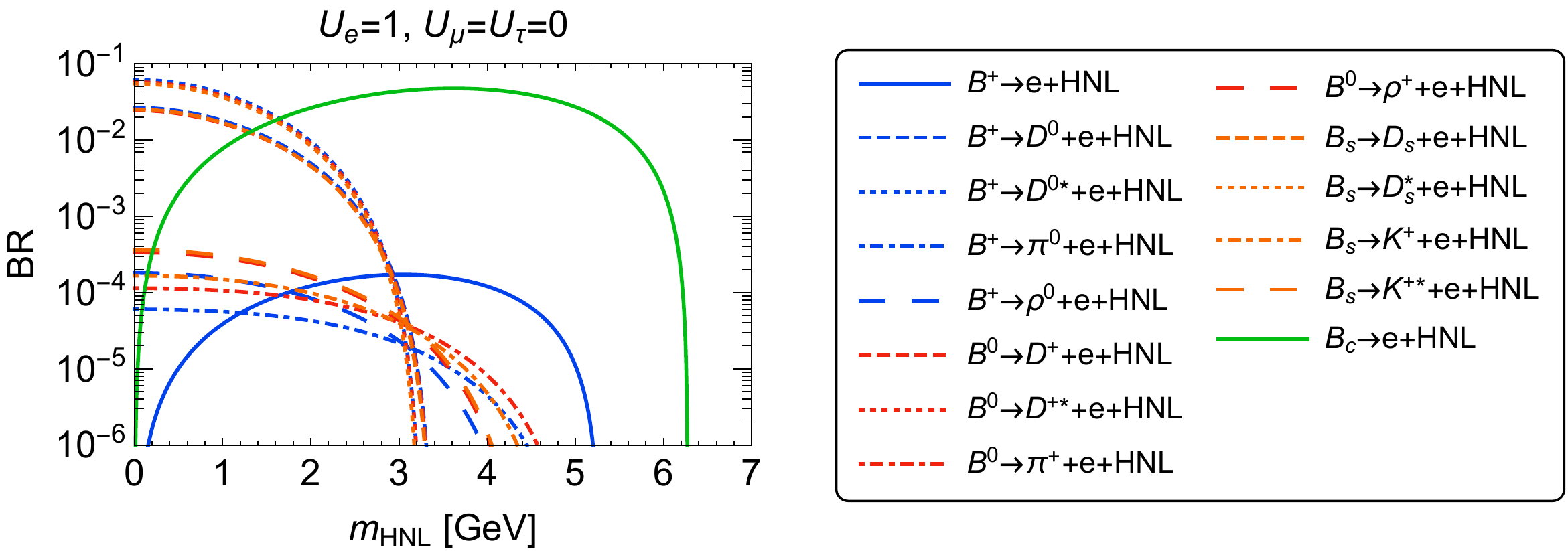}
\caption{
The same plot as Fig.~\ref{fig:brUeli} but for beauty mesons.
}
\label{fig:brUeB}
\end{figure*}

\begin{figure*}
\centering
\includegraphics[width=0.7\textwidth]{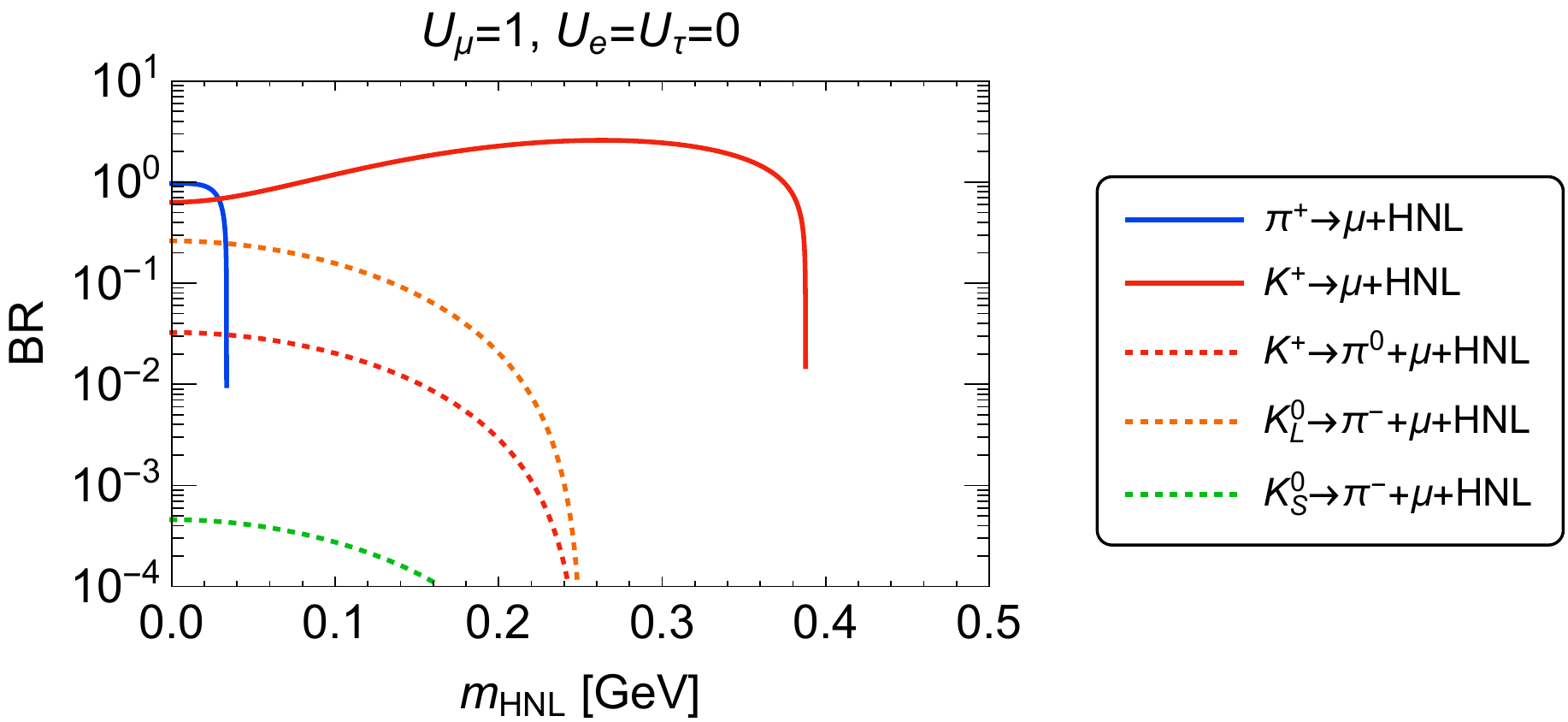}
\caption{
The same plot as Fig.~\ref{fig:brUeli} but for $U_{\mu}=1$, $U_{e}=U_{\tau}=0$.
}
\label{fig:brUmuli}
\end{figure*}

\begin{figure*}
\centering
\includegraphics[width=0.7\textwidth]{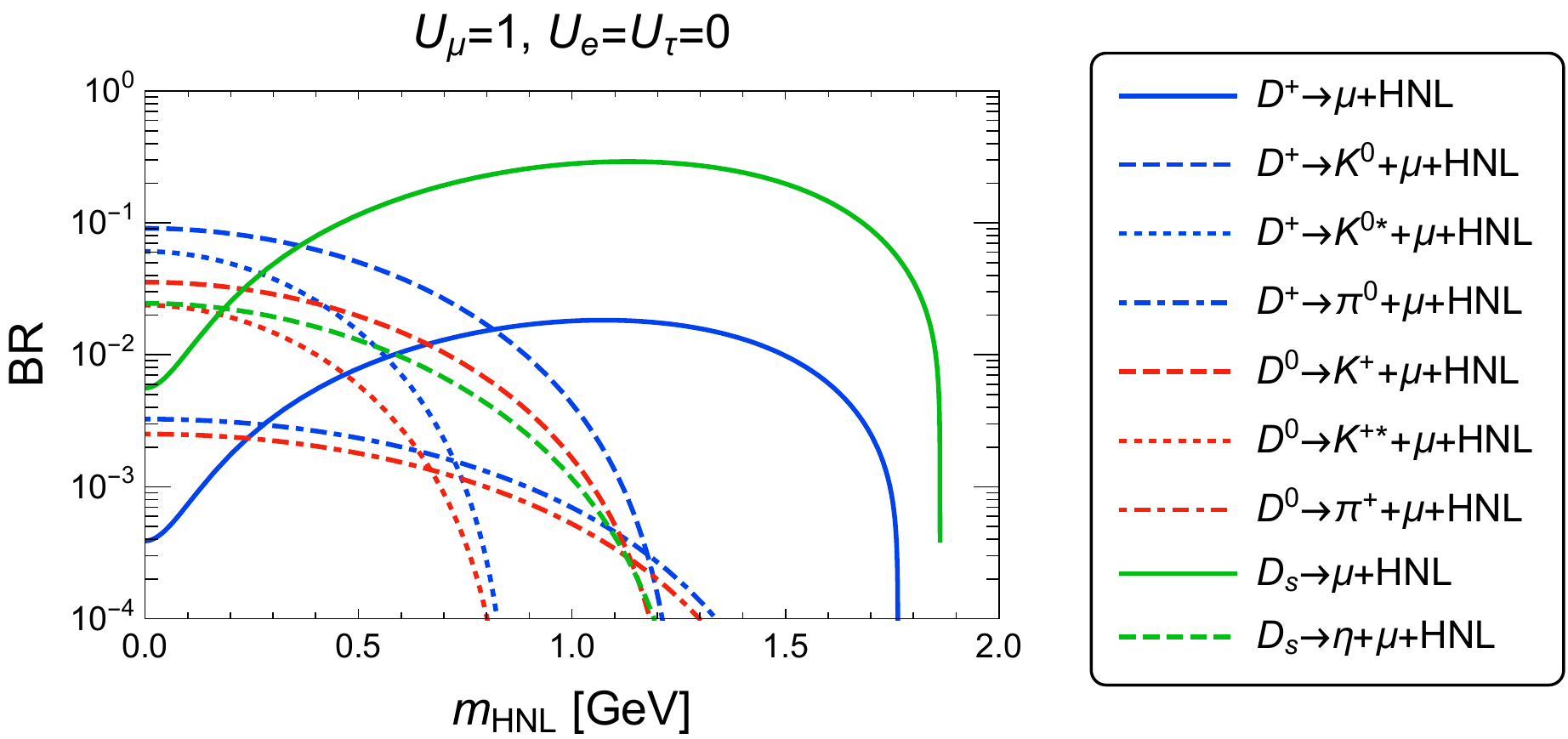}
\caption{The same plot as Fig.~\ref{fig:brUeD} but for $U_{\mu}=1$, $U_{e}=U_{\tau}=0$.
}
\label{fig:brUmuD}
\end{figure*}

\begin{figure*}
\centering
\includegraphics[width=0.9\textwidth]{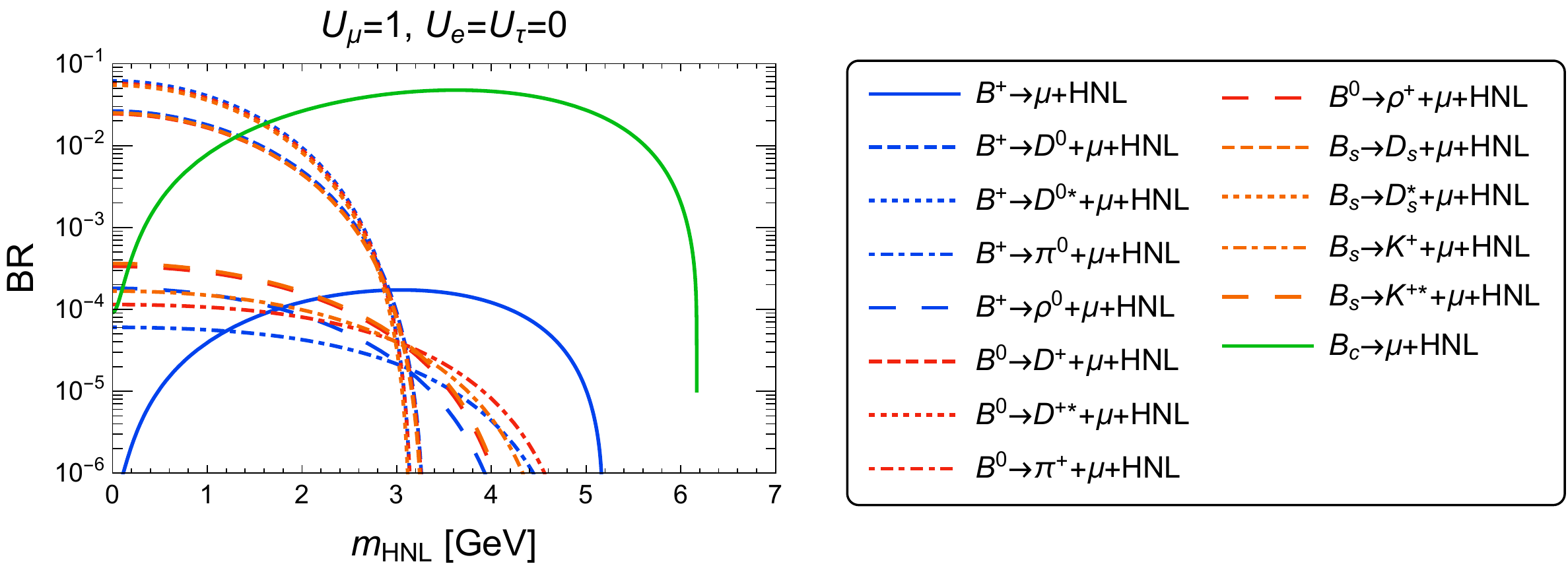}
\caption{The same plot as Fig.~\ref{fig:brUeB} but for $U_{\mu}=1$, $U_{e}=U_{\tau}=0$.
}
\label{fig:brUmuB}
\end{figure*}

\begin{figure*}
\centering
\includegraphics[width=0.53\textwidth]{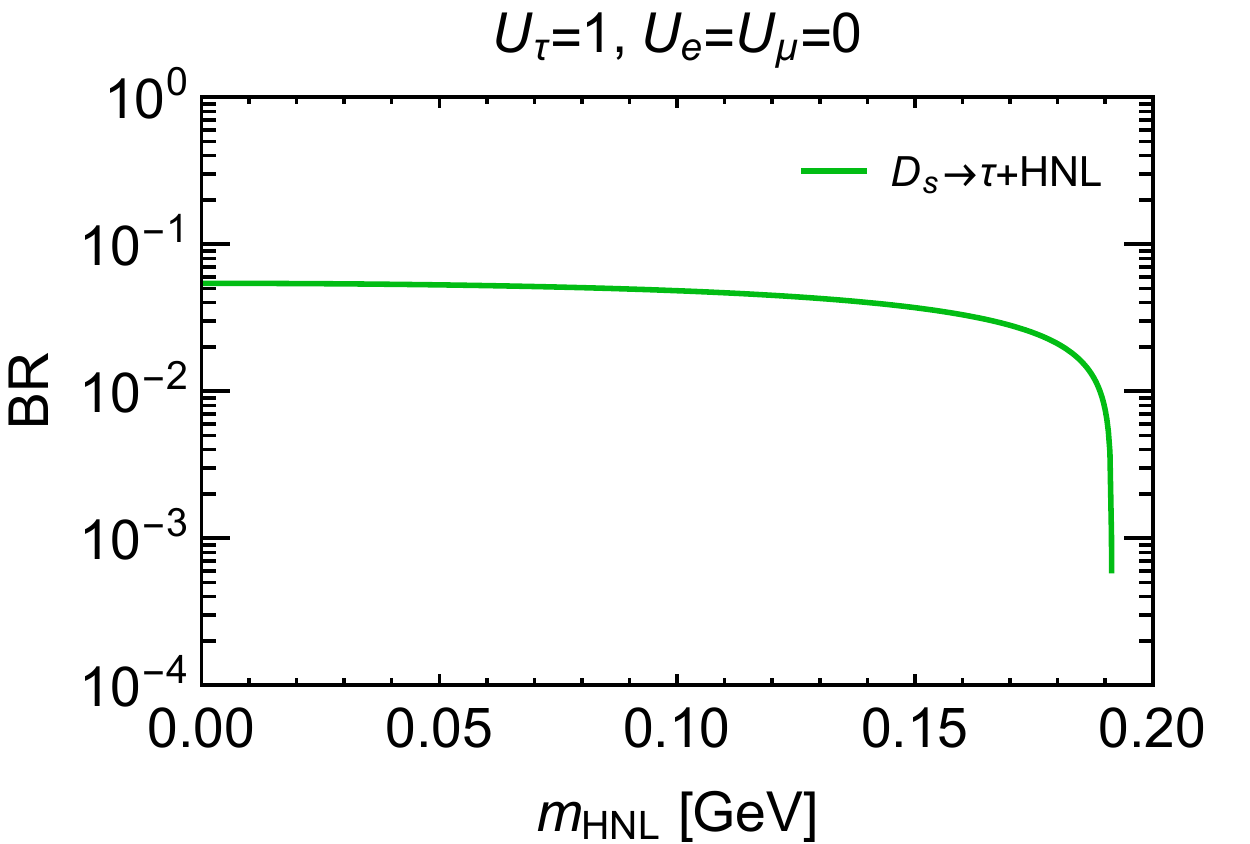}
\caption{The same plot as Fig.~\ref{fig:brUeD} but for $U_{\tau}=1$, $U_{e}=U_{\mu}=0$.
}
\label{fig:brUtauD}
\end{figure*}

\begin{figure*}
\centering
\includegraphics[width=1.0\textwidth]{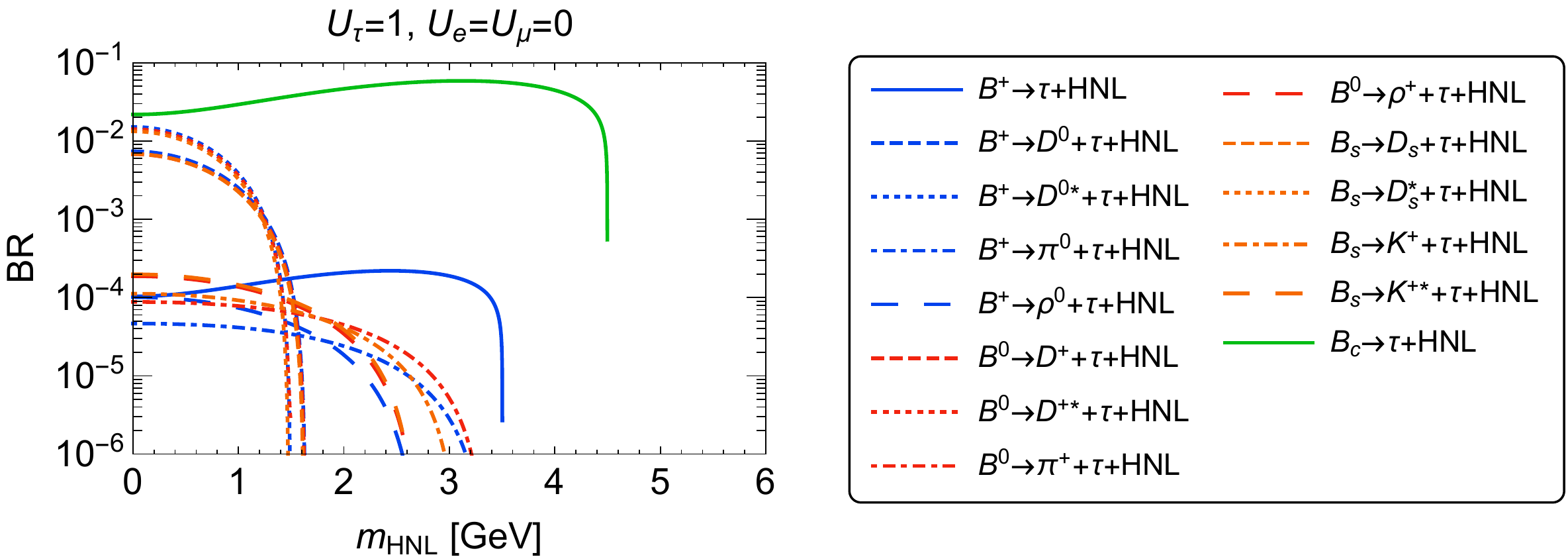}
\caption{The same plot as Fig.~\ref{fig:brUeB} but for $U_{\tau}=1$, $U_{e}=U_{\mu}=0$.
}
\label{fig:brUtauB}
\end{figure*}

For convenience, we summarize the branching ratios of the meson and $\tau$ lepton decays to produce an HNL.
The details of the branching ratio calculation are found in Ref.~\cite{2018}.  
The branching ratios for the light unflavored and strange mesons are shown in Fig.~\ref{fig:brUeli} and \ref{fig:brUmuli}.
The relevant mesons for the HNLs productions can be: $\pi^+ (u\bar{d},139.6)$, $K^+(u\bar{s},494)$, $K^0_S (d\bar{s},498)$, and $K^0_L (d\bar{s}, 498)$.
For $0.5~{\rm GeV}\lesssim m_{\rm HNL}\lesssim 2~{\rm GeV}$, the HNL production is dominated by the charmed mesons: $D^0(c\bar{u},1865)$, $D^+ (c\bar{d},1870)$, and $D_s(c\bar{s},1968)$.
The branching ratios for the charmed mesons are shown in Fig.~\ref{fig:brUeD}, \ref{fig:brUmuD}, and \ref{fig:brUtauD}.
For $2~{\rm GeV}\lesssim m_{\rm HNL}$, the decay of charmed mesons to the HNLs is forbidden, and the HNL production is dominated by the beauty mesons: $B^- (b\bar{u},5279)$, $B^0 (b\bar{d},5280)$, $B_s (b\bar{s},5367)$, $B_c (b\bar{c},6276)$. 
The branching ratios for the beauty mesons are shown in Fig.~\ref{fig:brUeB}, \ref{fig:brUmuB}, and \ref{fig:brUtauB}.

\begin{figure*}
\centering
\includegraphics[width=0.53\textwidth]{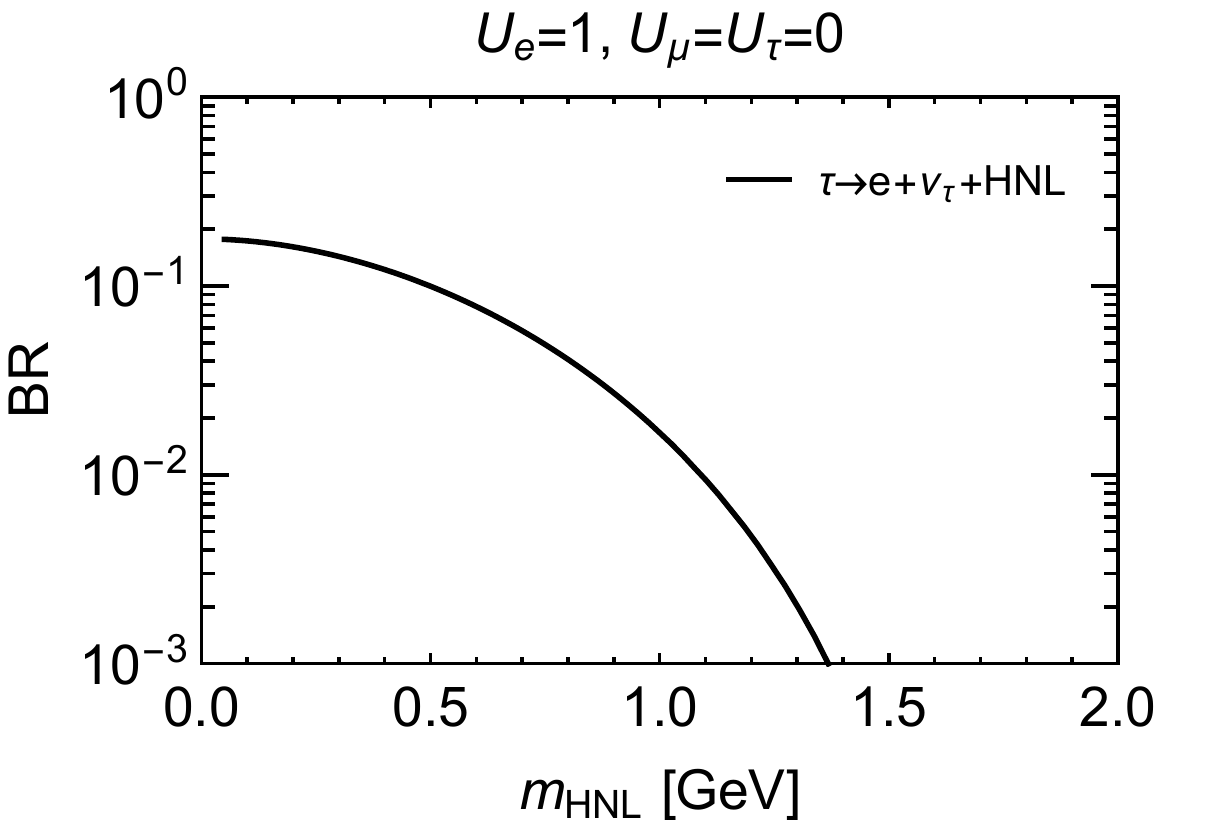}
\caption{The same plot as Fig.~\ref{fig:brUeli} but for $\tau$ lepton.
}
\label{fig:brUetau}
\end{figure*}

\begin{figure*}
\centering
\includegraphics[width=0.53\textwidth]{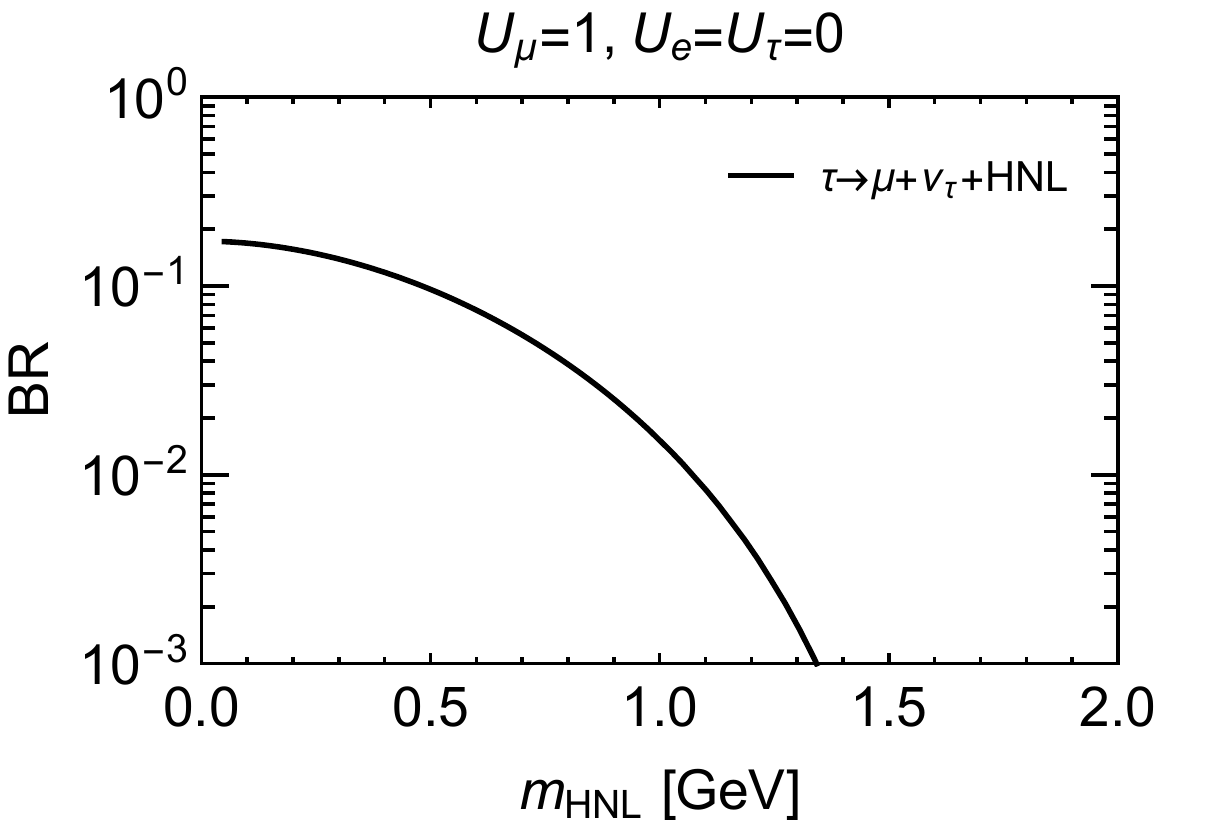}
\caption{The same plot as Fig.~\ref{fig:brUetau} but for $U_{\mu}=1$, $U_{e}=U_{\tau}=0$.
}
\label{fig:brUmutau}
\end{figure*}

\begin{figure*}
\centering
\includegraphics[width=0.7\textwidth]{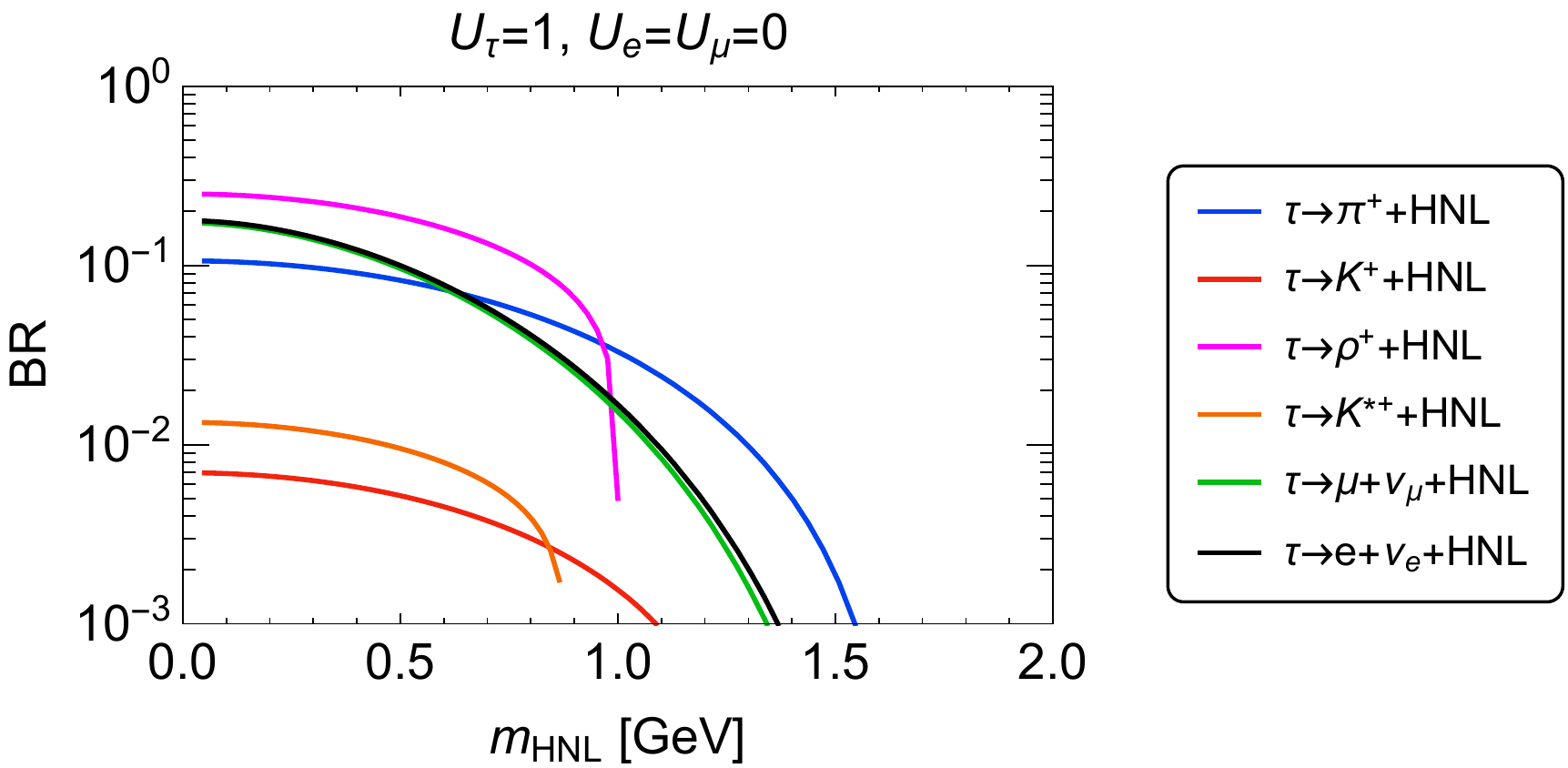}
\caption{The same plot as Fig.~\ref{fig:brUetau} but for $U_{\tau}=1$, $U_{e}=U_{\mu}=0$.
}
\label{fig:brUtautau}
\end{figure*}

In addition to the mesons, $\tau$ lepton can be produced by $D_s$ meson decays and $\tau$ pair productions in the electromagnetic showers. 
For $0.5~{\rm GeV}\lesssim m_{\rm HNL}\lesssim 2~{\rm GeV}$, the HNL production from the $\tau$ decays is significant effect.
The branching ratios for the $\tau$ lepton are shown in Fig.~\ref{fig:brUetau}, \ref{fig:brUmutau}, and \ref{fig:brUtautau}.

\clearpage

\section{Projected sensitivities for one year operation \label{app:one-year}}
We assumed 10-year operation in the main results. Here, for the reader's convenience, we add the projected sensitivities after one year of operation. The estimated background is expected to be a factor of 10 smaller than in the 10-year case. Therefore, background of 0.5 and 2 events are assumed at ILC-250 and ILC-1000, respectively. The 95\%CL sensitivities are shown as dashed lines, while the solid lines are for the 10-year operation for comparison.

\begin{figure*}[h]
\centering
\includegraphics[width=0.45\textwidth]{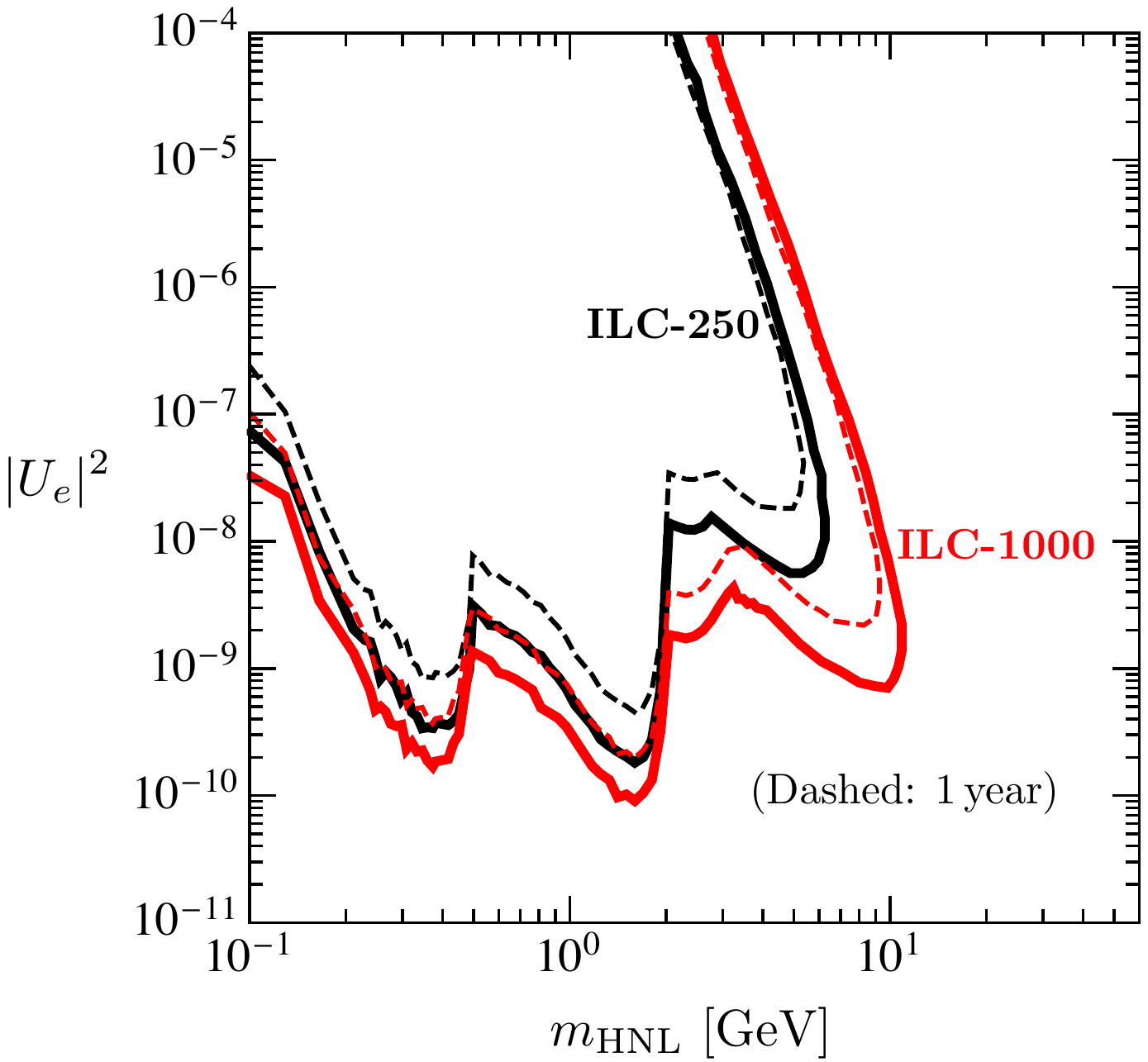}~~~
\includegraphics[width=0.45\textwidth]{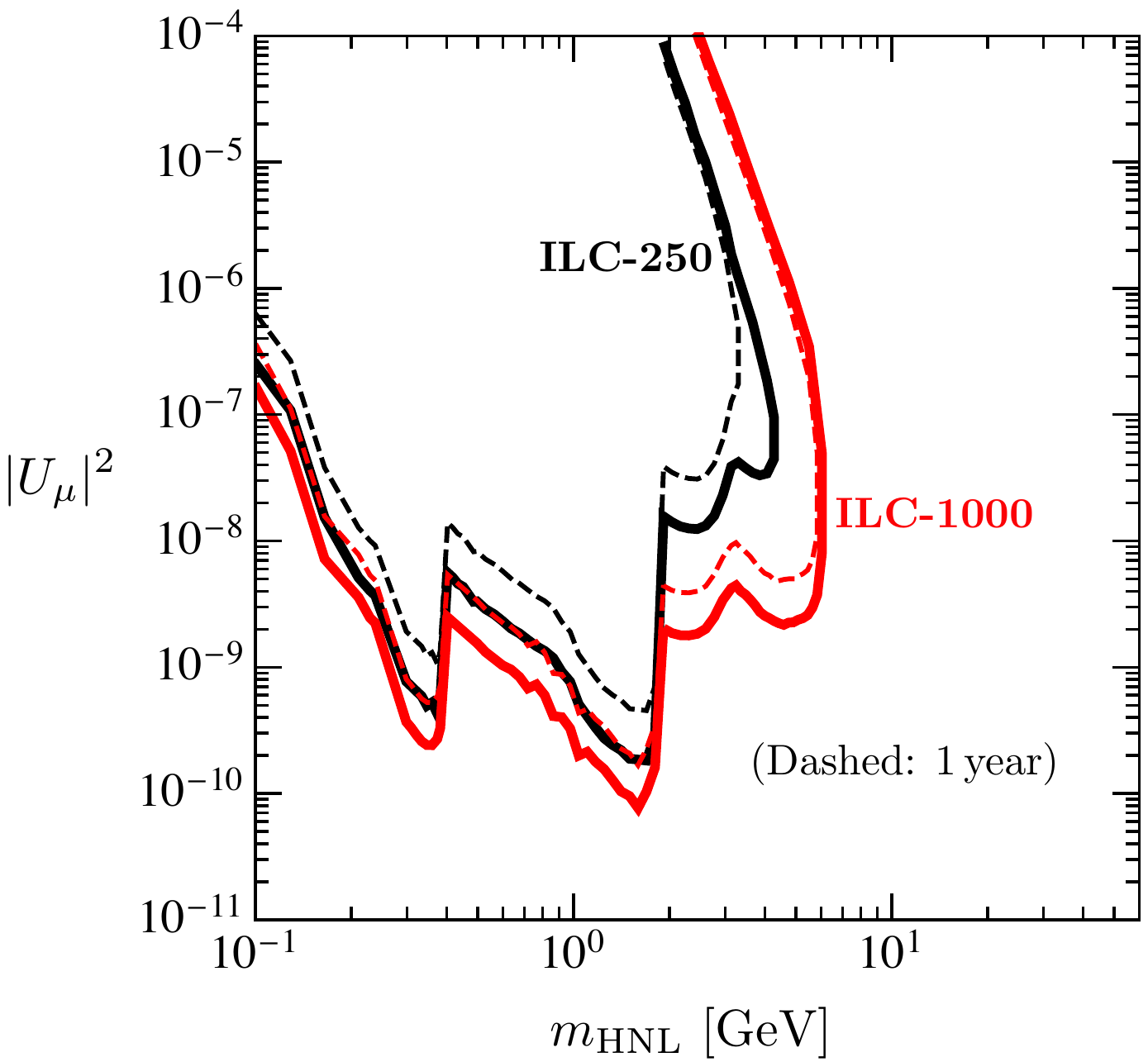}\\[5pt]
\includegraphics[width=0.45\textwidth]{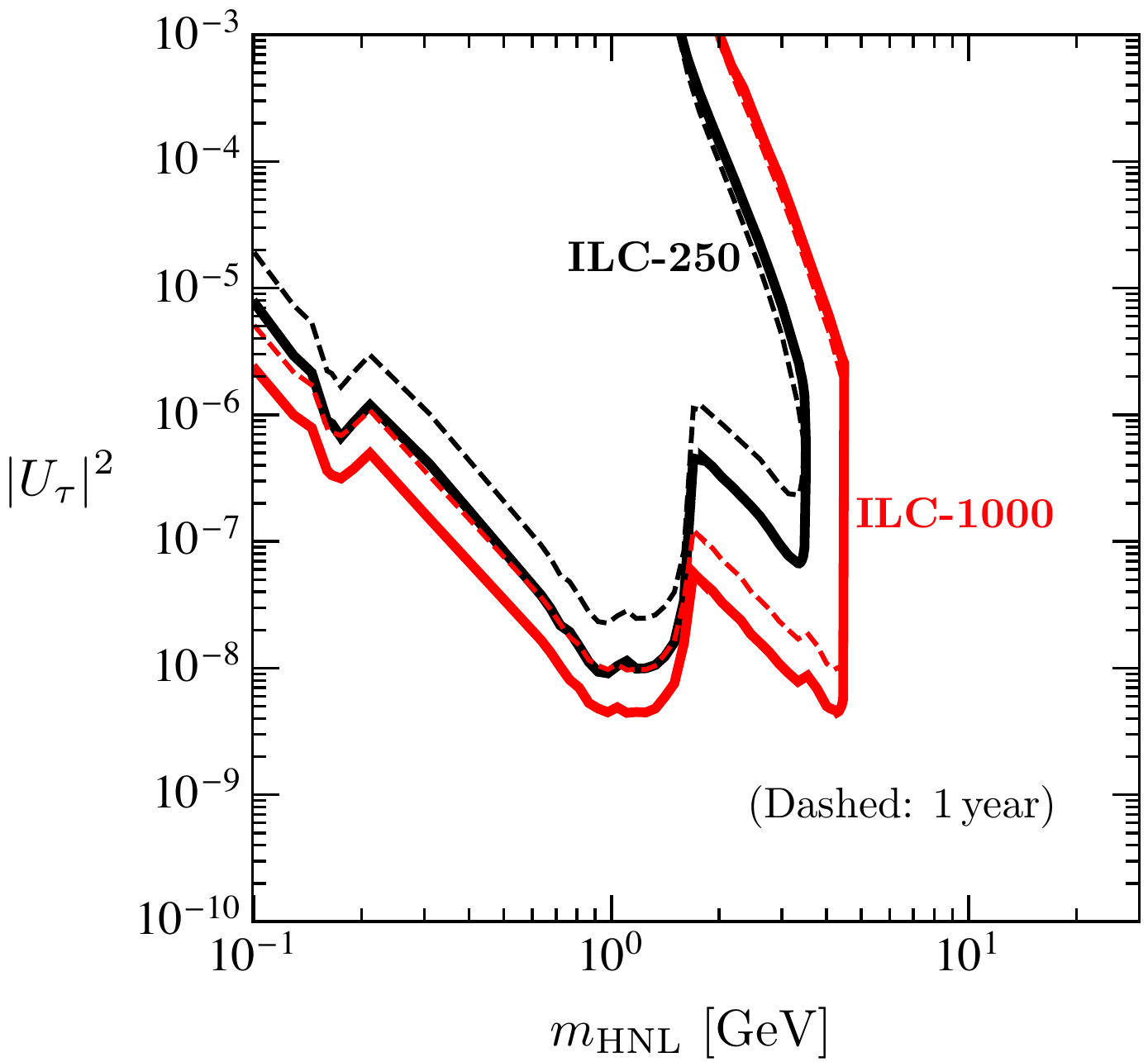}
\caption{Sensitivity plots for 1-year (dashed) and 10-year (solid) operations when $U_e$, $U_\mu$, and $U_\tau$ are dominant.}
\end{figure*}

\bibliographystyle{JHEP}
\bibliography{ILCdump.bib}

\providecommand{\href}[2]{#2}\begingroup\raggedright\begin{thebibliography}{10}

\bibitem{Beacham:2019nyx}
J.~Beacham et~al., {\it {Physics Beyond Colliders at CERN: Beyond the Standard
  Model Working Group Report}},  {\em J. Phys. G} {\bf 47} (2020), no.~1
  010501, [\href{http://arxiv.org/abs/1901.09966}{{\tt arXiv:1901.09966}}].

\bibitem{Izaguirre:2013uxa}
E.~Izaguirre, G.~Krnjaic, P.~Schuster, and N.~Toro, {\it {New Electron
  Beam-Dump Experiments to Search for MeV to few-GeV Dark Matter}},  {\em Phys.
  Rev. D} {\bf 88} (2013) 114015, [\href{http://arxiv.org/abs/1307.6554}{{\tt
  arXiv:1307.6554}}].

\bibitem{Kanemura:2015cxa}
S.~Kanemura, T.~Moroi, and T.~Tanabe, {\it {Beam dump experiment at future
  electron\textendash{}positron colliders}},  {\em Phys. Lett. B} {\bf 751}
  (2015) 25--28, [\href{http://arxiv.org/abs/1507.02809}{{\tt
  arXiv:1507.02809}}].

\bibitem{Sakaki:2020mqb}
Y.~Sakaki and D.~Ueda, {\it {Searching for new light particles at the
  international linear collider main beam dump}},  {\em Phys. Rev. D} {\bf 103}
  (2021), no.~3 035024, [\href{http://arxiv.org/abs/2009.13790}{{\tt
  arXiv:2009.13790}}].

\bibitem{Asai:2021ehn}
K.~Asai, S.~Iwamoto, Y.~Sakaki, and D.~Ueda, {\it {New physics searches at the
  ILC positron and electron beam dumps}},  {\em JHEP} {\bf 09} (2021) 183,
  [\href{http://arxiv.org/abs/2105.13768}{{\tt arXiv:2105.13768}}].

\bibitem{Asai:2021xtg}
K.~Asai, T.~Moroi, and A.~Niki, {\it {Leptophilic Gauge Bosons at ILC Beam Dump
  Experiment}},  {\em Phys. Lett. B} {\bf 818} (2021) 136374,
  [\href{http://arxiv.org/abs/2104.00888}{{\tt arXiv:2104.00888}}].

\bibitem{Moroi:2022qwz}
T.~Moroi and A.~Niki, {\it {Leptophilic Gauge Bosons at Lepton Beam Dump
  Experiments}},  \href{http://arxiv.org/abs/2205.11766}{{\tt
  arXiv:2205.11766}}.

\bibitem{Alekhin:2015byh}
S.~Alekhin et~al., {\it {A facility to Search for Hidden Particles at the CERN
  SPS: the SHiP physics case}},  {\em Rept. Prog. Phys.} {\bf 79} (2016),
  no.~12 124201, [\href{http://arxiv.org/abs/1504.04855}{{\tt
  arXiv:1504.04855}}].

\bibitem{Dasgupta:2021ies}
B.~Dasgupta and J.~Kopp, {\it {Sterile Neutrinos}},  {\em Phys. Rept.} {\bf
  928} (2021) 1--63, [\href{http://arxiv.org/abs/2106.05913}{{\tt
  arXiv:2106.05913}}].

\bibitem{Shaposhnikov:2006nn}
M.~Shaposhnikov, {\it {A Possible symmetry of the nuMSM}},  {\em Nucl. Phys. B}
  {\bf 763} (2007) 49--59, [\href{http://arxiv.org/abs/hep-ph/0605047}{{\tt
  hep-ph/0605047}}].

\bibitem{Akhmedov:1998qx}
E.~K. Akhmedov, V.~A. Rubakov, and A.~Y. Smirnov, {\it {Baryogenesis via
  neutrino oscillations}},  {\em Phys. Rev. Lett.} {\bf 81} (1998) 1359--1362,
  [\href{http://arxiv.org/abs/hep-ph/9803255}{{\tt hep-ph/9803255}}].

\bibitem{Asaka:2005pn}
T.~Asaka and M.~Shaposhnikov, {\it {The $\nu$MSM, dark matter and baryon
  asymmetry of the universe}},  {\em Phys. Lett. B} {\bf 620} (2005) 17--26,
  [\href{http://arxiv.org/abs/hep-ph/0505013}{{\tt hep-ph/0505013}}].

\bibitem{Sato:2018imy}
T.~Sato et~al., {\it {Features of Particle and Heavy Ion Transport code System
  (PHITS) version 3.02}},  {\em J. Nucl. Sci. Tech.} {\bf 55} (2018), no.~6
  684--690.

\bibitem{iwamoto2017benchmark}
Y.~Iwamoto, T.~Sato, S.~Hashimoto, T.~Ogawa, T.~Furuta, S.-i. Abe, T.~Kai,
  N.~Matsuda, R.~Hosoyamada, and K.~Niita, {\it Benchmark study of the recent
  version of the phits code},  {\em Journal of Nuclear Science and Technology}
  {\bf 54} (2017), no.~5 617--635.

\bibitem{matsuda2011benchmarking}
N.~Matsuda, Y.~Iwamoto, Y.~Sakamoto, H.~Nakashima, H.~Iwase, and K.~Niita, {\it
  Benchmarking of phits on pion production for medium-energy physics},  {\em
  Progress in Nuclear Science and Technology} {\bf 2} (2011) 927--930.

\bibitem{Bierlich:2022pfr}
C.~Bierlich et~al., {\it {A comprehensive guide to the physics and usage of
  PYTHIA 8.3}},  \href{http://arxiv.org/abs/2203.11601}{{\tt
  arXiv:2203.11601}}.

\bibitem{Hirayama:2005zm}
H.~Hirayama, Y.~Namito, A.~Bielajew, S.~Wilderman, and W.~Nelson, {\it {The
  EGS5 code system}}, . SLAC-R-730, KEK-2005-8, KEK-REPORT-2005-8.

\bibitem{Nara:1999dz}
Y.~Nara, N.~Otuka, A.~Ohnishi, K.~Niita, and S.~Chiba, {\it {Study of
  relativistic nuclear collisions at AGS energies from p + Be to Au + Au with
  hadronic cascade model}},  {\em Phys. Rev. C} {\bf 61} (2000) 024901,
  [\href{http://arxiv.org/abs/nucl-th/9904059}{{\tt nucl-th/9904059}}].

\bibitem{PhysRevC.52.2620}
K.~Niita, S.~Chiba, T.~Maruyama, T.~Maruyama, H.~Takada, T.~Fukahori,
  Y.~Nakahara, and A.~Iwamoto, {\it Analysis of the
  (n,${\mathit{xn}}^{\ensuremath{'}}$) reactions by quantum molecular dynamics
  plus statistical decay model},  {\em Phys. Rev. C} {\bf 52} (Nov, 1995)
  2620--2635.

\bibitem{PhysRevC.87.014606}
A.~Boudard, J.~Cugnon, J.-C. David, S.~Leray, and D.~Mancusi, {\it New
  potentialities of the li\`ege intranuclear cascade model for reactions
  induced by nucleons and light charged particles},  {\em Phys. Rev. C} {\bf
  87} (Jan, 2013) 014606.

\bibitem{ATIMA}
``Atima.'' \url{https://web-docs.gsi.de/~weick/atima/}.

\bibitem{SLACHybridFacilityPhoton:1985baw}
{\bf SLAC Hybrid Facility Photon} Collaboration, K.~Abe et~al., {\it
  {Lifetimes, Cross-sections and Production Mechanisms of Charmed Particles
  Produced by 20-{GeV} Photons}},  {\em Phys. Rev. D} {\bf 33} (1986) 1.

\bibitem{TaggedPhotonSpectrometer:1989bpi}
{\bf Tagged Photon Spectrometer} Collaboration, J.~C. Anjos et~al., {\it {Charm
  photoproduction}},  {\em Phys. Rev. Lett.} {\bf 62} (1989) 513--516.

\bibitem{Caldwell:1978ik}
D.~O. Caldwell et~al., {\it {Measurement of Shadowing in Photon - Nucleus Total
  Cross-sections From 20-{GeV} to 185-{GeV}}},  {\em Phys. Rev. Lett.} {\bf 42}
  (1979) 553.

\bibitem{Kopeliovich:2012kw}
B.~Z. Kopeliovich, J.~G. Morfin, and I.~Schmidt, {\it {Nuclear Shadowing in
  Electro-Weak Interactions}},  {\em Prog. Part. Nucl. Phys.} {\bf 68} (2013)
  314--372, [\href{http://arxiv.org/abs/1208.6541}{{\tt arXiv:1208.6541}}].

\bibitem{2018}
K.~Bondarenko, A.~Boyarsky, D.~Gorbunov, and O.~Ruchayskiy, {\it Phenomenology
  of gev-scale heavy neutral leptons},  {\em Journal of High Energy Physics}
  {\bf 2018} (Nov, 2018).

\bibitem{Tsai:1973py}
Y.-S. Tsai, {\it {Pair Production and Bremsstrahlung of Charged Leptons}},
  {\em Rev. Mod. Phys.} {\bf 46} (1974) 815. [Erratum: Rev.Mod.Phys. 49,
  421--423 (1977)].

\bibitem{Sakaki:2020cux}
Y.~Sakaki, Y.~Namito, T.~Sanami, H.~Iwase, and H.~Hirayama, {\it
  {Implementation of muon pair production in PHITS and verification by
  comparing with the muon shielding experiment at SLAC}},  {\em Nucl. Instrum.
  Meth. A} {\bf 977} (2020) 164323,
  [\href{http://arxiv.org/abs/2004.00212}{{\tt arXiv:2004.00212}}].

\bibitem{ParticleDataGroup:2020ssz}
{\bf Particle Data Group} Collaboration, P.~A. Zyla et~al., {\it {Review of
  Particle Physics}},  {\em PTEP} {\bf 2020} (2020), no.~8 083C01.

\bibitem{Boyarsky:2020dzc}
A.~Boyarsky, M.~Ovchynnikov, O.~Ruchayskiy, and V.~Syvolap, {\it {Improved big
  bang nucleosynthesis constraints on heavy neutral leptons}},  {\em Phys. Rev.
  D} {\bf 104} (2021), no.~2 023517,
  [\href{http://arxiv.org/abs/2008.00749}{{\tt arXiv:2008.00749}}].

\bibitem{Bonivento:2013jag}
W.~Bonivento et~al., {\it {Proposal to Search for Heavy Neutral Leptons at the
  SPS}},  \href{http://arxiv.org/abs/1310.1762}{{\tt arXiv:1310.1762}}.

\bibitem{SHiP:2015vad}
{\bf SHiP} Collaboration, M.~Anelli et~al., {\it {A facility to Search for
  Hidden Particles (SHiP) at the CERN SPS}},
  \href{http://arxiv.org/abs/1504.04956}{{\tt arXiv:1504.04956}}.

\bibitem{Shrock:1980vy}
R.~E. Shrock, {\it {New Tests For, and Bounds On, Neutrino Masses and Lepton
  Mixing}},  {\em Phys. Lett. B} {\bf 96} (1980) 159--164.

\bibitem{Shrock:1980ct}
R.~E. Shrock, {\it {General Theory of Weak Leptonic and Semileptonic Decays. 1.
  Leptonic Pseudoscalar Meson Decays, with Associated Tests For, and Bounds on,
  Neutrino Masses and Lepton Mixing}},  {\em Phys. Rev. D} {\bf 24} (1981)
  1232.

\bibitem{Shrock:1981wq}
R.~E. Shrock, {\it {General Theory of Weak Processes Involving Neutrinos. 2.
  Pure Leptonic Decays}},  {\em Phys. Rev. D} {\bf 24} (1981) 1275.

\bibitem{Johnson:1997cj}
L.~M. Johnson, D.~W. McKay, and T.~Bolton, {\it {Extending sensitivity for low
  mass neutral heavy lepton searches}},  {\em Phys. Rev. D} {\bf 56} (1997)
  2970--2981, [\href{http://arxiv.org/abs/hep-ph/9703333}{{\tt
  hep-ph/9703333}}].

\bibitem{Gorbunov:2007ak}
D.~Gorbunov and M.~Shaposhnikov, {\it {How to find neutral leptons of the
  $\nu$MSM?}},  {\em JHEP} {\bf 10} (2007) 015,
  [\href{http://arxiv.org/abs/0705.1729}{{\tt arXiv:0705.1729}}]. [Erratum:
  JHEP 11, 101 (2013)].

\bibitem{Formaggio:2012cpf}
J.~A. Formaggio and G.~P. Zeller, {\it {From eV to EeV: Neutrino Cross Sections
  Across Energy Scales}},  {\em Rev. Mod. Phys.} {\bf 84} (2012) 1307--1341,
  [\href{http://arxiv.org/abs/1305.7513}{{\tt arXiv:1305.7513}}].

\bibitem{Lynch:1990sq}
G.~R. Lynch and O.~I. Dahl, {\it {Approximations to multiple Coulomb
  scattering}},  {\em Nucl. Instrum. Meth. B} {\bf 58} (1991) 6--10.

\bibitem{Coloma:2020lgy}
P.~Coloma, E.~Fern\'andez-Mart\'\i{}nez, M.~Gonz\'alez-L\'opez,
  J.~Hern\'andez-Garc\'\i{}a, and Z.~Pavlovic, {\it {GeV-scale neutrinos:
  interactions with mesons and DUNE sensitivity}},  {\em Eur. Phys. J. C} {\bf
  81} (2021), no.~1 78, [\href{http://arxiv.org/abs/2007.03701}{{\tt
  arXiv:2007.03701}}].

\bibitem{Kling:2018wct}
F.~Kling and S.~Trojanowski, {\it {Heavy Neutral Leptons at FASER}},  {\em
  Phys. Rev. D} {\bf 97} (2018), no.~9 095016,
  [\href{http://arxiv.org/abs/1801.08947}{{\tt arXiv:1801.08947}}].

\bibitem{Curtin:2018mvb}
D.~Curtin et~al., {\it {Long-Lived Particles at the Energy Frontier: The
  MATHUSLA Physics Case}},  {\em Rept. Prog. Phys.} {\bf 82} (2019), no.~11
  116201, [\href{http://arxiv.org/abs/1806.07396}{{\tt arXiv:1806.07396}}].

\bibitem{DELPHI:1996qcc}
{\bf DELPHI} Collaboration, P.~Abreu et~al., {\it {Search for neutral heavy
  leptons produced in Z decays}},  {\em Z. Phys. C} {\bf 74} (1997) 57--71.
  [Erratum: Z.Phys.C 75, 580 (1997)].

\bibitem{Behnke:2013lya}
H.~Abramowicz et~al., {\it {The International Linear Collider Technical Design
  Report - Volume 4: Detectors}},  \href{http://arxiv.org/abs/1306.6329}{{\tt
  arXiv:1306.6329}}.

\bibitem{Bertholet:2021hjl}
E.~Bertholet, S.~Chakraborty, V.~Loladze, T.~Okui, A.~Soffer, and K.~Tobioka,
  {\it {Heavy QCD axion at Belle II: Displaced and prompt signals}},  {\em
  Phys. Rev. D} {\bf 105} (2022), no.~7 L071701,
  [\href{http://arxiv.org/abs/2108.10331}{{\tt arXiv:2108.10331}}].

\bibitem{Blondel:2014bra}
{\bf FCC-ee study Team} Collaboration, A.~Blondel, E.~Graverini, N.~Serra, and
  M.~Shaposhnikov, {\it {Search for Heavy Right Handed Neutrinos at the
  FCC-ee}},  {\em Nucl. Part. Phys. Proc.} {\bf 273-275} (2016) 1883--1890,
  [\href{http://arxiv.org/abs/1411.5230}{{\tt arXiv:1411.5230}}].

\bibitem{Abazajian:2012ys}
K.~N. Abazajian et~al., {\it {Light Sterile Neutrinos: A White Paper}},
  \href{http://arxiv.org/abs/1204.5379}{{\tt arXiv:1204.5379}}.

\bibitem{Alonso-Alvarez:2022uxp}
G.~Alonso-\'Alvarez and J.~M. Cline, {\it {Sterile neutrino production at small
  mixing in the early universe}},  \href{http://arxiv.org/abs/2204.04224}{{\tt
  arXiv:2204.04224}}.

\bibitem{Bondarenko:2021cpc}
K.~Bondarenko, A.~Boyarsky, J.~Klaric, O.~Mikulenko, O.~Ruchayskiy, V.~Syvolap,
  and I.~Timiryasov, {\it {An allowed window for heavy neutral leptons below
  the kaon mass}},  {\em JHEP} {\bf 07} (2021) 193,
  [\href{http://arxiv.org/abs/2101.09255}{{\tt arXiv:2101.09255}}].

\bibitem{CHARM:1983ayi}
{\bf CHARM} Collaboration, F.~Bergsma et~al., {\it {A Search for Decays of
  Heavy Neutrinos}},  {\em Phys. Lett. B} {\bf 128} (1983) 361.

\bibitem{CHARM:1985nku}
{\bf CHARM} Collaboration, F.~Bergsma et~al., {\it {A Search for Decays of
  Heavy Neutrinos in the Mass Range 0.5-{GeV} to 2.8-{GeV}}},  {\em Phys. Lett.
  B} {\bf 166} (1986) 473--478.

\bibitem{Boiarska:2021yho}
I.~Boiarska, A.~Boyarsky, O.~Mikulenko, and M.~Ovchynnikov, {\it {Constraints
  from the CHARM experiment on heavy neutral leptons with tau mixing}},  {\em
  Phys. Rev. D} {\bf 104} (2021), no.~9 095019,
  [\href{http://arxiv.org/abs/2107.14685}{{\tt arXiv:2107.14685}}].

\bibitem{NuTeV:1999kej}
{\bf NuTeV, E815} Collaboration, A.~Vaitaitis et~al., {\it {Search for neutral
  heavy leptons in a high-energy neutrino beam}},  {\em Phys. Rev. Lett.} {\bf
  83} (1999) 4943--4946, [\href{http://arxiv.org/abs/hep-ex/9908011}{{\tt
  hep-ex/9908011}}].

\bibitem{WA66:1985mfx}
{\bf WA66} Collaboration, A.~M. Cooper-Sarkar et~al., {\it {Search for Heavy
  Neutrino Decays in the {BEBC} Beam Dump Experiment}},  {\em Phys. Lett. B}
  {\bf 160} (1985) 207--211.

\bibitem{Bernardi:1987ek}
G.~Bernardi et~al., {\it {FURTHER LIMITS ON HEAVY NEUTRINO COUPLINGS}},  {\em
  Phys. Lett. B} {\bf 203} (1988) 332--334.

\bibitem{T2K:2019jwa}
{\bf T2K} Collaboration, K.~Abe et~al., {\it {Search for heavy neutrinos with
  the T2K near detector ND280}},  {\em Phys. Rev. D} {\bf 100} (2019), no.~5
  052006, [\href{http://arxiv.org/abs/1902.07598}{{\tt arXiv:1902.07598}}].

\bibitem{Coloma:2019htx}
P.~Coloma, P.~Hern\'andez, V.~Mu\~noz, and I.~M. Shoemaker, {\it {New
  constraints on Heavy Neutral Leptons from Super-Kamiokande data}},  {\em Eur.
  Phys. J. C} {\bf 80} (2020), no.~3 235,
  [\href{http://arxiv.org/abs/1911.09129}{{\tt arXiv:1911.09129}}].

\bibitem{Super-Kamiokande:2017yvm}
{\bf Super-Kamiokande} Collaboration, K.~Abe et~al., {\it {Atmospheric neutrino
  oscillation analysis with external constraints in Super-Kamiokande I-IV}},
  {\em Phys. Rev. D} {\bf 97} (2018), no.~7 072001,
  [\href{http://arxiv.org/abs/1710.09126}{{\tt arXiv:1710.09126}}].

\bibitem{PIENU:2017wbj}
{\bf PIENU} Collaboration, A.~Aguilar-Arevalo et~al., {\it {Improved search for
  heavy neutrinos in the decay $\pi\rightarrow e\nu$}},  {\em Phys. Rev. D}
  {\bf 97} (2018), no.~7 072012, [\href{http://arxiv.org/abs/1712.03275}{{\tt
  arXiv:1712.03275}}].

\bibitem{PIENU:2019usb}
{\bf PIENU} Collaboration, A.~Aguilar-Arevalo et~al., {\it {Search for heavy
  neutrinos in $\pi \to \mu\nu$ decay}},  {\em Phys. Lett. B} {\bf 798} (2019)
  134980, [\href{http://arxiv.org/abs/1904.03269}{{\tt arXiv:1904.03269}}].

\bibitem{NA62:2020mcv}
{\bf NA62} Collaboration, E.~Cortina~Gil et~al., {\it {Search for heavy neutral
  lepton production in $K^+$ decays to positrons}},  {\em Phys. Lett. B} {\bf
  807} (2020) 135599, [\href{http://arxiv.org/abs/2005.09575}{{\tt
  arXiv:2005.09575}}].

\bibitem{NA62:2021bji}
{\bf NA62} Collaboration, E.~Cortina~Gil et~al., {\it {Search for $K^+$ decays
  to a muon and invisible particles}},  {\em Phys. Lett. B} {\bf 816} (2021)
  136259, [\href{http://arxiv.org/abs/2101.12304}{{\tt arXiv:2101.12304}}].

\bibitem{E949:2014gsn}
{\bf E949} Collaboration, A.~V. Artamonov et~al., {\it {Search for heavy
  neutrinos in $K^+\to\mu^+\nu_H$ decays}},  {\em Phys. Rev. D} {\bf 91}
  (2015), no.~5 052001, [\href{http://arxiv.org/abs/1411.3963}{{\tt
  arXiv:1411.3963}}]. [Erratum: Phys.Rev.D 91, 059903 (2015)].

\bibitem{Yamazaki:1984sj}
T.~Yamazaki et~al., {\it {Search for Heavy Neutrinos in Kaon Decay}},  {\em
  Conf. Proc. C} {\bf 840719} (1984) 262.

\bibitem{Belle:2013ytx}
{\bf Belle} Collaboration, D.~Liventsev et~al., {\it {Search for heavy
  neutrinos at Belle}},  {\em Phys. Rev. D} {\bf 87} (2013), no.~7 071102,
  [\href{http://arxiv.org/abs/1301.1105}{{\tt arXiv:1301.1105}}]. [Erratum:
  Phys.Rev.D 95, 099903 (2017)].

\bibitem{Dib:2019tuj}
C.~O. Dib, J.~C. Helo, M.~Nayak, N.~A. Neill, A.~Soffer, and J.~Zamora-Saa,
  {\it {Searching for a sterile neutrino that mixes predominantly with
  $\nu_\tau$ at $B$ factories}},  {\em Phys. Rev. D} {\bf 101} (2020), no.~9
  093003, [\href{http://arxiv.org/abs/1908.09719}{{\tt arXiv:1908.09719}}].

\bibitem{ATLAS:2019kpx}
{\bf ATLAS} Collaboration, G.~Aad et~al., {\it {Search for heavy neutral
  leptons in decays of $W$ bosons produced in 13 TeV $pp$ collisions using
  prompt and displaced signatures with the ATLAS detector}},  {\em JHEP} {\bf
  10} (2019) 265, [\href{http://arxiv.org/abs/1905.09787}{{\tt
  arXiv:1905.09787}}].

\bibitem{ATLAS:2022atq}
{\bf ATLAS} Collaboration, {\it {Search for heavy neutral leptons in decays of
  $W$ bosons using a dilepton displaced vertex in $\sqrt{s}=13$ TeV $pp$
  collisions with the ATLAS detector}},
  \href{http://arxiv.org/abs/2204.11988}{{\tt arXiv:2204.11988}}.

\bibitem{CMS:2018jxx}
{\bf CMS} Collaboration, A.~M. Sirunyan et~al., {\it {Search for heavy Majorana
  neutrinos in same-sign dilepton channels in proton-proton collisions at $
  \sqrt{s}=13 $ TeV}},  {\em JHEP} {\bf 01} (2019) 122,
  [\href{http://arxiv.org/abs/1806.10905}{{\tt arXiv:1806.10905}}].

\bibitem{CMS:2018iaf}
{\bf CMS} Collaboration, A.~M. Sirunyan et~al., {\it {Search for heavy neutral
  leptons in events with three charged leptons in proton-proton collisions at
  $\sqrt{s} =$ 13 TeV}},  {\em Phys. Rev. Lett.} {\bf 120} (2018), no.~22
  221801, [\href{http://arxiv.org/abs/1802.02965}{{\tt arXiv:1802.02965}}].

\bibitem{CMS:2022fut}
{\bf CMS} Collaboration, A.~Tumasyan et~al., {\it {Search for long-lived heavy
  neutral leptons with displaced vertices in proton-proton collisions at
  $\sqrt{s}$ =13 TeV}},  \href{http://arxiv.org/abs/2201.05578}{{\tt
  arXiv:2201.05578}}.

\bibitem{FASER:2018eoc}
{\bf FASER} Collaboration, A.~Ariga et~al., {\it {FASER\textquoteright{}s
  physics reach for long-lived particles}},  {\em Phys. Rev. D} {\bf 99}
  (2019), no.~9 095011, [\href{http://arxiv.org/abs/1811.12522}{{\tt
  arXiv:1811.12522}}].

\bibitem{Antusch:2017hhu}
S.~Antusch, E.~Cazzato, and O.~Fischer, {\it {Sterile neutrino searches via
  displaced vertices at LHCb}},  {\em Phys. Lett. B} {\bf 774} (2017) 114--118,
  [\href{http://arxiv.org/abs/1706.05990}{{\tt arXiv:1706.05990}}].

\bibitem{Cvetic:2019shl}
G.~Cveti\v{c} and C.~S. Kim, {\it {Sensitivity bounds on heavy neutrino mixing
  $|U_{\mu N}|^2$ and $|U_{\tau N}|^2$ from LHCb upgrade}},  {\em Phys. Rev. D}
  {\bf 100} (2019), no.~1 015014, [\href{http://arxiv.org/abs/1904.12858}{{\tt
  arXiv:1904.12858}}].

\bibitem{Aielli:2019ivi}
G.~Aielli et~al., {\it {Expression of interest for the CODEX-b detector}},
  {\em Eur. Phys. J. C} {\bf 80} (2020), no.~12 1177,
  [\href{http://arxiv.org/abs/1911.00481}{{\tt arXiv:1911.00481}}].

\bibitem{Batell:2020vqn}
B.~Batell, J.~A. Evans, S.~Gori, and M.~Rai, {\it {Dark Scalars and Heavy
  Neutral Leptons at DarkQuest}},  {\em JHEP} {\bf 05} (2021) 049,
  [\href{http://arxiv.org/abs/2008.08108}{{\tt arXiv:2008.08108}}].

\bibitem{Coloma:2017ppo}
P.~Coloma, P.~A.~N. Machado, I.~Martinez-Soler, and I.~M. Shoemaker, {\it
  {Double-Cascade Events from New Physics in Icecube}},  {\em Phys. Rev. Lett.}
  {\bf 119} (2017), no.~20 201804, [\href{http://arxiv.org/abs/1707.08573}{{\tt
  arXiv:1707.08573}}].

\bibitem{Cheung:2020buy}
K.~Cheung, Y.-L. Chung, H.~Ishida, and C.-T. Lu, {\it {Sensitivity reach on
  heavy neutral leptons and $\tau$-neutrino mixing $|U_{\tau N}|^2 $ at the
  HL-LHC}},  {\em Phys. Rev. D} {\bf 102} (2020), no.~7 075038,
  [\href{http://arxiv.org/abs/2004.11537}{{\tt arXiv:2004.11537}}].

\bibitem{Boyarsky:2022epg}
A.~Boyarsky, O.~Mikulenko, M.~Ovchynnikov, and L.~Shchutska, {\it {Exploring
  the potential of FCC-hh to search for particles from $B$ mesons}},
  \href{http://arxiv.org/abs/2204.01622}{{\tt arXiv:2204.01622}}.

\bibitem{Giffin:2022rei}
P.~Giffin, S.~Gori, Y.-D. Tsai, and D.~Tuckler, {\it {Heavy Neutral Leptons at
  Beam Dump Experiments of Future Lepton Colliders}},
  \href{http://arxiv.org/abs/2206.13745}{{\tt arXiv:2206.13745}}.

\bibitem{Albright:1974ts}
C.~H. Albright and C.~Jarlskog, {\it {Neutrino Production of m+ and e+ Heavy
  Leptons. 1.}},  {\em Nucl. Phys. B} {\bf 84} (1975) 467--492.

\bibitem{Kamal:1979ev}
A.~N. Kamal and J.~N. Ng, {\it {CONSTRAINTS ON HEAVY LEPTON MIXINGS FROM DEEP
  INELASTIC CHARGED LEPTON SCATTERING}},  {\em Phys. Rev. D} {\bf 21} (1980)
  1224.

\bibitem{Anselmino:1993tc}
M.~Anselmino, P.~Gambino, and J.~Kalinowski, {\it {Polarized deep inelastic
  scattering at high-energies and parity violating structure functions}},  {\em
  Z. Phys. C} {\bf 64} (1994) 267--274,
  [\href{http://arxiv.org/abs/hep-ph/9401264}{{\tt hep-ph/9401264}}].

\bibitem{Grover:2018ggi}
D.~Grover, K.~Saraswat, P.~Shukla, and V.~Singh, {\it {Charged current deep
  inelastic scattering of $\nu_\mu$ off $^{56}$Fe}},  {\em Phys. Rev. C} {\bf
  98} (2018), no.~6 065503, [\href{http://arxiv.org/abs/1808.00287}{{\tt
  arXiv:1808.00287}}].

\bibitem{Kretzer:2002fr}
S.~Kretzer and M.~H. Reno, {\it {Tau neutrino deep inelastic charged current
  interactions}},  {\em Phys. Rev. D} {\bf 66} (2002) 113007,
  [\href{http://arxiv.org/abs/hep-ph/0208187}{{\tt hep-ph/0208187}}].

\bibitem{Callan:1969uq}
C.~G. Callan, Jr. and D.~J. Gross, {\it {High-energy electroproduction and the
  constitution of the electric current}},  {\em Phys. Rev. Lett.} {\bf 22}
  (1969) 156--159.

\bibitem{Martin:2009iq}
A.~D. Martin, W.~J. Stirling, R.~S. Thorne, and G.~Watt, {\it {Parton
  distributions for the LHC}},  {\em Eur. Phys. J. C} {\bf 63} (2009) 189--285,
  [\href{http://arxiv.org/abs/0901.0002}{{\tt arXiv:0901.0002}}].

\end{thebibliography}\endgroup

\end{document}